\documentclass[sigplan,screen]{acmart}

\usepackage{multirow}
\usepackage{array} 
\usepackage{tcolorbox}
\usepackage[edges]{forest}
\useforestlibrary{edges} 
\usepackage{booktabs}
\usepackage{graphicx}
\usepackage[normalem]{ulem}
\useunder{\uline}{\ul}{}
\usepackage{multirow}

\setcopyright{none}
\renewcommand\footnotetextcopyrightpermission[1]{}
\AtBeginDocument{%
  }

\begin{document}

\title{Synergizing RAG and Reasoning: A Systematic Review}

\author{Yunfan Gao}

\affiliation{%
  \institution{Shanghai Research Institute for Intelligent Autonomous Systems, Tongji University}
  \country{China}
}
\email{gaoyunfan1602@gmail.com}

\author{Yun Xiong}
\affiliation{%
  \institution{Shanghai Key Laboratory of Data Science, School of Computer Science, Fudan University}
    \country{China}
    % \country{China}
  % \city{Hekla}
  % \country{Iceland}}
  }
\email{yunx@fudan.edu.cn}

\author{Yijie Zhong}
\affiliation{%
  \institution{College of Design and Innovation, Tongji University}
    \country{China}
    % \country{China}
    }
\email{dun.haski@gmail.com}

\author{Yuxi Bi}
\affiliation{%
  \institution{College of Design and Innovation, Tongji University}
    \country{China}
    % \country{China}
    }
\email{yuxibi@gmail.com}

\author{Ming Xue}
\affiliation{%
  \institution{Percena AI}
    \country{China}
    % \country{China}
    }
\email{mxue@percena.co}

\author{Haofen Wang}
\authornote{Corresponding Author}
\affiliation{%
  \institution{College of Design and Innovation, Tongji University}
    \country{China}
    % \country{China}
    }
\email{carter.whfcarter@gmail.com}

% \author{}
% \affiliation{%

% \author{}
% \affiliation{%

\renewcommand{\shortauthors}{Gao et al.}


\begin{abstract}

Recent breakthroughs in large language models (LLMs), particularly in reasoning capabilities, have propelled Retrieval-Augmented Generation (RAG) to unprecedented levels. By synergizing retrieval mechanisms with advanced reasoning, LLMs can now tackle increasingly complex problems. This paper presents a systematic review of the collaborative interplay between RAG and reasoning, clearly defining "reasoning" within the RAG context. It construct a comprehensive taxonomy encompassing multi-dimensional collaborative objectives, representative paradigms, and technical implementations, and analyze the bidirectional synergy methods. Additionally, we critically evaluate current limitations in RAG assessment, including the absence of intermediate supervision for multi-step reasoning and practical challenges related to cost-risk trade-offs. To bridge theory and practice, we provide practical guidelines tailored to diverse real-world applications. Finally, we identify promising research directions, such as graph-based knowledge integration, hybrid model collaboration, and RL-driven optimization. Overall, this work presents a theoretical framework and practical foundation to advance RAG systems in academia and industry, fostering the next generation of RAG solutions.

\end{abstract}

\maketitle

\begin{figure*}[htbp]
    \centering
    \includegraphics[width=1\linewidth]{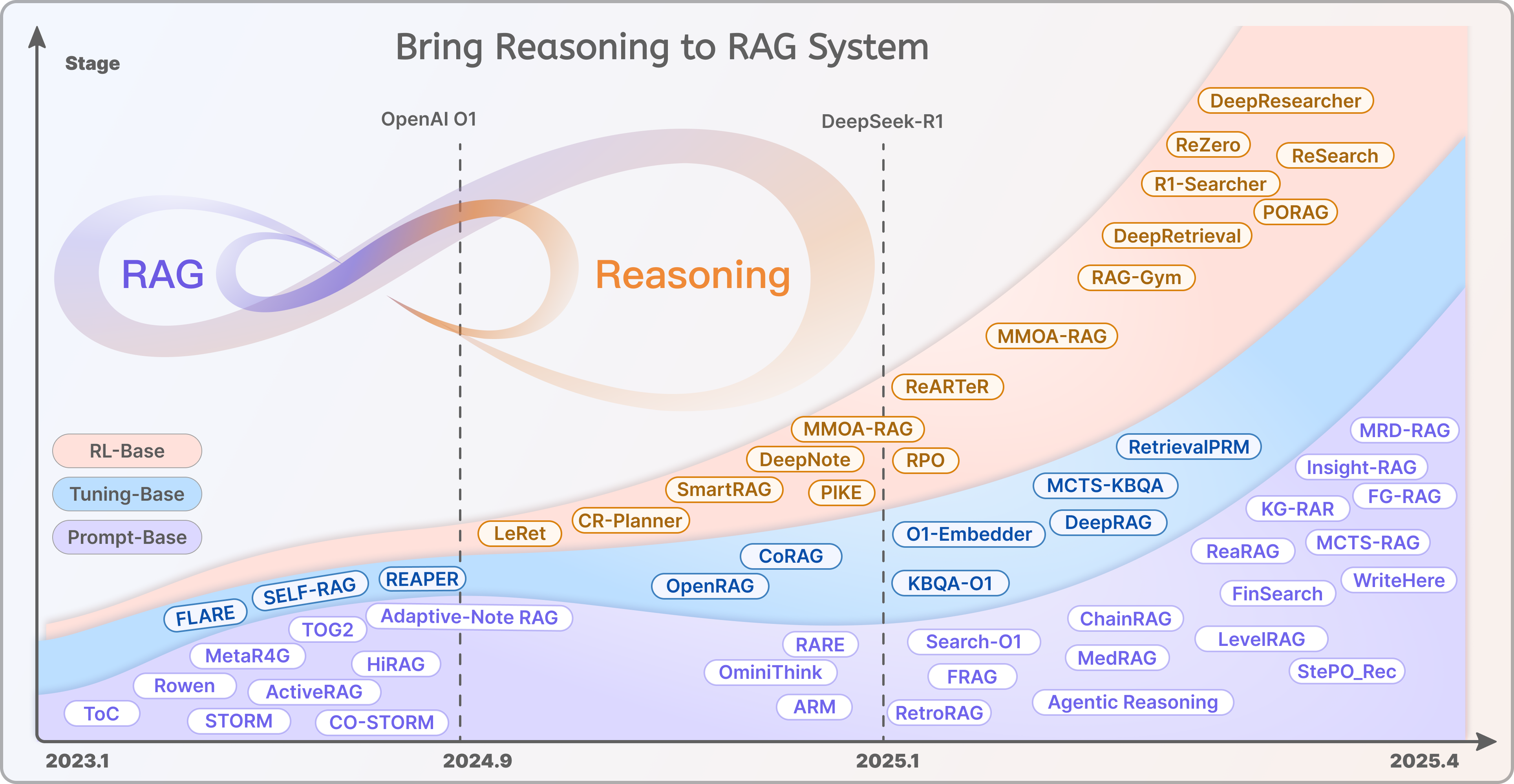}
    \caption{Timeline of studies on RAG-reasoning synergy. From a technical perspective, the approaches can be categorized into Prompt-Based, Tuning-Based, and RL-Based methods. A notable trend is the increasing use of Reinforcement Learning to enhance RAG systems, particularly following the prosperity of test-time scaling. Meanwhile, Prompt-Based and Tuning-Based methods continue to evolve in parallel, demonstrating that there are multiple pathways to integrating reasoning capabilities into RAG systems.}
    \label{fig:timeline}
\end{figure*}
\section{Introduction}

Recent breakthroughs in Large Language Models (LLMs) like OpenAI O1~\cite{O1} and DeepSeek-R1~\cite{deepseek-r1} have shifted the paradigm from "pre-training scaling" to "test-time scaling"~\cite{s1}. Unlike traditional language models that improve via corpus accumulation during pre-training, these models enhance performance in complex tasks—such as mathematical derivation and code generation~\cite{olympiadbench}—through post-training innovations during the inference phase (e.g., Long-CoT thinking~\cite{long_cot}). This shift has led to the emergence of "Large Reasoning Models" (LRMs)~\cite{lrm_survey} with advanced internal reasoning abilities.

These advancements have not only boosted basic model capabilities but also opened new avenues for application technologies like Retrieval-Augmented Generation (RAG)~\cite{retrieval_survey}. Serving as a key link between language models and external knowledge, RAG overcomes traditional LLMs' limits in knowledge freshness, domain specificity, and factual accuracy by retrieving real-time non-parametric information and integrating it into the context. This enhances information processing and reduces hallucination risks in knowledge-intensive tasks.

Technological evolution is advancing RAG architectures through innovations like query rewriting~\cite{rrr}, re-ranking~\cite{rankify}, and hybrid retrieval~\cite{searching}, creating an Advanced RAG paradigm focused on pre-retrieval optimization and post-retrieval refinement. Modular RAG~\cite{modularRAG} further breaks down these systems into component-based, service-oriented architectures, using orchestration to tackle practical challenges.

Despite improvements in query intent recognition and knowledge use, challenges of RAG remain in demanding tasks like deep research and complex decision-making. Key issues include: 1) difficulty capturing intent from ambiguous queries; 2) poor logical coherence in multi-hop reasoning; 3) efficiency limits of traditional retrieval in open domains; and 4) degraded generation quality from noisy retrieved data.

Models like DeepSeek-R1, with strong reasoning capabilities, inspire new directions for RAG systems. As shown in Figure~\ref{fig:timeline}, recent research explores integrating formal reasoning frameworks with knowledge retrieval. This approach optimizes retrieval through logic-driven query reformulation and uses reasoning to analyze and validate retrieved knowledge, creating cognitive synergy between retrieval and generation. This paradigm aims to overcome conventional limitations, enabling intelligent systems with rigorous logic and reliable knowledge use. From a trend perspective, an increasing number of methods combine reasoning and retrieval abilities through reinforcement learning (RL), marking a new direction in the LRM era. Meanwhile, prompt-based approaches continue to rapidly evolve, with researchers aiming to achieve results through workflow design while keeping model parameters frozen. Notably, sole reliance on tuning methods is steadily decreasing, suggesting limited improvements from additional fine-tuning at this stage.

\begin{figure*}[htbp]
    \centering
    \includegraphics[width=1\linewidth]{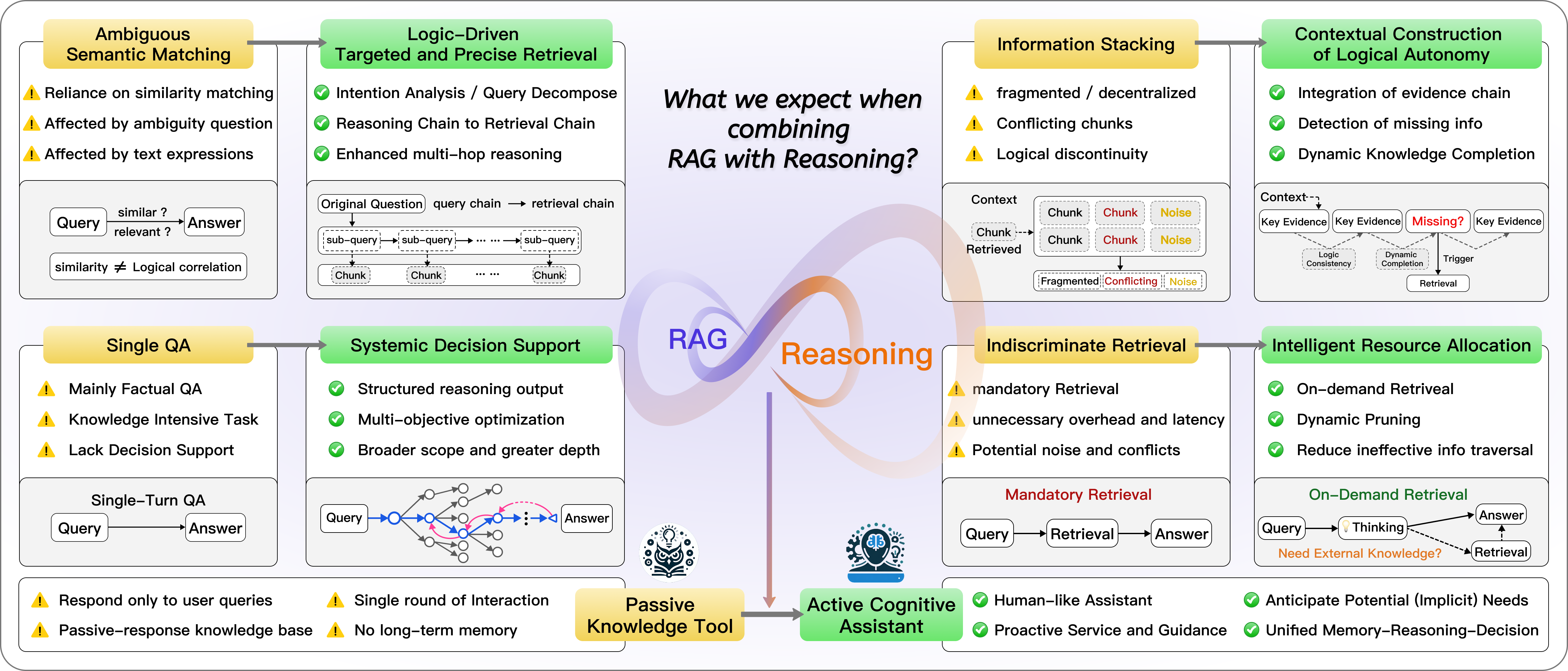}
    \caption{Advantages of Combining RAG with Reasoning}
    \label{fig:advantage}
\end{figure*}

Traditional RAG is limited by its unidirectional flow (retrieval → generation). Integrating reasoning capabilities grants the system greater autonomy, unlocking new possibilities. As shown in Figure~\ref{fig:advantage}, this integration is poised to drive major breakthroughs, enabling practical use in complex real-world scenarios.

\textbf{1) From Ambiguous Semantic Matching to Logic-Driven Targeted Retrieval.} Traditional RAG relies on semantic similarity for retrieval; however, it is sensitive to phrasing variations. Advanced reasoning allows deep logical analysis of queries (e.g., causal links, conditional constraints) to dynamically refine retrieval strategies~\cite{deeprag}. For example, to answer "How to reduce postoperative infection risks in diabetes patients?", the system prioritizes retrieving "blood glucose control thresholds" and "antibiotic usage guidelines" over simply matching "diabetes postoperative care". This approach supports multi-hop retrieval by breaking down complex queries into sequential sub-queries while preserving cross-document coherence through reasoning chains.

\textbf{2) From  Simple Information Aggregation to Logically Coherent  Context Construction.} Current RAG systems input all retrieved document chunks into context directly, often causing fragmented or contradictory information that confuses LLMs. Reasoning-enhanced systems integrate evidence chains by logically verifying and inferring causality in retrieved content, filtering conflicts and forming coherent explanations~\cite{activerag}. They also use dynamic knowledge completion to detect missing logical links, prompting iterative retrieval or inference to fill gaps~\cite{search_o1}.

\textbf{3) From Simple and Single-Turn QA to Systemic Decision Support.} Traditional RAG performs well in factual QA~\cite{kilt} but struggles with multi-step and complex decision-making. Reasoning-integrated systems produce structured reasoning output, enhancing multi-objective optimization to balance retrieval breadth and solution feasibility under various constraints. For example, multiple constraints under different conditions in engineering construction plans~\cite{li2025deepsolution}, and the formulation of diagnosis and treatment plans for various diseases in the medical field~\cite{medrag}.

\textbf{4) From Indiscriminate  Retrieval to Intelligent Resource Allocation.} Traditional RAG retrieves documents for all queries, regardless of complexity. Reasoning-enhanced systems use on-demand retrieval, handling simple queries with direct generation and complex ones with multi-round retrieval to reduce latency~\cite{smartrag}. Dynamic retrieval pruning uses pre-reasoning predictions to target key information, minimizing unnecessary document and graph  traversal~\cite{adaptiveRAG}.

\textbf{5) From Passive Knowledge Tool to Proactive Cognitive Assistant.} Advancing beyond reactive knowledge retrieval, reasoning-enhanced systems can proactively serve users by asking clarifying questions and anticipating implicit needs. This shift enables human-like assistants that integrate memory, reasoning, and decision-making, proving especially valuable for complex tasks such as deep research~\cite{co-storm}, business analytics~\cite{FinSearch}, personal assistant~\cite{zhong2025meta} and urban planning~\cite{wang2024decoding}.

However, the synergistic pathway between RAG and reasoning requires more than simply replacing conventional generative LLMs with LRM modules. It necessitates deep integration of technological evolution insights from LRM - achieved through reconstructing knowledge retrieval mechanisms and strengthening reasoning-generation collaborative linkages - to enable system-level enhancement of cognitive capabilities within the RAG architecture.

Therefore, this paper aims to address the pivotal and forward-looking research question of "how RAG systems can synergize with reasoning capabilities". We systematically review current studies after 2024  while establishing explicit definitions for reasoning within RAG contexts. Building on this foundation, we provide an in-depth taxonomy and analysis of the objectives, typical patterns, and  implementations underlying RAG-reasoning integration, clarifying key technological trajectories and critical breakthroughs.

As RAG technology enters its next developmental phase, downstream task complexity has escalated significantly - particularly evident in emerging challenges like Deep Research~\cite{deepresearcher}. These advanced applications not only demand enhanced reasoning capacities but also drive RAG's expansion into multimodal, cross-domain, and dynamic environments. However, while the integration of reasoning capabilities demonstrably improves complex task performance, existing research frequently overlooks associated computational overheads and potential risks. Through systematic examination of these operational constraints and analysis of industry applications, we propose practical guidelines for multiple real-world scenarios with diverse requirements.

Finally, we outline future research directions grounded in current technological evolution, including: 1) RAG-graph architecture integration, 2) coordinated multimodal reasoning frameworks, 3) hybrid  model collaboration, and 4) RL optimization specifically designed for RAG systems. This work establishes both theoretical foundations and practical roadmaps for subsequent research in this evolving field.

The contributions of this paper can be summarized as follows:
\begin{itemize}
    \item \textbf{Pioneering Review}. This work represents the first comprehensive survey focusing on the integration of RAG with reasoning, offering novel insights and forward-looking guidance for advancing this emerging research frontier.
    \item \textbf{Systematic Taxonomy}. We present a multi-dimensional framework to systematically examine the objectives, paradigms, and methodologies for combining RAG with reasoning capabilities, establishing clear classification criteria across technical dimensions.
    \item \textbf{Practical Guidance}. Beyond theoretical exploration, we critically discuss the additional cost and potential risks associated with the introduction of reasoning, accompanied by an actionable Practical Guide for real-world scenarios.
    \item \textbf{Open Resource Platform}\footnote{\url{https://openrag.notion.site/open-rag-base?pvs=4}} Through the OpenRAG platform, we provide a rich, multi-dimensional review of related work, which allows readers to quickly search and compare different methods.
\end{itemize}

\section{Overview}
This chapter establishes a conceptual framework for the paper along two key dimensions. First, it formally defines "reasoning" and distinguishes it from "inference." Second, it organizes a taxonomy of synergy mechanisms between "RAG and Reasoning." To construct a clear cognitive pathway, we address three progressive research questions:
\begin{itemize}
    \item \textbf{\textit{Why synergize RAG and reasoning?}}
    \item \textbf{\textit{What are their typical collaboration paradigms?}}
    \item \textbf{\textit{How can this integration be realized?}}
\end{itemize}

\subsection{Definition}
The definition of reasoning in modern AI systems remains an evolving construct, particularly within the context of LRMs exemplified by DeepSeek R1 and OpenAI O1. Here, under the scope of LLMs, we formalize reasoning as \textbf{a structured, multi-step process that dynamically decomposes complex problems, generates intermediate hypotheses, and iteratively refines solutions through logical and evidence-based transformations.} Mathematically, let a reasoning process \( \mathcal{R} \) be defined as a tuple \( \langle \mathcal{K}_p, \mathcal{K}_r, \mathcal{S}_t, \Phi \rangle \), where \( \mathcal{K}_p \) denotes parametric knowledge embeddings, \( \mathcal{K}_r \) represents retrieved contextual knowledge, \( \mathcal{S}_t = \{s_0, s_1, \ldots, s_n\} \) constitutes the evolving state sequence with \( s_0 \) as the initial query and \( s_n \) as the final response, and \( \Phi: \mathcal{S}_i \times \mathcal{K}_p \times \mathcal{K}_r \rightarrow \mathcal{S}_{i+1} \) defines the state transition function.  

The reasoning process exhibits three defining characteristics. First, it is inherently \textbf{multi-step}, systematically decomposing complex problems into intermediate cognitive states (e.g., sub-question generation or temporary conclusions) rather than pursuing direct input-output mapping. Second, it generates \textbf{novel knowledge or facts} — synthesizing implicit relationships, deriving latent constraints, or reformulating problems in ways not explicitly present in the initial input or parametric memory (e.g., transforming "Is A greater than B?" into comparative subquestions about A and B’s attributes). Crucially, these representations are not merely retrieved but dynamically constructed through the reasoning trajectory. Third, the process is \textbf{teleological} — its architecture and termination conditions are explicitly optimized for complex problem resolution, where complexity is measured by the necessity of state transitions or the insufficiency of direct retrieval from either parametric (\( \mathcal{K}_p \)) or external (\( \mathcal{K}_r \)) knowledge sources. This stands in stark contrast to atomic inference, which lacks such deliberate state construction and goal-aware iteration.

The distinction between reasoning and inference manifests most saliently in their computational signatures. While inference \( \mathcal{I} \) constitutes a single-step conditional probability computation \( P(y|x) = \prod_{t=1}^T P(y_t|x, y_{<t}) \), reasoning \( \mathcal{R} \) implements a meta-process coordinating multiple inference calls through explicit state management \( \mathcal{R}(x) = \Phi_1 \circ \Phi_2 \circ \cdots \circ \Phi_n(x) \). This multi-phase architecture enables systematic error correction through backtracking mechanisms and dynamic retrieval refinement — features fundamentally absent in conventional inference pipelines. The operational boundary emerges when state transitions involve explicit symbolic manipulation (equation restructuring in mathematical reasoning) or knowledge graph traversal (temporal reasoning over retrieved events), distinguishing true reasoning from mere multi-step inference.

\begin{figure*}[ht]
 \centering
        \tiny
 \begin{forest}
  for tree={
   forked edges,
   grow'=0,
   draw,
   rounded corners,
    node options={align=center},
   calign=edge midpoint,
            font=\scriptsize,
  }
[The Synergy of RAG and Reasoning, fill=black!10,rotate=90,text width=6cm,font=\large
    [Purpose §\ref{sec:purpose}, text width=2.5cm, for tree={fill=red!20},font=\large
        [ Reasoning-Augmented Retrieval\\ (Better Retrieval) §\ref{subsec:better_retrieval}, text width=3.5cm
            [ ARM~\cite{ARM}; AdaptiveRAG~\cite{adaptiveRAG};
            FinSearch~\cite{FinSearch}; LevelRAG~\cite{levelrag}; OmniThink~\cite{omnithink}; PlanRAG~\cite{planrag}; SmartRAG~\cite{smartrag}; UAR~\cite{UAR}; O1-Embedeer~\cite{o1_embedder}; ITER-RETGEN~\cite{iter-retgen}; 
            HiRAG~\cite{HiRAG};
            MetaRAG~\cite{metaRAG};
            MMOA-RAG~\cite{mmoa_rag};
            ReZero~\cite{ReZero};
            DeepResearcher~\cite{deepresearcher};
            ReaRAG~\cite{rearag};
            FRAG~\cite{frag};
            PORAG~\cite{PORAG};
            Insight-RAG~\cite{Insight-RAG};
            ChainRAG~\cite{ChainRAG};
            FG-RAG~\cite{FG-RAG};
            MCTS-RAG~\cite{MCTS-RAG};
            REAPER~\cite{reaper};
            DeepNote~\cite{DeepNote},              
        text width=10.0cm, node options={align=left}
            ]
         ]
        [ Retrieval-Augmented Reasoning\\(Better Reasoning) §\ref{subsec:better_resaoning}, text width=3.5cm
            [ ActiveRAG~\cite{activerag}; AgenticReasoning~\cite{agentic_reasoning};
            CoRAG~\cite{CoRAG}; CR-planner~\cite{CR-planner}; DeepRAG~\cite{deeprag}; Deepsolution~\cite{li2025deepsolution}; KBQA-O1~\cite{kbqa_o1}; OpenRAG~\cite{openrag}; PIKE~\cite{pike}; R1-Seacher~\cite{R1-Searcher}; RAG-Gym~\cite{rag-gym}; ReARTeR~\cite{rearter};
            ReSearch~\cite{ReSearch};
            MedRAG~\cite{medrag};
            RARE~\cite{rare};
            \textit{RARE}\cite{rare2};
            RetroRAG~\cite{RetroRAG};
            KG-RAR~\cite{KG-RAR};
            RetrievalPRM~\cite{RetrievalPRM};
            MRD-RAG~\cite{MRD-RAG};
            Search-O1~\cite{search_o1};
            StePO-Rec~\cite{StePO-Rec},
                    text width=10.0cm, node options={align=left}
            ]
        ]   
   ]
    [Paradigms §\ref{sec:pattern}, text width=2.5cm, for tree={fill=orange!20},font=\large
        [Pre-defined Workflow §~\ref{subsec:pre_defined}, text width=2.8cm
            [Reasoning in Pre-Retrieval,text width=2.1cm
                [
                LeReT~\cite{LeReT}; O1-Embedder~\cite{o1_embedder}; planRAG~\cite{planrag}; UAR~\cite{UAR}; 
                MMOA-RAG~\cite{mmoa_rag};
                MedRAG~\cite{medrag};
                FRAG~\cite{frag};
                Insight-RAG~\cite{Insight-RAG};
                ChainRAG~\cite{ChainRAG};
                RetrievalPRM~\cite{RetrievalPRM};
                REAPER~\cite{reaper};
                AdaptiveRAG~\cite{adaptiveRAG},
                    text width=8.2cm, node options={align=left}
                ]
            ]
            [Reasoning in Post-Retrieval,text width=2.1cm
                [
                ActiveRAG~\cite{activerag}; ARM~\cite{ARM};
                MRD-RAG~\cite{MRD-RAG};
                FG-RAG~\cite{FG-RAG};
                ToG2~\cite{ToG2},
                    text width=8.2cm, node options={align=left}
                ]
            ]
            [Hybrid Reasoning,text width=2.1cm
                [
                IR-COT~\cite{IRCoT};
                FinSearch~\cite{FinSearch}; 
                ITER-RETGEN~\cite{iter-retgen}; 
                LevelRAG~\cite{levelrag};
                HiRAG~\cite{HiRAG};
                MetaRAG~\cite{metaRAG};
                RetroRAG~\cite{RetroRAG};
                KG-RAR~\cite{KG-RAR};
                DeepNote~\cite{DeepNote},
                    text width=8.2cm, node options={align=left}
                ]
            ]
        ]
        [Dynamic Workflow §\ref{subsec:dynamic}, text width=2.8cm
            [Proactivity-Driven Reasoning,text width=2.2cm
                [
                AgenticReasoning~\cite{agentic_reasoning}; 
                DeepRAG~\cite{deeprag};
                CoRAG~\cite{CoRAG}; 
                Co-STORM~\cite{co-storm}; 
                PIKE~\cite{pike}; 
                Search-O1~\cite{search_o1}; 
                R1-Searcher~\cite{R1-Searcher},
                    text width=8.1cm, node options={align=left}
                ]
            ]
            [Reflection-Driven Reasoning,text width=2.2cm
                [
                Flare~\cite{Flare};
                OpenRAG~\cite{openrag}; 
                WriteHere~\cite{WriteHere};
                ReaRAG~\cite{rearag};
                Self-RAG~\cite{self_rag},
                    text width=8.1cm, node options={align=left}
                    ]
                ]
            [Feedback-Driven Reasoning,text width=2.2cm
                [
                SmartRAG~\cite{smartrag}; 
                CR-Planner~\cite{CR-planner};
                MCTS-KBQA~\cite{mcts-KBQA}; 
                DeepSolution~\cite{li2025deepsolution};
                RAG-Gym~\cite{rag-gym}; 
                ReZero~\cite{ReZero};
                ReSearch~\cite{ReSearch};
                RARE~\cite{rare};
                DeepResearcher~\cite{deepresearcher};
                PORAG~\cite{PORAG};
                MCTS-RAG~\cite{MCTS-RAG};
                KBQA-O1~\cite{kbqa_o1},
                    text width=8.1cm, node options={align=left}
                ]
            ]                
        ]
   ]    
   [Implementation §\ref{sec:imple}, text width=2.5cm, for tree={fill=blue!20},font=\large
        [Resoning Method §\ref{subsec:method}, text width=2.8cm,
            [LLM/CoT, text width=2.0cm
                [
                ActiveRAG~\cite{activerag};
                AdaptiveRAG~\cite{adaptiveRAG};
                O1-Embedder~\cite{o1_embedder};
                DeepNote~\cite{DeepNote};
                HiRAG~\cite{HiRAG};
                MetaRAG~\cite{metaRAG};
                MMOA-RAG~\cite{mmoa_rag};
                RetroRAG~\cite{RetroRAG};
                PORAG~\cite{PORAG};
                KG-RAR~\cite{KG-RAR};
                MRD-RAG~\cite{MRD-RAG};
                PlanRAG~\cite{planrag},
                        text width=8.2cm, node options={align=left}
                ]
            ]
            [Special Token Prediction, text width=2.0cm
                [
                Self-RAG~\cite{self_rag};
                SmartRAG~\cite{smartrag};
                OpenRAG~\cite{openrag};
                R1-Searcher~\cite{R1-Searcher};
                ReZero~\cite{ReZero};
                ReSearch~\cite{ReSearch};
                ReaRAG~\cite{rearag};
                Search-O1~\cite{search_o1},
                        text width=8.2cm, node options={align=left}
                ]
            ]
            [Search-Driven Reasoning, text width=2.0cm
                [
                        OminiThink~\cite{omnithink};
                        DeepRAG~\cite{deeprag};
                        CoRAG~\cite{CoRAG};
                        DeepSolution~\cite{li2025deepsolution};
                        ReARTeR~\cite{rearter};
                        KBQA-O1~\cite{kbqa_o1};
                        WriteHere~\cite{WriteHere};
                        RARE~\cite{rare};
                        MCTS-KBQA~\cite{mcts-KBQA};
                        StePO-Rec~\cite{StePO-Rec},
                        text width=8.2cm, node options={align=left}
                ]
            ]
            [Reasoning on Graph, text width=2.0cm
                [
                FinSearch~\cite{FinSearch};
                ToG2~\cite{ToG2};
                LighRAG~\cite{lightrag};
                FRAG~\cite{frag};
                FG-RAG~\cite{FG-RAG};
                MedRAG~\cite{medrag},
                        text width=8.2cm, node options={align=left}
                ]
            ]
            [External Solver,text width=2.0cm
                [  
                ARM~\cite{ARM},
                text width=8.2cm, node options={align=left}
                ]
            ]
        ]
        [Optimization §\ref{subsec:opt}, text width=2.8cm, for tree={fill=blue!20}
            [Prompt-Based, text width=2.0cm
                [
                       Co-STORM~\cite{co-storm};
                       Agentic Reasoning~\cite{agentic_reasoning};
                       FinSearch~\cite{FinSearch};
                       PlanRAG~\cite{planrag};
                       HiRAG~\cite{HiRAG};
                       MetaRAG~\cite{metaRAG};
                       MedRAG~\cite{medrag};
                       RARE~\cite{rare};
                       ReaRAG~\cite{rearag};
                       RetroRAG~\cite{RetroRAG};
                       FRAG~\cite{frag};
                       KG-RAR~\cite{KG-RAR};
                       MRD-RAG~\cite{MRD-RAG};
                       FG-RAG~\cite{FG-RAG};
                       MCTS-RAG~\cite{MCTS-RAG};
                       DeepSolution~\cite{li2025deepsolution};
                       StePO-Rec~\cite{StePO-Rec},
                        text width=8.2cm, node options={align=left}
                ]
            ]
            [Tuning-Based, text width=2.0cm
                [
                    KBQA-Q1~\cite{kbqa_o1};
                    O1-Embedder~\cite{o1_embedder};
                    DeepRAG~\cite{deeprag};
                    CoRAG~\cite{CoRAG};
                    MCTS-KBQA~\cite{mcts-KBQA};
                    RetrievalPRM~\cite{RetrievalPRM};
                    REAPER~\cite{reaper};
                    \textit{RARE}\cite{rare2};
                    UAR~\cite{UAR},
                    text width=8.2cm, node options={align=left}
                ]
            ]
            [RL-Based, text width=2.0cm
                [PIKE~\cite{pike};
                LeReT~\cite{LeReT};
                RAG-Gym~\cite{rag-gym};
                ReARTeR~\cite{rearter};
                SmartRAG~\cite{smartrag};
                CR-Planner~\cite{CR-planner};
                DeepRetrieval~\cite{deepretrieval};
                DeepNote~\cite{DeepNote};
                MMOA-RAG~\cite{mmoa_rag};
                ReZero~\cite{ReZero};
                ReSearch~\cite{ReSearch};
                DeepResearcher~\cite{deepresearcher};
                PORAG~\cite{PORAG};
                R1-Search~\cite{R1-Searcher},
                        text width=8.2cm, node options={align=left}
                ]
            ]
        ]
    ]        
]   
\end{forest}
 \caption{A structured taxonomy of synthesizing RAG and Reasoning.}
\label{fig:taxonomy}
\end{figure*}
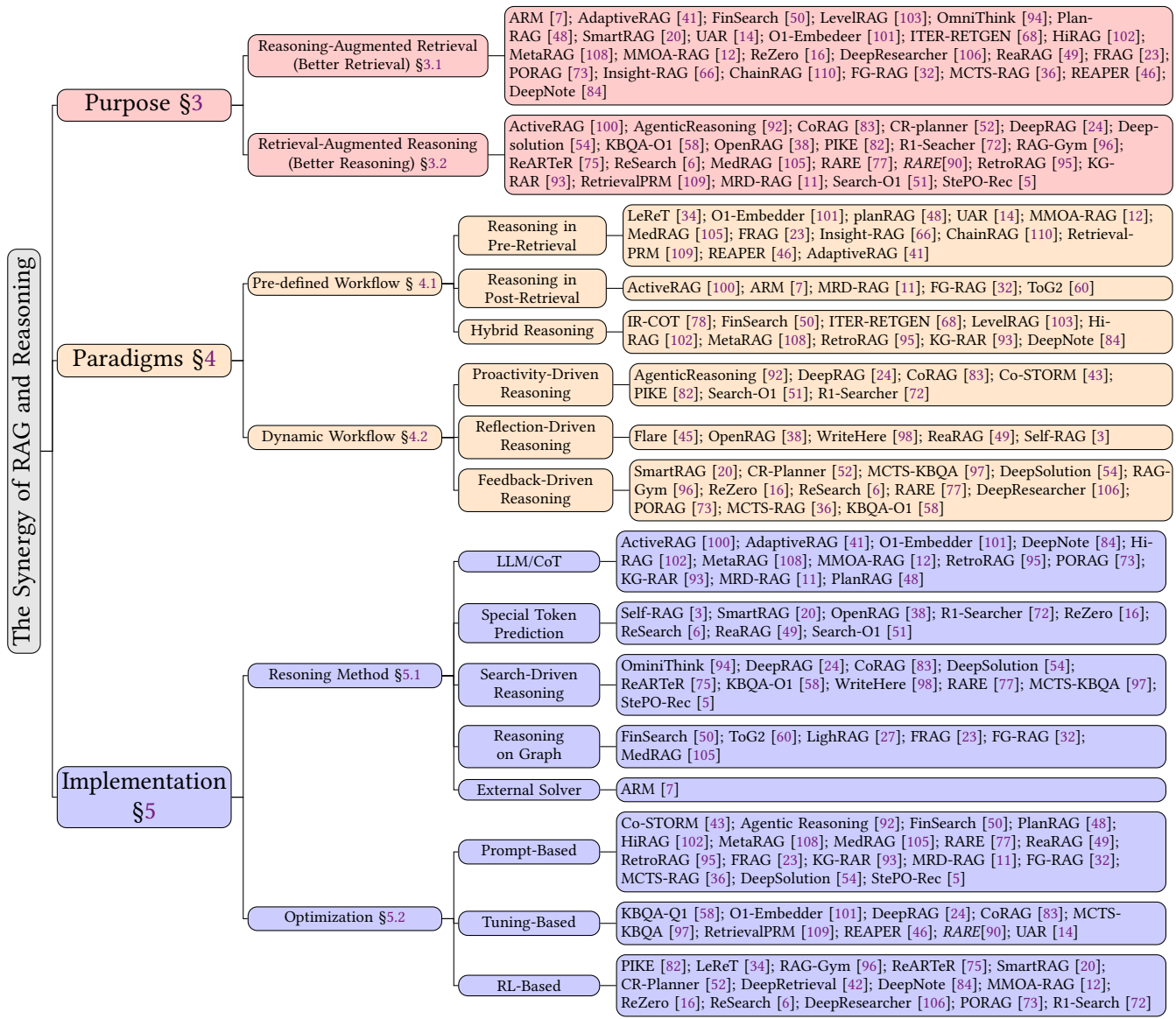

\subsection{Taxonomy}
Integrating RAG with reasoning marks a paradigm shift in tackling complex knowledge-intensive tasks. This work develops a hierarchical taxonomy (Figure~\ref{fig:taxonomy}) based on three key questions: \textbf{why }reasoning is needed with RAG (\textbf{Purpose}), how they structurally interact (\textbf{Paradigm}), and \textbf{what} methods enable effective integration (\textbf{Implementation}). This framework guides readers through the technical innovations in later chapters, providing a clear conceptual path without premature technical details, and highlighting the field’s evolutionary logic, avoiding delving prematurely into specific technical details.

\subsubsection{Synergy Purpose}
Integrating reasoning with RAG addresses the limitations of traditional RAG systems, which struggle with multi-step logic, contextual adaptation, and implicit knowledge synthesis due to reliance on superficial semantic matching and fixed knowledge limits. Adding reasoning enables dynamic retrieval planning, logical verification of evidence, and insight generation beyond retrieved data through abductive or counterfactual reasoning. At the same time, the introduction of external knowledge retrieval also helps alleviate reasoning interruptions caused by the knowledge limitations of LRM and reduces the likelihood of hallucinations. This integration occurs in two main ways: \textbf{Reasoning-Augmented Retrieval,} where inference drives context-aware information gathering; \textbf{Retrieval-Augmented Reasoning}, where external knowledge supports and expands the model’s deductive abilities.

\subsubsection{Synergy Paradigms}
Building upon the above necessity, our taxonomy categorizes RAG+Reasoning systems along the axis of procedural dynamism. \textbf{Pre-defined workflows} employ fixed  templates that systematically alternate between retrieval and reasoning phases, with intervention points predetermined at \emph{pre-retrieval reasoning} (e.g., query decomposition), \emph{post-retrieval reasoning} (e.g., evidence synthesis), or \emph{hybrid stages}. While offering operational transparency, these architectures exhibit limited adaptability to emergent task complexities. In contrast, \textbf{dynamic workflows} implement state-contingent reasoning processes where retrieval actions are conditionally triggered through continuous system introspection. This paradigm further branches into \emph{Proactivity-Driven} strategies (self-initiated knowledge requests), \emph{Reflection-driven} mechanisms (error-corrective retrieval based on intermediate result analysis), and \emph{Feedback-driven} approaches (environmental reward signals or external model evaluations). The progression from static to dynamic architectures reflects the field’s maturation toward human-like contextual adaptation in open-world problem-solving.

\subsubsection{Synergy Implementation}
Operationalizing these synergies requires innovations across reasoning and retrieval strategies. Foundational reasoning architectures span\textbf{ LLM-Based} like COT, \textbf{search-based }hypothesis generation (tree search, Monte Carlo methods), \textbf{symbolic solver integration}, and \textbf{graph-structured} multi-hop inference. These capabilities are further enhanced through three principal augmentation strategies: \textbf{prompt-based} techniques that utilize natural language templates and special token (e.g., <Plan>, <Verify>) to steer model behavior, \textbf{tuning-based} methods that inject domain-specific knowledge or distill reasoning capability, and \textbf{RL-based} frameworks that optimize retrieval-reasoning policies through outcome reward models (ORM) or process reward models (PRM). The alignment between these methodologies and the proposed taxonomy is critical—static workflows predominantly rely on predictable prompt-guided reasoning chains, whereas dynamic systems increasingly integrate search-based exploration or solver-augmented strategies to navigate evolving state spaces.

Overall, this tripartite taxonomy—motivational drivers, architectural paradigms, and implementation methodologies—establishes a unified lens for analyzing RAG+Reasoning systems. Subsequent chapters will elaborate on each stratum, progressively revealing how these conceptual distinctions translate into technical innovations that push the boundaries of machine intelligence.

\section{The purpose of the synergy}
\label{sec:purpose}

\begin{figure*}[htbp]
    \centering
    \includegraphics[width=1\linewidth]{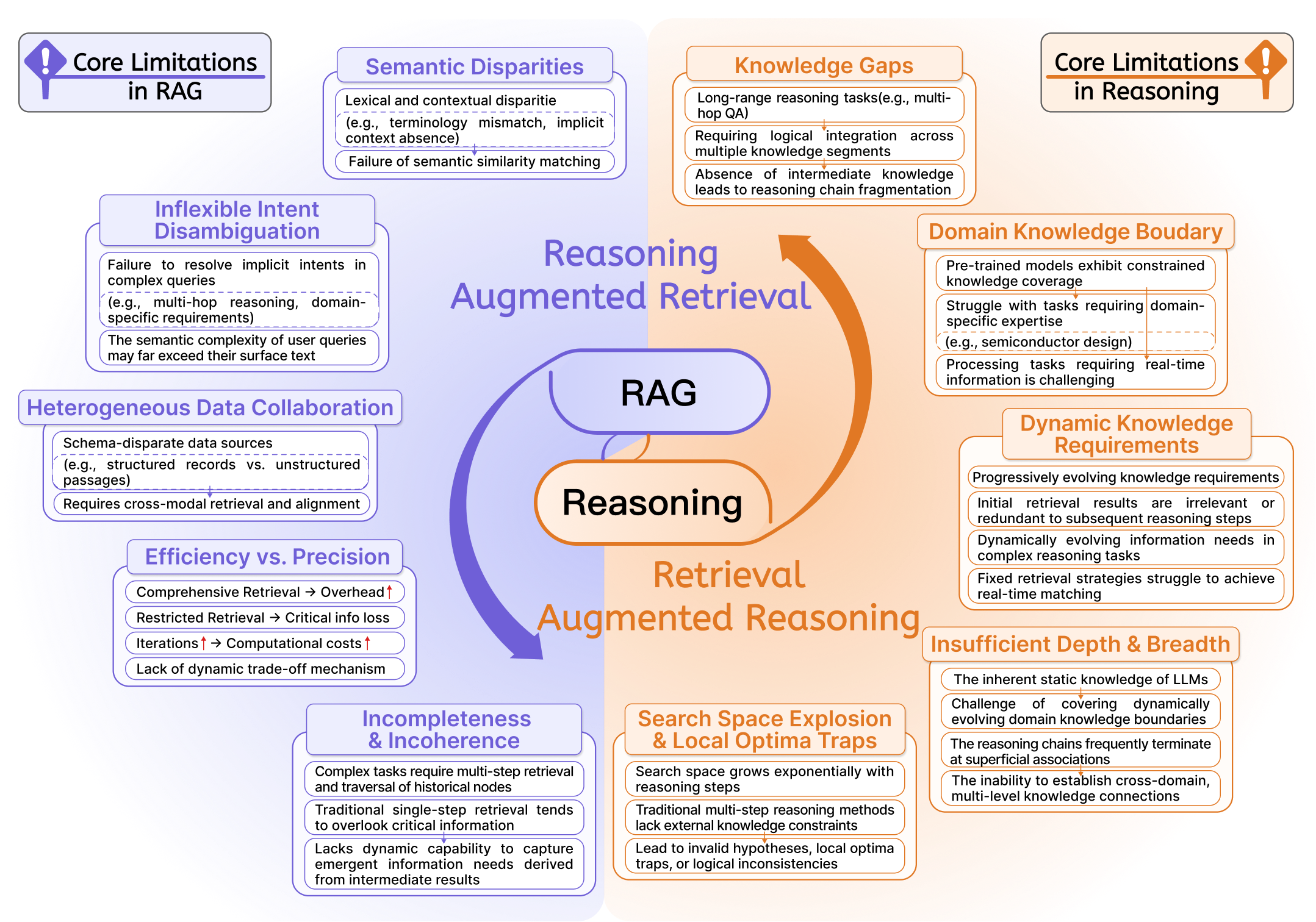}
    \caption{The purpose of the synergy between RAG and reasoning}
    \label{fig:purpose}
\end{figure*}

The integration of RAG and reasoning marks a crucial advancement in enhancing LLMs' problem-solving abilities. Their true potential lies not in isolated use but in their synergy, which overcomes key limitations in retrieval and reasoning. This section explains the main motivations for combining RAG with reasoning, emphasizing two primary benefits: (1) enhancing retrieval accuracy and flexibility through reasoning, and (2) reinforcing complex reasoning by using context-rich retrieved knowledge. Figure~\ref{fig:purpose} illustrates these collaborative aims and the limitations they address.

The first key benefit is \textbf{Reasoning-Augmented Retrieval}, where reasoning improves the retrieval process. Traditional RAG systems struggle with query formulation, relevance assessment, and iterative refinement—tasks needing logical and contextual analysis. Reasoning enables adaptive retrieval through dynamic query expansion, ambiguity resolution, and multi-hop evidence aggregation, overcoming the limits of keyword- or embedding-based methods and aligning retrieval with the task’s reasoning demands.

The second benefit is \textbf{Retrieval-Augmented Reasoning}, where external knowledge supplements the limitations of purely parametric LLM reasoning. Even advanced models face hallucination, knowledge gaps, and compositional challenges alone. Retrieval grounds reasoning in up-to-date, domain-specific, or rare information absent from model weights, crucial for explainability, multi-step deduction, and integrating diverse sources.

Together, combining RAG and reasoning fills fundamental gaps in both techniques. By enhancing retrieval via reasoning and strengthening reasoning through retrieval, it broadens LLMs’ capacity to address complex real-world problems.

\subsection{Reasoning-Augmented Retrieval}
\label{subsec:better_retrieval}

Reasoning-Augmented Retrieval (RAR) represents a significant advancement in information retrieval by integrating multi-step reasoning to dynamically enhance retrieval quality. Unlike traditional methods that depend on static semantic matching, RAR creates a cognitive feedback loop mimicking human iterative reasoning, surpassing the limitations of simple "query-document" interactions. 

RAR’s effectiveness stems from several key features. It often uses on-demand retrieval, where reasoning—evaluating intent clarity, knowledge state, and temporal factors—guides adaptive search initiation, reducing redundancies present in fixed triggers (e.g., UAR’s classifier~\cite{UAR}). It improves semantic alignment by inferring implicit query logic such as business rules or entity relationships to generate precise retrieval requests aligned with data schemas (e.g., PlanRAG’s plan-retrieval loops~\cite{planrag}). RAR also applies multi-step iterative refinement, using intermediate reasoning outputs (e.g., chain-of-thought, partial answers~\cite{IRCoT}) to recursively reformulate queries in a closed-loop system essential for resolving multi-hop dependencies~\cite{iter-retgen}. Furthermore, it adapts to specific domains by tailoring retrieval to vertical contexts (e.g.,  financial  or medical) and balances efficiency and precision through lightweight reasoning strategies (e.g., AdaptiveRAG’s complexity-based selection~\cite{adaptiveRAG}).

Traditional retrieval systems, effective for simple queries, struggle with complex information needs due to rigid designs favoring static matching over dynamic reasoning, limiting their adaptability to changing contexts and diverse data. RAR primarily addresses five core challenges inherent in these conventional methods.

\subsubsection{Semantic Disparities Between Queries and Documents}

A key challenge lies in the mismatch between user queries and documents—whether due to differing expression styles (professional jargon vs. casual language) or implicit contextual gaps—making direct semantic matching unreliable. Importantly, high similarity does not guarantee true relevance, as documents may share keywords or surface features without addressing the underlying intent or logic of the query. Retrieval models must therefore understand deeper semantics beyond superficial similarity.Domain adaptation further complicates this issue. To overcome these gaps, approaches such as reasoning-augmented embeddings (O1-Embedder~\cite{o1_embedder} enriches queries with inferred “thinking” text), feedback-driven rewriting (SmartRAG~\cite{smartrag} dynamically refines queries based on retrieved results), and pre-planning (PlanRAG~\cite{planrag} extracts business rules to generate SQL queries aligned with database schemas) help better capture domain-specific semantics and ensure relevance beyond mere similarity.

\subsubsection{Inflexible Intent Disambiguation }

Traditional RAG methods rely on fixed  embedding similarity strategies , which fail to dynamically interpret the implicit intent behind complex queries (e.g., multi-hop reasoning or domain-specific requirements). User queries often exhibit semantic complexity that far exceeds their surface text—for instance, a request to "optimize supply chain costs" may require correlating disparate database fields not explicitly mentioned. 
Static retrieval methods lack the adaptability to capture such dynamically evolving information needs. A critical limitation lies in \textit{intent dynamicity}: as contextual understanding expands, traditional systems generate fixed retrieval results based solely on the initial query. Furthermore, \textit{semantic representation limitations} of dense retrieval models (e.g., BERT-based models) hinder their ability to encode intricate semantic relationships (e.g., irony, metaphors), leading to misaligned results. Current approaches attempt to mitigate these issues through \textit{multi-step intent decomposition} (e.g., LevelRAG’s high-level searcher breaks complex queries into multi-hop sub-queries~\cite{levelrag}) and \textit{dynamic query reformulation} (e.g., LeReT’s reinforcement learning generates diversified query candidates~\cite{LeReT}), iteratively refining retrieval strategies to align with document content.

\subsubsection{Inefficient Coordination of Multi-Source Heterogeneous Data}

Retrieval from diverse sources—text, tables, graphs, web, and APIs—often produces fragmented results due to a lack of global reasoning. The key challenge is modal heterogeneity: different retrieval techniques (dense retrieval for text, SQL for tables, GQL for graphs) operate independently without unified coordination. For example, experiments show standard RAG methods (like dense retrieval with query decomposition) yield only 32.7\% perfect recall and 40.9\% F1 on the OTT-QA dataset. These outcomes reveal the limitations of traditional approaches in aligning textual queries with structured tables—such as failing to link concepts like "K-12 student free rates" in text to related "education expenditure" columns when not explicitly mentioned. Additionally, disconnected entity matching (e.g., relating "company revenue" in text to financial tables) worsens inefficiencies, as conventional methods depend on semantic similarity and overlook domain-specific relationships and exact-value matches. Advanced techniques—such as reasoning-driven alignment (ARM’s N-gram constraints for cross-modal entity decoding~\cite{ARM}) and unified semantic spaces (LevelRAG’s shared multi-modal representations~\cite{levelrag})—enable more effective, integrated retrieval.

\subsubsection{Incompleteness and Incoherence in Complex Retrieval Tasks}

Single-step retrieval systems fall short in complex multi-hop reasoning tasks, such as deducing entity chains or conducting decision analysis. Traditional static retrieval conflicts with multi-step cognitive needs, resulting in three main issues: 1) Path dependency, where later retrievals rely on information from earlier steps (e.g., finding "the most populous county in California" before its education policies), but conventional systems lack state management; 2) Error propagation,early retrieval errors cause mistakes in intermediate results, which then affect the next round of retrieval; 3) Semantic inflexibility of fixed queries, which cannot adapt to dynamic concepts like entity aliases or relational predicates.

Advanced methods address these flaws through integrated strategies. PlanRAG uses iterative "plan-retrospect-replan" cycles to trigger sub-queries when gaps arise.  Reinforcement learning in LeReT improves query generation via reward-driven path selection.  Likewise, ITER-RETGEN rebuilds follow-up queries using intermediate answers (e.g., "award recipient’s height") to resolve multi-hop dependencies.

\subsubsection{Trade-offs Between Retrieval Efficiency and Precision}

Complex scenarios face a tension between exhaustive retrieval, which is computationally costly, and restricted retrieval, which risks information loss. Expanding retrieval blindly inflates costs (e.g., LLM API calls) without ensuring relevance. Simple queries suffer from unnecessary multi-step retrieval, wasting resources, while complex queries face quality risks if retrieval is too limited. Adaptive approaches like complexity-aware routing (Adaptive-RAG’s lightweight classifier allocates retrieval budgets~\cite{adaptiveRAG}) and cost-sensitive training (SmartRAG’s reinforcement learning balances quality and steps~\cite{smartrag}) dynamically manage this trade-off. 

In summary, Reasoning-Augmented Retrieval overcomes traditional RAG’s limitations in dynamic triggering, semantic alignment, multi-hop support, domain adaptation, and efficiency trade-offs by deeply integrating reasoning into the retrieval process. Its key innovation is a bidirectional enhancement between reasoning and retrieval—reasoning refines retrieval strategies, while retrieval supports iterative reasoning—jointly boosting accuracy and efficiency in complex information tasks.

\subsection{Retrieval-Augmented Reasoning}
\label{subsec:better_resaoning}

Retrieval-Augmented Reasoning (ReAR) combines external knowledge retrieval with inherent model reasoning to overcome failures from knowledge gaps or logical discontinuities in complex tasks. Unlike traditional RAG methods that retrieve information once, ReAR uses an iterative, context-sensitive retrieval that continuously provides relevant data to support multi-step reasoning. This approach is crucial for tasks needing strict logic, such as mathematical proofs, where intermediate steps require specific theorems or lemmas. By making retrieval an adaptive, ongoing process rather than a one-time step, ReAR strengthens each reasoning stage with accurate, current information, improving the overall inference’s reliability and robustness.

ReAR’s core feature is dynamic knowledge supplementation, generating retrieval queries in real-time based on the evolving reasoning context. This overcomes the limits of single-round retrieval by enabling knowledge refinement at each step, as seen in process supervision frameworks like RAG-Gym~\cite{rag-gym}. ReAR also improves reasoning paths using methods like search space compression—for example, MCTS-guided heuristics in KBQA—and structured feedback from diverse sources like knowledge graphs~\cite{mcts-KBQA}. These techniques maintain logical consistency while reducing irrelevant or conflicting information. Importantly, ReAR adapts well across domains, supporting precise knowledge retrieval and tool use for specialized tasks such as industrial problem-solving in PIKE~\cite{pike} or scientific reasoning~\cite{deepresearcher}.

By integrating retrieval as an active part of the reasoning loop, ReAR addresses LLMs' temporal and depth constraints, ensuring adherence to domain-specific and time-sensitive requirements. This close coupling turns external knowledge into an on-demand resource, creating a closed-loop system that enhances the model’s ability to handle complex, knowledge-intensive problems. Specifically, ReAR seeks to address the following limitations and challenges:

\subsubsection{Knowledge Gap in Multi-step Reasoning}

In long-range reasoning, missing intermediate knowledge often breaks logical chains, especially in industrial and scientific contexts requiring multi-source data integration (e.g., text, tables, time-series). Static retrieval methods worsen this by not adapting to the reasoning process’s changing needs. ReAR techniques address this with chained retrieval, as in CoRAG~\cite{CoRAG}, which breaks multi-hop questions into sequential sub-queries (e.g., retrieving "event causes" then their "impacts"), systematically linking knowledge. Reasoning-state-aware retrieval, used in FLARE~\cite{Flare}, predicts future information needs by generating interim prompts (e.g., "the next step requires discussion of ..."), enabling dynamic query construction that preserves coherence. Together, these approaches resolve the conflict between discrete retrieval and continuous reasoning.

\subsubsection{Reasoning Discontinuity Caused by Domain Knowledge Boundaries}

Reasoning discontinuity arises from LLMs' limited knowledge, struggling with specialized domains (e.g., semiconductor design in PIKE~\cite{pike}) and real-time data (e.g., medical parameters in Agentic Reasoning~\cite{agentic_reasoning}). End-to-end models often produce factual errors, while traditional RAG methods fail to retrieve deep professional knowledge due to coarse retrieval, especially with complex data like tables,charts and images.

ReAR addresses this with two complementary solutions: knowledge atomization and structural organization, as in PIKE’s decomposition of documents into fine-grained units and multi-layer knowledge graphs for semantic and logical retrieval; and dynamic tool integration, as in Agentic Reasoning’s real-time data acquisition via code execution and API calls to compute critical indicators (e.g., medical FiO2). These innovations overcome the challenges of specialized knowledge depth and timely information relevance that limit conventional methods.

\subsubsection{Search Space Explosion and Local Optima Traps}

The main challenge in multi-step reasoning is the exponential growth of the search space, where methods like Chain-of-Thought (CoT) often yield suboptimal or inconsistent results due to unconstrained hypotheses. Traditional approaches like CoT and Tree-of-Thought (ToT) lack external knowledge constraints, causing invalid assumptions, while purely symbolic reasoning falls short in open-domain tasks. To address this, two strategies are used: knowledge base-anchored heuristic search (KBQA-O1~\cite{kbqa_o1}), which limits reasoning actions to subgraphs in knowledge graphs, and a retrieval-verification mechanism (Search-o1~\cite{search_o1}) that prunes unsupported reasoning paths using evidence from the knowledge base. Together, these reduce the search space and preserve reasoning coherence.

\subsubsection{Dynamic Knowledge Requirements in Multi-Step Reasoning}

Complex multi-step reasoning tasks face the challenge of continuously changing knowledge requirements. This is evident in cases like multi-hop reasoning and engineering planning, where each stage generates new sub-problems (e.g., moving from "architectural design" to "material cost estimation"). Static knowledge bases or one-time retrieval methods cannot meet this evolving demand. This manifests in two ways: initial knowledge may miss later needs, causing gaps; and fixed knowledge sets may include irrelevant information, reducing reasoning accuracy. To address this, new retrieval-augmented reasoning approaches introduce dynamic solutions: process supervision (e.g., reward models in RAG-Gym~\cite{rag-gym}) detects knowledge gaps in real time, atomic decision-making (e.g., step decomposition in DeepRAG~\cite{deeprag}) triggers retrieval as needed, and tree-like expansions (e.g., multi-path retrieval in DeepSolution~\cite{li2025deepsolution}) enable parallel exploration. By integrating knowledge retrieval within reasoning, these methods let the system identify, supplement, and verify knowledge dynamically—much like a human expert—greatly enhancing the reliability and completeness of complex reasoning.

\subsubsection{Insufficient Depth and Breadth of Reasoning}

This issue is prominent in expert tasks like medical diagnosis, legal analysis, and research report generation. LLMs’ static knowledge often fails to capture the evolving scope of domain knowledge, resulting in shallow reasoning that misses multi-level, cross-domain connections. For example, when assessing "Company A is affected by economic recession," traditional methods rely on superficial statistical patterns and cannot systematically follow the deeper logical chain from "Company A → industry supply chain → macroeconomic policy → international political landscape," leading to reasoning that lacks causal depth.

To overcome this, recent advances use structured, retrieval-enhanced frameworks. ToG2.0~\cite{ToG2} models Knowledge Graph relational paths as retrieval guidance vectors, enabling targeted queries along entity paths, surpassing the limits of keyword-based retrieval. This approach complements CR-Planner’s~\cite{CR-planner} iterative expansion, which triggers retrieval of specialized knowledge (e.g., textbook proofs of algorithm complexity) at critical reasoning points, ensuring accurate domain knowledge integration via multi-round validation. Addressing cross-domain knowledge linkage, CO-STORM~\cite{co-storm} employs a multi-agent system whose host module generates cross-modal retrieval commands by analyzing potential semantics in uncited documents.

\section{Patterns of synergy}
\label{sec:pattern}

\begin{figure*}[hbtp]
    \centering
    \includegraphics[width=1\linewidth]{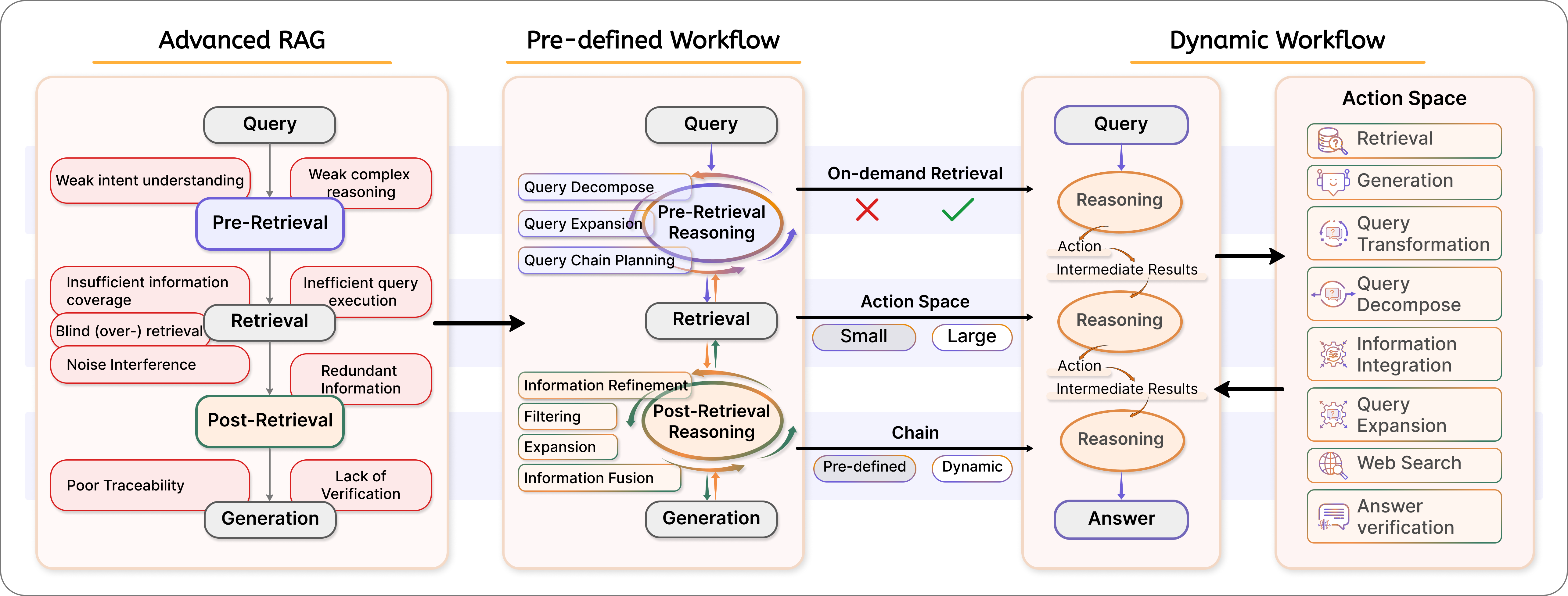}
    \caption{Patterns of Synergy between RAG and Reasoning}
    \label{fig:pattern}
\end{figure*}

Section~\ref{sec:purpose} detailed the need and motivation for integrating RAG with reasoning. Building on this, this section presents two core implementation patterns for RAG-reasoning synergy (Figure~\ref{fig:pattern}): (1) the \textbf{Pre-defined Workflow}, which uses logical architectures with preset rules for coordination, and (2) \textbf{Dynamic Workflow}, which relies on context-aware, adaptive coordination via real-time decision engines. These patterns illustrate current frameworks combining knowledge retrieval and multi-step reasoning from deterministic and flexible perspectives.

\subsection{Pre-defined workflow}
\label{subsec:pre_defined}

Pre-defined workflow is a multi-step reasoning approach with a fixed architecture and sequential execution, emphasizing process clarity and operational determinism. It consists of predefined iterative stages, each with strict input-output rules and no dynamic changes based on intermediate results. This modular design ensures controllability and structured reasoning for complex tasks. All steps are executed regardless of intermediate outcomes, guaranteeing repeatability and stability while avoiding uncertainties from dynamic decisions. Although it sacrifices adaptability, this approach offers procedural predictability and is well-suited for scenarios demanding clear reasoning paths, albeit with possible computational redundancy due to lack of real-time adjustments.

Mathematically, the pre-defined RAG workflow can be formalized as a deterministic multi-step operational chain. Given an input query \( Q \) and a predefined sequence of \( N \) reasoning steps and the final decision output $D$, the complete workflow is expressed as:  

\begin{equation}
    D = f_N \circ \cdots \circ f_2 \circ f_1(Q)
\end{equation}

where each \( f_i \in \{ \Psi, R, \Gamma \} \) denotes strictly defined functions for reasoning (\(\Psi\)), retrieval (\(R\)), or decision-making (\(\Gamma\)), with \( \circ \) representing function composition. This formulation adheres to the fixed mapping sequence \( Q \mapsto \Psi(Q) \mapsto R(\Psi(Q)) \mapsto \Gamma(R(\Psi(Q))) \), exhibiting Markovian properties where \( f_{t+1} \) depends solely on \( f_t \)'s output while remaining independent of historical states \( \{f_{<t}\} \). The chained composition guarantees process closure and reproducibility, though constrained by the static combinatorial nature of \( \{f_i\}_{i=1}^N \).

In the pre-defined pipeline, based on the position where reasoning is introduced, it can be further divided into Pre-Retrieval, Post-Retrieval, and Hybrid.

\subsubsection{Pre-Retrieval Reasoning}

For pre-retrieval methods, the sequence is explicitly defined as 
\begin{equation}
    D = \Gamma \circ \mathcal R \circ \Psi(Q)
\end{equation}

where \( \Psi \) denotes a reasoning operator that systematically transforms or enriches the query prior to retrieval. 
This paradigm enhances retrieval precision by resolving ambiguities, inferring implicit intents, or optimizing query representations.  Current research identifies four principal methodological categories for designing \( \Psi \):  

\textbf{Query Optimization }focuses on generating and selecting query variants to maximize retrieval relevance. Mathematically, this is formalized as $\text{Candidates} = \text{Generate}(Q, C)$, $\quad \Psi_{\text{Optimize}}(Q, C) = \arg\max_{\text{candidate} \in \text{Candidates}} \text{Score}(\text{candidate})$, where ($\text{Generate}$) produces candidate queries and ($\arg\max$) selects optimal variants based on contrastive training or reinforcement learning. Representative implementations, such as LeReT~\cite{LeReT}, leverage iterative sampling and optimization to balance query diversity and specificity.  

\textbf{Attribute Judgment} employs classification mechanisms to dynamically regulate retrieval triggers. This is modeled as \( \Psi_{\text{Classify}}(Q) = \text{Classify}(Q) \), where \( \text{Classify} \) evaluates query attributes (e.g., temporal sensitivity, intent complexity) against predefined criteria. Frameworks like UAR~\cite{UAR} and  AdaptiveRAG~\cite{adaptiveRAG} exemplify this approach by integrating multi-stage classifiers to minimize unnecessary retrievals.  

\textbf{Plan Generation} decomposes complex queries into structured sub-task sequences to guide retrieval direction. Formulated as \( \Psi_{\text{Plan}}(Q) = \text{Plan}(Q) \), the operator \( \text{Plan} \) generates hierarchical task decompositions, as seen in PlanRAG~\cite{planrag} , which utilizes chain-of-thought reasoning to align retrieval targets with multi-step problem-solving requirements.  

\textbf{Semantic Enhancement} enriches query representations using domain-specific or task-aware embeddings. Expressed as \( \Psi_{\text{Enhance}}(Q) = \text{Encode}(Q, \mathcal{K}) \), where \( \mathcal{K} \) denotes auxiliary knowledge (e.g., reasoning trajectories), methods like O1-Embedder~\cite{o1_embedder} integrate latent reasoning patterns into query embeddings to improve retrieval robustness.  

Collectively, these methodologies demonstrate that pre-retrieval reasoning serves as a systematic interface to mitigate semantic gaps between raw queries and knowledge bases, establishing a critical component for precision-driven RAG architectures.

\subsubsection{Post-Retrieval Reasoning}

In pre-defined RAG systems with multi-step reasoning pipelines, the post-retrieval reasoning paradigm represents a critical advancement where cognitive processing occurs after information retrieval from external sources. This approach addresses inherent limitations in conventional RAG, particularly in managing knowledge conflicts, mitigating information insufficiency, and enhancing logical consistency across complex reasoning tasks. Mathematically, this process can be formalized as a deterministic function composition: 

\begin{equation}
    D = \Gamma \circ \Psi \circ \mathcal R(Q)
\end{equation}
$\mathcal R$ denotes the retrieval operator, $\Psi$ implements the reasoning transformation, and $\Gamma$ represents the final decision function. 

The core characteristic of Post-Retrieval Reasoning lies in its execution of the reasoning process after retrieval, with the reasoning target being the retrieved content. ToG2.0~\cite{ToG2} proposes an iterative multi-step reasoning framework that alternates between graph retrieval and context retrieval, integrating the reasoning judgment of LLMs to progressively expand entities and prune irrelevant information, ultimately generating accurate answers. This approach dynamically addresses the issue of insufficient information through iterative refinement while establishing a dual-evidence verification mechanism via knowledge graph relation pruning and entity-guided context retrieval. Its graph-structured reasoning module transforms the connectivity validation of triple paths into a constraint satisfaction problem, effectively mitigating logical inconsistencies between text fragments and thereby significantly improving the quality of complex question answering.  

ActiveRAG~\cite{activerag}, on the other hand, employs a predefined three-stage process (Self-Inquiry → Knowledge Assimilation → Thought Accommodation) to structurally comprehend and calibrate retrieved knowledge, resolving conflicts between parametric memory and external knowledge. During the Knowledge Assimilation stage, ActiveRAG enhances the corrective effect of external knowledge on the internal representations of LLMs through multi-instruction fine-tuning strategies (e.g., counterfactual comparison and anchor association), substantially reducing the likelihood of hallucination generation. ARM's~\cite{ARM} structural alignment and self-verification stages also demonstrate optimization for post-retrieval reasoning. By incorporating domain knowledge via mixed-integer programming (MIP) solvers, ARM ensures the rationality and coverage of retrieval results, providing a scalable optimization framework for multi-source data compatibility and thereby enabling globally optimal cross-modal retrieval.

\subsubsection{Hybrid Reasoning}

The Hybrid pattern of pre-defined process forms a composite processing paradigm by integrating pre-retrieval reasoning with post-retrieval reasoning. The  essence is formalized as a multi-round recursive iterative process, where each iteration cycle strictly comprises three phases: Retrieval, Generation, and Reasoning, executed as structured composite operations. Let the total number of iterations be \( T \); the workflow is defined as:  

\begin{equation}
    Q_{T} = \left( \bigcirc_{t=1}^{T} \mathcal{R_t} \circ \Gamma_t \circ \Psi_t\right)(Q_0)
\end{equation}

Here, each iterative unit is indexed by \( t \). The process terminates when a predefined condition \( \mathcal{T}(Q_t, D_t, C_t) \) is met, yielding the final response \( \Gamma_{\text{final}}(C_T) \). This recursive mechanism enables dynamic synergy between knowledge acquisition and semantic inference, overcoming the linear limitations of single-cycle retrieval-generation frameworks.

IR-CoT~\cite{IRCoT} leverages chain-of-thought reasoning to iteratively construct intermediate logic chains, enabling multi-hop retrieval guided by progressively refined contextual cues. FinSearch~\cite{FinSearch} introduces a dual-phase architecture that first generates structured search graphs to model temporal and entity dependencies, followed by dynamic query rewriting to optimize financial data retrieval. LevelRAG employs hierarchical validation mechanisms, aggregating multi-granular retrieval results and triggering supplementary retrievals based on context completeness assessments. ITER-RETGEN~\cite{iter-retgen} utilizes generation-enhanced feedback loops to iteratively refine query representations, enhancing semantic alignment between retrieval and generation phases.  

These approaches share a common foundation in structured recursion while diverging in operational mechanisms. By enforcing deterministic iteration cycles, they balance controlled workflow execution with adaptive semantic exploration, addressing challenges such as multi-step reasoning, temporal coherence, and cross-domain knowledge synthesis. The hybrid paradigm’s strength lies in its capacity to decompose complex queries into iterative retrieval-generation units, systematically bridging knowledge gaps while maintaining interpretability and robustness in open-domain problem-solving scenarios.

\subsection{Dynamic RAG Workflow}
\label{subsec:dynamic}

The RAG with dynamic workflow represents an autonomous reasoning architecture centered around LLMs, characterized by the integration of non-deterministic operational workflows and real-time decision-making capabilities. Unlike pre-defined pipelines, this architecture enables continuous monitoring of reasoning states to dynamically trigger retrieval, generation, or verification operations. The LLM actively evaluates contextual demands during reasoning processes, autonomously determining optimal moments for invoking external tools or resources through a hybrid feedback coordination mechanism. By eliminating fixed iterative units and pre-determined tool-calling sequences, the framework achieves dynamic evolution of execution pathways, demonstrating superior adaptability in complex cognitive tasks through real-time adjustment of computational workflows based on intermediate reasoning outcomes. 

This dynamic architecture manifests three principal characteristics: 1) Operator invocation is governed by the LLM's contextual state analysis, exemplified through special token prediction (e.g., `[Web-Search]` or `<begin\_of\_query>`) to initiate external operations; 2) Reasoning trajectories exhibit high flexibility, allowing dynamic query reformulation and sub-problem generation to overcome limitations of static workflows; 3) Context-driven decision mechanisms prioritize real-time reasoning states over predefined rules, enhancing systemic responsiveness to emergent task complexities while improving  precision.

Defining the reasoning state at time \( t \) as \( S_t = (H_t, C_t) \), where \( H_t \) denotes historical information aggregation and \( C_t \) represents contextual embedding vectors, the decision process is modeled as a stochastic system:

\begin{equation}
a_{t+1} \sim \pi(S_t; \Theta)
\end{equation}

\begin{equation}
S_{t+1} = \delta(S_t, \mathcal{T}_{a_{t+1}}(S_t))
\end{equation}

Here, \( \pi: \mathcal{S} \to \Delta(\mathcal{A}) \) constitutes the policy function mapping states to probability distributions over action space \( \mathcal{A} \) (retrieval, generation, verification, etc.), while \( \mathcal{T}_a \) denotes state transition functions corresponding to action \( a \). The non-Markovian nature of the system emerges from \( S_{t+1} \)'s dependence on complete historical trajectories \( \{S_{\leq t}\} \), with dynamic adaptability ensured through extensible action spaces \( \mathcal{A} \) and online optimization of policy parameters \( \Theta \). This formulation enables context-sensitive state updates via \( \delta: \mathcal{S} \times \mathcal{O} \to \mathcal{S} \), establishing a theoretical foundation for open-ended reasoning processes in complex problem domains.

Based on the mode of reasoning initiation, agentic RAG with dynamic workflows can be further categorized into three distinct types: Proactivity-driven, Reflection-driven, and Feedback-driven mechanisms. 
The LLM proactivity-driven approach is characterized by the model's autonomous triggering of actions based on internal assessments, executing operations without external intervention through mechanisms analogous to human intuitive decision-making—for instance, when the model independently identifies insufficient evidentiary support in the current reasoning process, it proactively generates retrieval requests to supplement information. The reflection-driven mode emphasizes self-examination of the reasoning process, dynamically initiating subsequent operations through quantitative evaluation of intermediate result quality (e.g., triggering actions when the calculated reasoning support score of 0.7 exceeds a predefined threshold of 0.6), which simulates the self-optimization logic of expert systems, enabling the model to adjust reasoning pathways through introspection. The feedback-driven mechanism incorporates external intervention, employing independent models or rule-based systems to perform real-time scoring of intermediate states (e.g., an external reward model assigning a 2.5/5 score to reasoning steps) while providing corrective suggestions, operating similarly to a mentor-guided mode that continuously calibrates the reasoning workflow through external feedback signals.

% \begin{table}
% \centering

% \begin{tabular}{l l l l}
% \textbf{驱动类型} & \textbf{核心机制} & \textbf{决策依据} & \textbf{比喻} \\
% \textbf{LLM主动驱动} & 模型自主判断触发动作 & 内部预测 & 直觉驱动（人类直觉）。例如，运动员直觉反应，在短跑中突然调整呼吸节奏 \\
% \textbf{反思驱动} & 自我评估中间结果质量 & 内部反思或概率阈值 & 自我审视（专家系统），例如厨师味觉校准，熬汤时定期尝味决定加盐 \\
% \textbf{反馈驱动} & 外部模型或规则评分修正 & 外部反馈信号（如奖励模型评分） & 导师指导（实时优化）。例如驾校教练纠偏，倒车时教练拍窗要求调整方向 \\

% \end{tabular}

% \end{table}

\subsubsection{Proactivity-Driven Reasoning}

The core innovation of Proactivity-driven Reasoning lies in enabling LLMs to fully govern the reasoning process through self-triggered prediction mechanisms. This active control manifests through three key mechanisms: (1) direct tool invocation via model-generated special tokens (e.g., [Web-Search]),  without external intervention, (2) context-aware decision making based on real-time knowledge gaps or hypothesis verification requirements, and (3) Markov Decision Process (MDP)-based dynamic path optimization. 

Formally, the reasoning process can be modeled as a state sequence \( S = \{s_0, s_1, \dots, s_t\} \), where each state \( s_t \) encapsulates the current reasoning context. At each step \( t \), the LLM selects an action \( a_t \in \{ \text{retrieve}, \text{generate}, \text{terminate} \} \) based on \( s_t \), executes the corresponding operation (e.g., document retrieval or answer generation), and updates its state through transition function \( s_{t+1} = \delta(s_t, a_t, o_t) \) where \( o_t \) represents action outcomes. This MDP framework enables dynamic path adjustment through real-time feedback until termination (\( a_T = \text{terminate} \)) and final answer generation.

Recent advancements demonstrate significant improvements over conventional RAG approaches. The Agentic Reasoning framework achieves granular control through dynamic tool invocation, eliminating predefined execution sequences. DeepRAG~\cite{deeprag} optimizes cost-accuracy tradeoffs via MDP-based imitation learning, addressing the retrieval-generation disconnection in traditional systems. CoRAG~\cite{CoRAG} introduces hybrid-driven mechanisms combining LLM-initiated subqueries with external policy control, enhancing error tolerance for complex queries. Collectively, these approaches establish a paradigm shift from fixed pipelines to context-sensitive, self-optimizing reasoning architectures.

\subsubsection{Reflection-Driven Reasoning}

The reflection-driven mechanism represents a dynamic reasoning framework that enables iterative self-evaluation and revision of intermediate outputs through model introspection. Common methods include: (1) a  evaluation system combining explicit token prediction and implicit confidence scoring, (2) self-monitoring capabilities through grounding tokens for content-document consistency verification and utility tokens for answer effectiveness assessment, and (3) adaptive routing mechanisms that automatically select single-hop or multi-hop reasoning paths based on contextual complexity. The mathematical formalism of this process can be expressed as:

\begin{equation}
    \mathcal{P} = \bigcup_{t=1}^T \left[ G(\mathbf{C}_t) \rightarrow E(\mathbf{H}_t, \mathcal{D}) \rightarrow \psi(\phi(\mathbf{e}_t), \tau) \right]
\end{equation}

where $G$ denotes the generation function operating on current context $\mathbf{c}_t$, $E$ represents the evaluation function that assesses hidden states $\mathbf{h}_t$ against external knowledge base $\mathcal{D}$, $\phi$ serves as the confidence mapping function, $\tau$ is the decision threshold, and $\psi$ functions as the branch selector. 

In practical implementations like Self-RAG~\cite{self_rag}, this framework generates candidate responses alongside reflection tokens, computes passage relevance scores ($\text{ISREL} \in [0,1]$) and factual support metrics ($\text{ISSUP}$), and employs weighted aggregation of token probabilities in $\phi$ to determine retrieval activation or generation revision through threshold-based $\delta$ operations. Meanwhile, Open-RAG~\cite{openrag} incorporates hybrid threshold mechanisms and Mixture-of-Experts architecture to enforce counterfactual verification through non-retrieval confidence scoring ($\text{Pr}_{\text{NoRT}}$), enabling dynamic expansion of complex reasoning capabilities while preserving base model efficiency. ReaRAG~\cite{rearag} utilizes knowledge-guided reasoning chains combined with external knowledge sources to perform reflection-driven reasoning. In each iteration, it adjusts the reasoning path through the "Thought-Action-Observation" paradigm, effectively preventing error propagation and improving answer accuracy.

The paradigm's innovation lies in reconstructing traditional sequential processes into conditional Markov decision processes, where state transition probabilities $P(s_{t+1}|s_t)$ are dynamically determined by model self-evaluation outcomes. Compared to proactive LLM-driven methods (e.g., Toolformer's direct API invocation), the reflection-driven approach establishes closed-loop control through explicit evaluation stages (function $E$), effectively mitigating hallucination risks while maintaining computational efficiency.

\subsubsection{Feedback-Driven Reasoning}

The feedback-driven dynamic RAG system establishes closed-loop control over reasoning processes through external signals, formally modeled as a Partially Observable Markov Decision Process. The system state $s_t = (q_t, \mathcal{K}_t, H_t)$ evolves through iterative interactions, comprising the current query representation $q_t$, dynamic knowledge base $\mathcal{K}_t$, and historical trajectory $\mathcal{H}_t$. Initialized with $q_0$ and $\mathcal{K}_0 = \emptyset$, the policy function $\pi(a_t|s_t)$ generates actions from the operational space $\mathcal{A} = \{\text{Retrieve}, \text{Reason}, \text{Verify}, \text{Answer}, \emptyset\}$. State transitions follow $s_{t+1} = \delta(s_t, a_t)$ with knowledge base updates
\begin{equation}
    \mathcal{K}_{t+1} = \mathcal{K}_t \oplus \text{Retrieve}(q_t) \cdot \mathbb{I}(a_t=\text{Retrieve})
\end{equation}
where $\oplus$ denotes incremental updates and $\mathbb{I}$ represents an indicator function. The reward function $R(s_t,a_t,s_{t+1}) \rightarrow r_t$ drives policy optimization through
\begin{equation}
    \pi_{t+1} = \Omega(\pi_t, \nabla_{\theta} \mathbb{E}_{a \sim \pi_t}[R(s_t,a,s_{t+1})])
\end{equation}
forming an adaptive control loop. Three distinct feedback mechanisms emerge within this framework.

\textbf{Explicit reward feedback} employs specialized models $\pi_{\text{reward}}$ for quantitative evaluation, exemplified by RAG-Gym's process rewards~\cite{rag-gym}. The reward function combines immediate and terminal rewards: 
\begin{equation}
    r_t = \lambda_1 \pi_{\text{reward}}(s_t) + \lambda_2 \mathbb{E}_{s_{t+k}}[\gamma^k R_{\text{terminal}}]
\end{equation}

with discount factor $\gamma$. SmartRAG extends this through policy gradient optimization 
\begin{equation}
    \nabla_\theta J(\theta) = \mathbb{E}_{\tau \sim \pi_\theta}[\sum_{t=0}^T \nabla_\theta \log \pi_\theta(a_t|s_t) \hat{A}_t]
\end{equation}

where the advantage function $\hat{A}_t$ integrates temporal feedback.

\textbf{Implicit environmental feedback} derives from knowledge base validation, as implemented in KBQA-o1's SPARQL verification and SolutionRAG's pruning mechanisms~\cite{kbqa_o1}. This feedback is formalized as $r_t = \mathbb{I}(\mathcal{K}_t \models q_0) \cdot c_{\text{valid}} - \mathbb{I}(\bot \in \mathcal{K}_t) \cdot c_{\text{invalid}}$ with validation function $\mathbb{I}(\cdot)$ and penalty coefficients $c$. ReARTeR~\cite{rearter} introduces threshold-triggered correction: when $r_t < \tau$, it activates refinement loops $\mathcal{K}_{t+1} = \text{PEM}(\mathcal{K}_t, q_0) \oplus \text{Retrieve}(\text{PRM}(s_t))$.

\textbf{Structured rule feedback} encodes domain knowledge through differentiable scoring functions. MCTS-KBQA~\cite{mcts-KBQA} implements depth-attenuated rewards 
\begin{equation}
    r_t = \frac{1}{1+\alpha d_t} \sum_{i=1}^{n} \text{LLM}_{\text{score}}(a_t^{(i)})
\end{equation}
with search depth $d_t$and decay coefficient $\alpha$. CR-Planner's hierarchical critique combines subgoal and execution scores: $r_t^{\text{total}} = \beta_1 \pi_{\text{sub}}(s_t) + \beta_2 \pi_{\text{exec}}(a_t|s_t)$ through weighted fusion.

These feedback mechanisms interact through a unified strategy update framework, where external feedback-driven approaches achieve controllable optimization of the reasoning process through interpretable feedback signals while maintaining the generative capabilities of LLMs. Overall, the dynamic process of RAG, by endowing the model with autonomy in the reasoning process, not only enhances adaptability to complex tasks but also provides a new solution for efficient reasoning in resource-constrained environments.
\section{Implementation and Optimization}
\label{sec:imple}

Building upon preceding sections, this section systematically analyzes the concrete implementation and optimization strategies for reasoning within the RAG paradigm. In contrast to existing surveys that predominantly focus on post-training methodologies or isolated LLM reasoning mechanisms, our analysis maintains a dedicated focus on the synergistic integration of RAG with reasoning examining their co-adaptive implementations through a structural lens.

\begin{figure*}[htbp]
    \centering
    \includegraphics[width=1\linewidth]{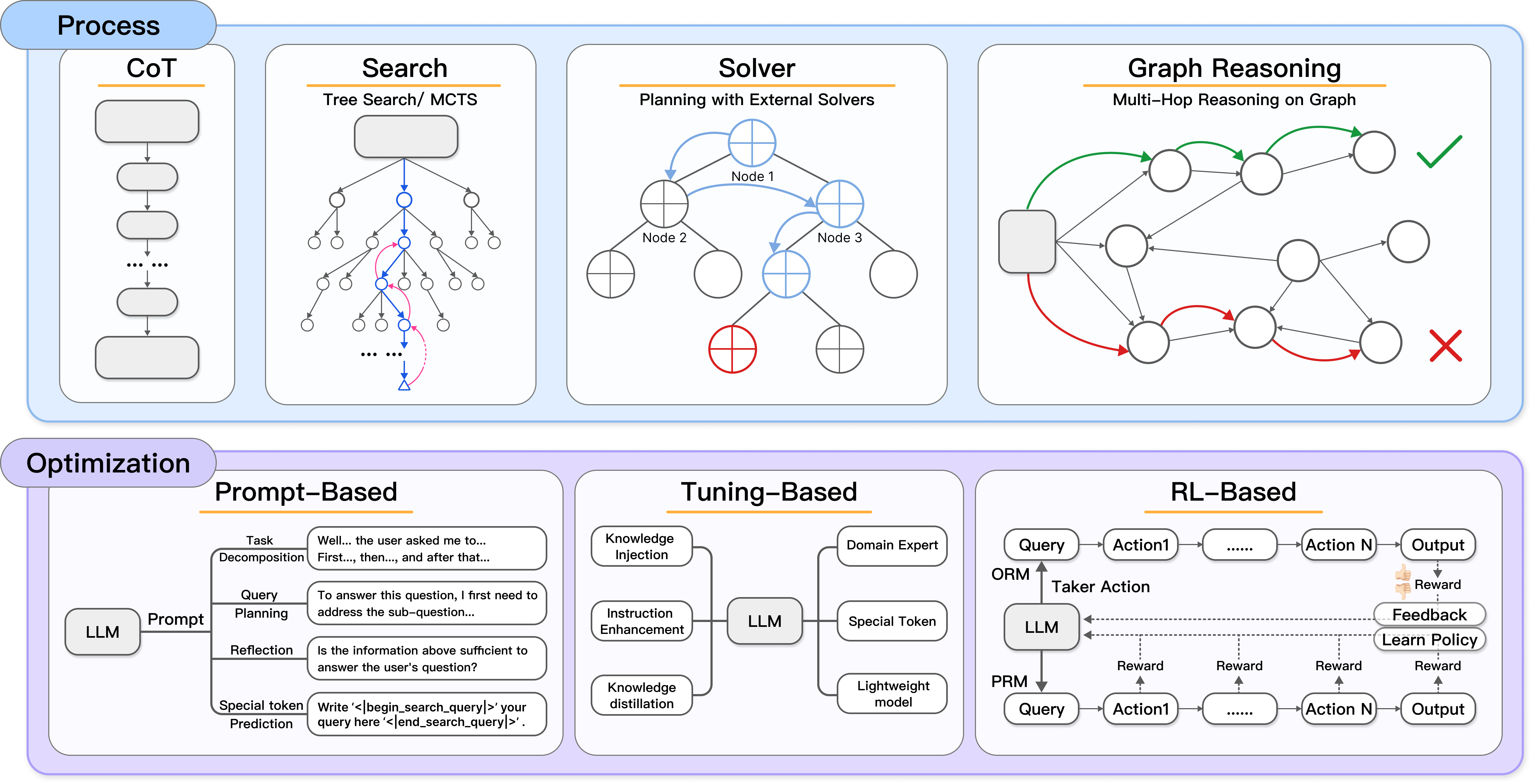}
    \caption{Implementation and optimization of the synergy between RAG and Reasoning}
    \label{fig:implementation}
\end{figure*}

\subsection{Reasoning Process}
\label{subsec:method}

\subsubsection{LLM CoT}

Integrating Chain-of-Thought (CoT) reasoning with LLMs is key to combining RAG with complex reasoning tasks. Research shows CoT enhances RAG systems by explicitly guiding multi-step reasoning and dynamically incorporating external knowledge. For example, ActiveRAG~\cite{activerag} uses a "Self-Inquiry → Knowledge Assimilation → Thought Accommodation" chain to align knowledge and reasoning: a knowledge assimilation agent merges external documents with LLM memory via operations like association and reflection, creating structured knowledge. Meanwhile, a reasoning adaptation agent refines inference chains from Self-Inquiry to ensure answers align with retrieved knowledge and address reasoning gaps. Similarly, Adaptive-RAG~\cite{adaptiveRAG} alternates between CoT and retrieval, breaking down multi-hop reasoning into steps such as entity localization and document correlation, refining retrieval and generation based on prior results.

At the knowledge and reasoning level, O1-Embedder~\cite{o1_embedder} drives RAG through open-ended long-text reasoning, extending CoT beyond fixed triggers via coherent thought processes like problem decomposition. PlanRAG~\cite{planrag} explicitly uses CoT to produce executable multi-step plans, adjusting operations dynamically through a closed-loop "plan-execute-feedback" cycle. Despite different implementations, these methods share two CoT strengths: breaking down complex problems into clear intermediate steps and guiding external knowledge selection through reasoning states. Studies show these approaches outperform traditional RAG in multi-hop QA and knowledge-intensive tasks by enhancing both LLMs' reasoning and adaptability to external knowledge.

\subsubsection{Special Token Prediction}

Recent advances active RAG also highlight special token prediction as a key method for dynamically linking external knowledge retrieval with multi-step reasoning~\cite{ReZero}. By embedding domain- or action-specific tokens (e.g., `[Web-search]`, `[Retrieve=Yes]`, `<begin\_of\_query>`) into LLM vocabularies, models can autonomously trigger tools or self-reflect during text generation. Frameworks like Self-RAG~\cite{self_rag} and SmartRAG~\cite{smartrag} use dedicated tokens (`Retrieve`, `ISREL`, `[RETRIEVE]`) to manage retrieval activation, relevance checks, and output verification, turning static reasoning chains into conditional workflows. The innovation lies in predicting these tokens within generated sequences, segmenting tasks into retrieval initiation, document evaluation, and knowledge grounding phases.

Hybrid models such as Open-RAG~\cite{openrag} combine token control with mixture-of-experts (MoE) routing, sparsely activating experts aligned with token-predicted reasoning. Unlike traditional chain-of-thought or search tree methods, special token prediction offers finer control and interpretability by encoding decision logic explicitly in token sequences while maintaining end-to-end training. This approach also overcomes latency and inflexibility of preset retrieval schedules by enabling context-aware, on-demand tool use. For example, R1-Searcher~\cite{R1-Searcher} and Search-o1~\cite{search_o1} use token boundaries like `<end\_of\_query>` to coordinate retrieval pauses and resume generation after knowledge integration.

Together, these systems show that token-level prediction not only bridges reasoning and retrieval but also creates a scalable framework for tool-enhanced language agents, preserving generative fluency while enabling systematic external knowledge integration and procedural reasoning.

\subsubsection{Search-Driven Reasoning}

Recent advancements in search-driven reasoning have significantly improved RAG frameworks by employing structured search strategies for dynamic information exploration and multi-step reasoning with external knowledge. Current approaches mainly follow three paradigms: tree-based search, MCTS, and reinforcement learning-optimized policy networks.

Tree-based methods organize reasoning hierarchically through structured path exploration. For example,StePO-Rec~\cite{StePO-Rec} uses a multi-step tree-structured reasoning method that iteratively retrieves different outfit matching knowledge and user preferences at each node, ultimately achieving generative recommendations for complementary items. OmniThink~\cite{omnithink} uses an information tree to expand topic analysis by generating subqueries that guide breadth-first or depth-first retrievals. DeepRAG~\cite{deeprag} applies a binary tree search within a Markov decision process to explore parametric knowledge and retrieval paths in parallel, selecting optimal branches. DeepSolution’s~\cite{li2025deepsolution} bidirectional thinking tree alternates expanding solution and critique nodes with scoring for path pruning, aligning naturally with MCTS evaluation. These methods balance exploration efficiency with solution coverage through explicit tree structures.

MCTS enhances robustness by optimizing long-term decisions via simulation, evaluation, and backpropagation. CR-Planner~\cite{CR-planner} integrates MCTS with the UCB strategy to balance exploration and exploitation while estimating optimal subgoals through multi-step simulations. KBQA-O1~\cite{kbqa_o1} and MCTS-KBQA~\cite{mcts-KBQA} generate candidate actions using policy models and combine reward models to globally assess logical forms, reducing local optima. ReARTeR~\cite{rearter} innovatively merges MCTS with procedural reward models (PRMs), interleaving retrieval and reasoning steps, and filtering high-reward paths to form a closed-loop "reason-retrieve-reason" cycle. These methods probabilistically explore paths and use reinforcement learning feedback to improve global reasoning for complex tasks.

Reinforcement learning-optimized policy networks adaptively refine search strategies. LeReT~\cite{LeReT} replaces fixed search algorithms with reinforcement learning (e.g., IPO) to dynamically optimize query generation based on rewards like retrieval accuracy, implicitly learning optimal search patterns without explicit tree or graph structures, thus offering greater flexibility and scalability.

In summary, search-driven reasoning unites inference and retrieval through structured strategies, combining multi-path exploration, dynamic evaluation, and adaptive optimization to deliver interpretable, efficient solutions for knowledge-intensive tasks. Future work may focus on hybrid paradigms (e.g., integrating MCTS and reinforcement learning) and lightweight algorithms to balance performance with computational efficiency.

\subsubsection{Reasoning on Graph}

Graph-structured reasoning offers a novel approach for multi-hop inference in RAG systems by explicitly modeling knowledge interaction paths through topology. Current methods fall into two categories: query-flow-oriented search graphs (e.g. FinSearch~\cite{FinSearch}) and knowledge-association-based expansion graphs (ToG-2.0~\cite{ToG2}). FinSearch builds a directed acyclic graph (DAG) where nodes are atomic subqueries (e.g., stock prices, financial reports) and edges capture logical and temporal dependencies. A pre-planner breaks down queries into subquery sequences, using graph traversal to control information flow and dynamically adjust paths, such as backtracking when conflicts arise—substantially surpassing linear chain-of-thought methods in handling complex logic.

\subsubsection{External Solver}

The integration of RAG and reasoning is also can be achieved by incorporating external solvers, where specialized solvers, such as the Alignment-Oriented LLM-based Retrieval Method (ARM), are employed to handle the reasoning component. The retrieval process for complex problems is formulated as a global optimization task, leveraging external solvers like mixed-integer programming (MIP) to achieve structural alignment and joint optimization of data objects. Specifically, ARM first decomposes user queries into keywords that match N-grams in the dataset through an information alignment module, generating an initial set of retrieval candidates via constrained decoding. Subsequently, in the structural alignment phase, the MIP solver performs global filtering on candidate objects based on a predefined objective function that maximizes both the relevance of retrieved objects to the query and their mutual compatibility. This ensures that the selected objects not only cover the requirements of the query but also form a coherent information chain through entity or inter-table linkages. Finally, the self-verification mechanism of the LLM, combined with a beam search-based aggregation strategy, dynamically refines and consolidates multiple candidate sets, ultimately producing a retrieval collection that satisfies both semantic matching and the structural organization of the data.

ToG-2.0 achieves multi-hop expansion by integrating knowledge graphs with documents, starting from an initial entity and iteratively extending relevant entities and relations (such as corporate ownership chains and technology dependency networks) via the Edge function. This process constructs structured triple paths while simultaneously retrieving and verifying document content. By tuning the width and depth parameters, the method emulates human reasoning: broadly exploring potential associations before deeply verifying high-confidence paths. FRAG~\cite{frag} dynamically adjusts retrieval strategies by predicting the hop range of reasoning paths based solely on the query text, thereby enhancing retrieval quality without requiring additional fine-tuning or invocation of large language models, enabling flexible and efficient retrieval optimization. FG-RAG~\cite{FG-RAG} further expands entity coverage in graph retrieval through context-aware entity expansion, providing richer background information. Combined with query-level fine-grained summary generation, FG-RAG transforms coarse-grained graph information into highly relevant detailed content, effectively improving the performance of query-focused summarization tasks.

Although differing in design from workflow-based methods, ToG-2.0 shares key advantages with other graph-structured approaches: explicitly modeling reasoning state dependencies, supporting dynamic path generation and optimization, and enabling closed-loop interaction between retrieval and reasoning. This effectively overcomes the limitations of traditional RAG in implicit relation inference and counterfactual analysis, thereby establishing an interpretable theoretical and practical framework for knowledge reasoning.

\subsection{Reasoning Optimization}
\label{subsec:opt}

In the previous chapter, we focused on introducing several approaches to integrate reasoning with RAG. This chapter shifts attention to how to augment the reasoning capabilities, specifically including Prompt-Based, Tuning-Based, and RL-Based strategies.

\subsubsection{Prompt-Based}
Prompt-Based optimization is a key approach to improving RAG and reasoning system performance by using carefully designed natural language prompts. These prompts break down complex reasoning tasks into manageable steps and guide LLMs to follow specific logical structures during generation. The main advantage is that control over reasoning flow is achieved solely through prompt design, without parameter fine-tuning or reinforcement learning, preserving the model’s generalization while enhancing task-specific results.

This approach has three main features. First, \textbf{task structuring}: prompts explicitly decompose and control reasoning chains via zero-shot or templated designs. Techniques like Co-STORM~\cite{co-storm} and WriteHere~\cite{WriteHere} use role assignments, stage divisions, and operation-specific instructions to guide multi-step reasoning—such as proposal generation, knowledge retrieval, refinement, and validation—improving interpretability by representing intermediate steps clearly.

Second, \textbf{result reliability} is improved by standardizing outputs and reducing hallucinations. Strategies include requiring citation of retrieval results, enforcing specific output formats, and integrating reflection and calibration based on retrieved knowledge. Systems like FinSearch~\cite{FinSearch} and ActiveRAG~\cite{activerag} incorporate temporal weighting, deduplication, and domain rules through prompts, enhancing consistency and logical coherence, especially in complex domains.

Third, \textbf{interactive adaptability} allows dynamic prompt adjustments. Special tokens (e.g., \verb|<Search>|, \verb|[Web-search]|) enable models to trigger tools or revise queries in real time based on intermediate results. Methods such as Agentic Reasoning~\cite{agentic_reasoning} and PlanRAG~\cite{planrag} use context-sensitive prompts and feedback loops to refine reasoning paths dynamically, maintaining coherence and accuracy in multi-hop tasks and outperforming traditional RAG methods in complex, evolving scenarios.

In summary, prompt-based optimization offers an efficient, flexible, and reliable approach to enhancing RAG+Reasoning by emphasizing task structuring, result standardization, and interactive adaptability. Its non-intrusive and broadly applicable design has established it as a mainstream strategy for optimizing LLM reasoning and serves as a foundation for future hybrid methods integrating fine-tuning and reinforcement learning. By systematically optimizing reasoning without altering model parameters through semantic structures, dynamic feedback, and symbolic constraints, this paradigm effectively manages macro-level controls like task decomposition and knowledge integration while addressing key challenges such as generation consistency, logical coherence, and external knowledge alignment. This makes prompt-based optimization a lightweight yet powerful solution for complex reasoning tasks.

\begin{table*}[htbp]
\centering
\caption{Comparison of RL-based RAG with Reasoning Methods}
\label{tab:rag_comparison}
\resizebox{\textwidth}{!}{%
\begin{tabular}{lllllll}
\toprule
\textbf{Method} & \textbf{Base Model} & \textbf{RL} & \textbf{Parameter} & \textbf{Supervision } & \textbf{Reward Function} & \textbf{Policy Strategy} \\
\midrule
PORAG~\cite{PORAG} & Qwen2.5/Llama3.2 & GRPO & QLoRA  & ORM & 
\begin{tabular}[c]{@{}l@{}}Dual rewards:\\1. Retrieval fidelity ($ R_{\text{fid}}$)\\2. Response quality ($R_{\text{qual}}$)\\Combined: $R = \alpha R_{\text{fid}} + \beta R_{\text{qual}}$ \end{tabular} & 
\begin{tabular}[c]{@{}l@{}}• Group-based advantage normalization\\• PPO-style clipped objective\\• KL regularization \end{tabular} \\
\midrule

DeepResearcher~\cite{deepresearcher} & Qwen2.5-7B & GRPO & Full & ORM & 
\begin{tabular}[c]{@{}l@{}}Format compliance penalty (-1)\\ + Answer F1 score \end{tabular} & 
\begin{tabular}[c]{@{}l@{}}• Reference policy constraints\\• KL divergence penalty \end{tabular} \\
\midrule

ReSearch~\cite{ReSearch} & Qwen2.5-7B & GRPO & Full & ORM & 
\begin{tabular}[c]{@{}l@{}}Hybrid rewards:\\• Answer F1 (vs ground truth)\\• Format compliance check \end{tabular} & 
\begin{tabular}[c]{@{}l@{}}• GRPO with clip ratio 0.2\\• Group advantage normalization (G=5)\\• $\beta=0.001$ KL penalty \end{tabular} \\
\midrule

ReZero~\cite{ReZero} & Llama3.2-3B & GRPO & Full & ORM+PRM & 
\begin{tabular}[c]{@{}l@{}}• Answer correctness\\• Format compliance\\• Search diversity\\• Chunk matching\\• Retry behavior\\• Strategy compliance \end{tabular} & 
\begin{tabular}[c]{@{}l@{}}• Intra-group reward comparison\\• Noise-injected robustness training\\• KL constraints \end{tabular} \\
\midrule

MMOA-RAG~\cite{mmoa_rag} & Llama-3-8B & MAPPO & Full & ORM & 
\begin{tabular}[c]{@{}l@{}}Shared F1 reward + penalties:\\• Excessive sub-questions\\• Document ID errors\\• Answer verbosity \end{tabular} & 
\begin{tabular}[c]{@{}l@{}}• MAPPO actor-critic updates\\• Cosine learning rate scheduling \end{tabular} \\
\midrule

DeepNote~\cite{DeepNote} & Qwen2.5/Llama3.1 & DPO & Full & ORM & 
\begin{tabular}[c]{@{}l@{}}Implicit preference modeling\\via likelihood contrast \end{tabular} & 
\begin{tabular}[c]{@{}l@{}}• Direct Preference Optimization\\• Preference gap maximization \end{tabular} \\
\midrule

R1-Searcher~\cite{R1-Searcher} & Qwen2.5/Llama3.1 & Reinforce++ & Full & ORM & 
\begin{tabular}[c]{@{}l@{}}Two-stage rewards:\\1. Retrieval count + format\\2. F1 score + format penalty \end{tabular} & 
\begin{tabular}[c]{@{}l@{}}• RAG-based rollout\\• Retrieval-masked loss \end{tabular} \\
\midrule

KBQA-O1~\cite{kbqa_o1} & Llama3/Qwen2.5/Gemma2 & MCTS & DoRA & ORM+PRM & 
\begin{tabular}[c]{@{}l@{}}Composite reward:\\• Stepwise policy model score\\• Final reward model score \end{tabular} & 
\begin{tabular}[c]{@{}l@{}}• MCTS trajectory optimization\\• Q-value backpropagation \end{tabular} \\
\midrule

DeepRetrieval~\cite{deepretrieval} & Qwen2.5-3B & PPO & Full & ORM & 
\begin{tabular}[c]{@{}l@{}}Task metrics:\\• Recall@k/NDCG\\• Syntax validity \end{tabular} & 
\begin{tabular}[c]{@{}l@{}}• GAE advantage estimation\\• Distributed HybridFlow \end{tabular} \\
\midrule

LeReT~\cite{LeReT} & Llama3-8B/Gemma-9B& IPO & Full & PRM & 
\begin{tabular}[c]{@{}l@{}}Average Precision (AP)\\of retrieved documents \end{tabular} & 
\begin{tabular}[c]{@{}l@{}}• Identity Policy Optimization\\• Context distillation \end{tabular} \\
\midrule

SmartRAG~\cite{smartrag} & Flan-T5-L/Llama2-7B & PPO & Full/LoRA & ORM & 
\begin{tabular}[c]{@{}l@{}}Action-specific:\\• EM+F1 for answers\\• Cost penalty for retrievals \end{tabular} & 
\begin{tabular}[c]{@{}l@{}}• On-policy sampling\\• PPO updates \end{tabular} \\
\midrule

ReARTeR~\cite{rearter} & LLaMA3.1-8B & MCTS & LoRA & ORM+PRM & 
\begin{tabular}[c]{@{}l@{}}Monte Carlo step scoring\\+ TD look-ahead \end{tabular} & 
\begin{tabular}[c]{@{}l@{}}• Iterative preference optimization\\• KTO loss \end{tabular} \\
\midrule

DeepRAG~\cite{deeprag} & Qwen2.5-7B/Llama3.1-8B & Hybrid & Full & ORM+PRM & 
\begin{tabular}[c]{@{}l@{}}Cost-aware accuracy:\\$R = -C(o) \times T(s_t)$\\
$C(o)$: Answer correctness\\ $T(s_t)$: Total retrieval cost 

\end{tabular} & 
\begin{tabular}[c]{@{}l@{}}• Imitation + contrastive learning\\• PPO-like calibration \end{tabular} \\
\midrule

RAG-Gym~\cite{rag-gym} & LLaMA3.1-8B & Hybrid & LoRA & PRM & 
\begin{tabular}[c]{@{}l@{}}Triple criteria:\\• Sufficiency\\• Utility\\• Redundancy \end{tabular} & 
\begin{tabular}[c]{@{}l@{}}• SFT + DPO\\• PRM-guided selection \end{tabular} \\
\midrule

CR-Planner~\cite{CR-planner} & Skywork-Llama3.1-8B & MCTS & LoRA & PRM & 
\begin{tabular}[c]{@{}l@{}}Critic-estimated rewards:\\• Stepwise correctness\\• Global impact \end{tabular} & 
\begin{tabular}[c]{@{}l@{}}• MCTS simulation\\• Pairwise ranking loss \end{tabular} \\
\bottomrule
\end{tabular}%
}
\vspace{2mm}
\small
\textsuperscript{1}ORM: Outcome-based Reward Model; PRM: Process-based Reward Model.
\textsuperscript{2}Full: Full parameter tuning.
\end{table*}

\subsubsection{Tuning-Based}

The tuning-based approach improves the integration of  RAG and reasoning by optimizing model parameters to internalize the retrieval-augmented chain-of-thought mechanism within LLMs. Current research mainly targets three goals: \emph{retrieval pathway optimization}, \emph{structured generation enhancement}, and \emph{collaborative training with external modules}.

For retrieval pathway optimization, methods like CoRAG~\cite{CoRAG} and DeepRAG~\cite{deeprag} build end-to-end multistep reasoning frameworks through full parameter fine-tuning and multitask learning. CoRAG expands single-step QA datasets into retrieval-reasoning chains and jointly trains tasks such as sub-query generation, intermediate answer prediction, and final composition. This boosts the model’s ability to break down complex problems (e.g., multi-entity relational reasoning) and adapt retrieval strategies dynamically (e.g., query rewriting, error correction). DeepRAG combines imitation and contrastive learning with binary tree search to create efficient retrieval paths, using a DPO-style contrastive loss to reduce redundant retrieval while maintaining accuracy.

To improve structured generation, MCTS-KBQA~\cite{mcts-KBQA}and Self-RAG~\cite{self_rag} fine-tune models for precise special token generation. MCTS-KBQA uses supervised fine-tuning to make large language models output instructions that comply with knowledge graph protocols (e.g., SPARQL), modeling reasoning as executable tool-call sequences. Self-RAG enhances self-supervised generation control by expanding vocabulary and training the model to generate reflection tokens like retrieval triggers and relevance markers, preserving fluency and reducing factual errors. Additionally, O1-Embedder~\cite{o1_embedder} and Open-RAG~\cite{openrag} align semantic spaces via mixed fine-tuning: O1-Embedder combines generative and contrastive training with special tokens to separate generation from embedding tasks, enhancing multihop semantic understanding; Open-RAG uses QLoRA~\cite{qlora} quantized fine-tuning and Mixture of Experts (MoE) modules to specialize networks for single/multi-hop reasoning.

In collaborative optimization with external modules, AdaptiveRAG~\cite{adaptiveRAG} and CR-Planner~\cite{CR-planner} apply parameter isolation to balance generality and adaptability. AdaptiveRAG fine-tunes a lightweight classifier to select retrieval strategies dynamically. CR-Planner introduces a Critic model trained with contrastive loss on MCTS trajectory data to assess the long-term value of reasoning actions, prioritizing efficient solutions in tasks like mathematical reasoning.

Together, these tuning strategies restructure the parameter space to internalize retrieval-reasoning interactions effectively, enhancing the model’s ability to solve complex problems while ensuring computational efficiency and broad applicability across domains.

\subsubsection{RL-Based}

As shown in Table~\ref{tab:rag_comparison}, Reinforcement learning (RL) has recently become pivotal for tackling long-chain reasoning in modern inference models and optimizing RAG combined with reasoning tasks. Central to these advances is the use of dynamic reward mechanisms that guide LLMs  to balance knowledge retrieval and logical reasoning adaptively. RL optimization objectives generally fall into two categories: outcome-based reward modeling (ORM) and process-based reward modeling (PRM), with some hybrid approaches blending both to balance global goals and local optimizations.

The ORM paradigm focuses solely on the quality of the final output and its adherence to standards. For example, R1-Searcher~\cite{R1-Searcher} employs a two-stage Reinforce++~\cite{reinforce++} training where rewards in the first stage depend on correct retrieval calls and special token generation, while the second stage directly optimizes the F1 score of answers. This encourages the model to develop strategies maximizing knowledge integration, reducing hallucinations, and enhancing accuracy in multi-hop QA beyond traditional RAG methods. Similarly, KBQA-O1 ~\cite{kbqa_o1}uses MCTS with a policy network for candidate reasoning paths and a reward model evaluating logical consistency, effectively balancing exploration and exploitation in knowledge base QA.

Conversely, PRM emphasizes detailed supervision of intermediate reasoning steps. LeReT~\cite{LeReT} uses the Identity Policy Optimization (IPO) algorithm, optimizing query quality by rewarding average precision (AP) of retrieved documents, boosting retrieval recall and overall multi-hop task performance. ReARTeR~\cite{rearter} extends this with a step-level binary reward model, combining Monte Carlo scoring and temporal difference (TD) methods to evaluate reasoning paths proactively, reducing logical errors and redundant retrievals, and improving accuracy on benchmarks like HotpotQA.

Moreover, influenced by DeepSeek-R1, GRPO~\cite{grpo} is also gradually being applied in scenarios combining RAG and Reasoning. GRPO is a variant of the Proximal Policy Optimization (PPO) reinforcement learning algorithm that abandons the critic model and instead estimates the baseline from group scores, significantly reducing training resources. For example, ReZero~\cite{ReZero} uses GRPO to introduce a "retry" mechanism for LLMs, incentivizing LLMs to keep trying after an initial search failure by rewarding retry search queries. This mechanism simulates the human strategy of "if at first you don't succeed, try again" in information retrieval. PORAG~\cite{PORAG}, based on GRPO, directly optimizes retrieval quality, contextual relevance, and generation coherence through a dual reward mechanism (retrieval fidelity and response quality).

Hybrid methods merge ORM and PRM to optimize both final outcomes and intermediate steps via composite rewards. SmartRAG~\cite{smartrag} applies Proximal Policy Optimization (PPO), combining answer-level F1 rewards with penalties for excessive retrievals, balancing knowledge completeness and efficiency. RAG-Gym~\cite{rag-gym}advances this with multidimensional process rewards (sufficiency, utility, redundancy) and techniques like contrastive loss and Best-of-N sampling to promote efficient search decisions, even zero-shot. These hybrid strategies markedly lower retrieval costs while sustaining accuracy in complex tasks.

In addition, we can also observe that in current RL-based methods, academia focuses more on exploration with small-scale LLMs (<8B), among which the Qwen and Llama series are the most widely used. Overall, RL provides a flexible, scalable framework for integrating RAG and reasoning. ORM guides the discovery of globally optimal strategies, PRM enhances reasoning robustness via local refinements, and their combination addresses modular system limits. Future work may explore collaborative rewards in multi-agent settings, offline RL based on world models, and hierarchical reward decomposition for open-domain applications.

\section{Downstream Tasks and Evaluation}

While previous chapters focused on methodologies and advances in RAG combined with reasoning, this chapter shifts to tasks and evaluation. It provides a comprehensive overview and analysis of existing tasks, datasets, their current status, and emerging trends. By reviewing these resources, we highlight the landscape’s gaps and limitations in current evaluation methods. The chapter also explores key challenges in assessment frameworks, identifying shortcomings and suggesting potential improvements.

\begin{figure}[htbp]
    \centering
    \includegraphics[width=1\linewidth]{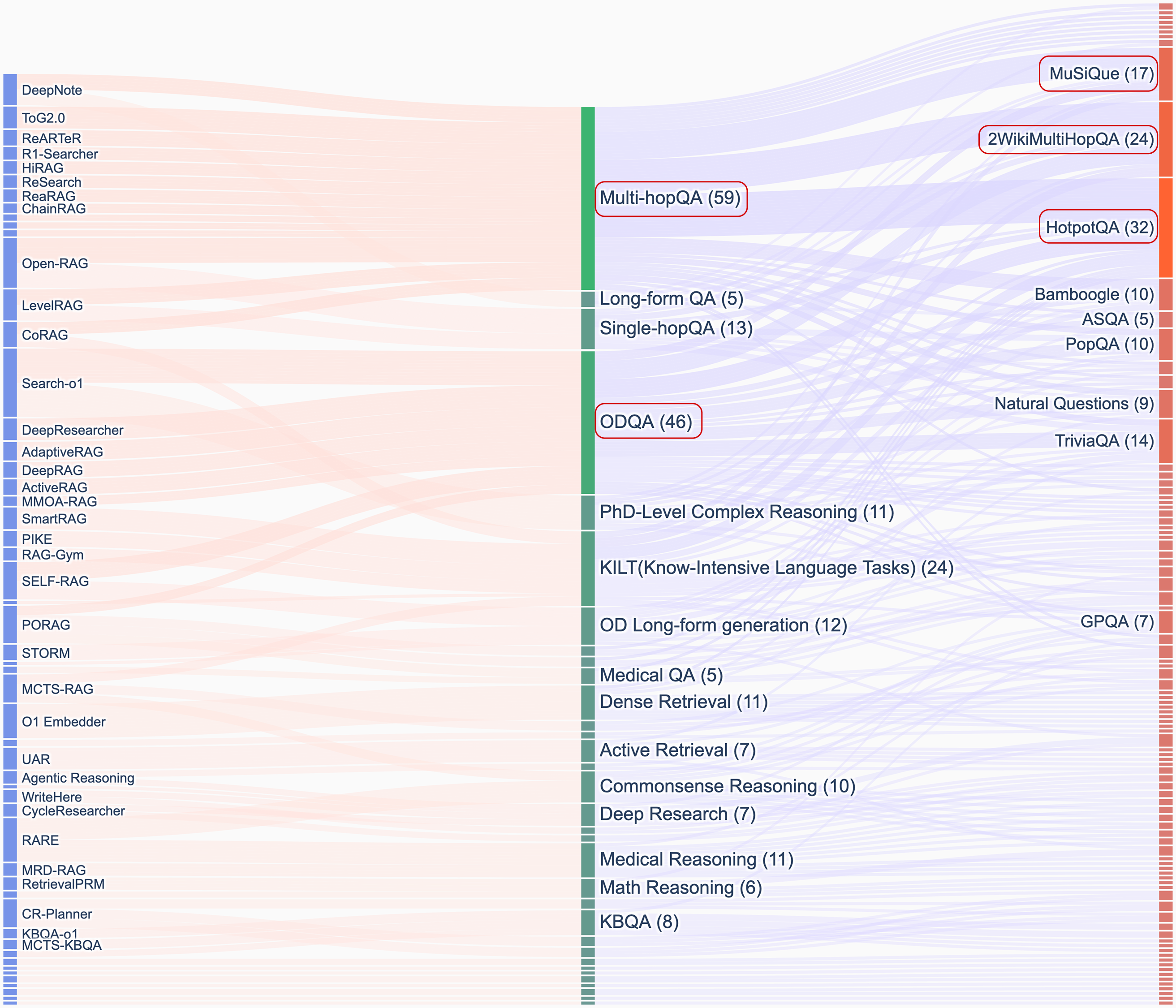}
    \caption{The current downstream tasks and datasets related to the combination of RAG and Reasoning show that multi-hop question answering tasks still dominate. Correspondingly, HotpotQA, 2WikiMultihopQA, and MuSiQue remain the most commonly used evaluation datasets.}
    \label{fig:sankey}
\end{figure}
\subsection{Knowledge-Intensive Tasks}

In the evaluation for RAG systems, knowledge-intensive question answering (QA) remains the primary focus (Figure~\ref{fig:sankey}). As LLMs improve in semantic understanding and reasoning, benchmarks have expanded to cover tasks from simple fact retrieval to complex multi-step reasoning. However, evaluation methods specifically designed for RAG lag behind due to the dual challenge of assessing both retrieval-generation coherence and adaptability to dynamic knowledge bases. For example, multi-hop QA requires integrating dispersed knowledge through multi-stage retrieval while verifying logical consistency between answers and retrieval paths. This complexity increases dataset construction costs compared to purely generative tasks, keeping research centered on knowledge-intensive QA subcategories such as open-domain QA, knowledge-base QA, and multi-hop QA.

Commonly used datasets include Natural Questions (NQ)~\cite{nq} for single-hop factual queries, HotpotQA, 2WikiMultiHopQA~\cite{2WikiMultiHopQA} and Musique~\cite{musique} for multi-hop QA. These benchmarks are mostly based on Wikipedia and fail to reflect the RAG demands and corresponding complexity in real-world scenarios. Some efforts have pushed evaluation boundaries, like CRUD-RAG’s~\cite{crud} operational metrics and DomainRAG’s~\cite{domainrag} domain-specific evaluations, but high costs and metric-task interdependencies limit progress. As a result, knowledge-intensive QA remains central for testing RAG robustness and practicality, highlighting a critical bottleneck: the need for innovative frameworks that balance retrieval flexibility and controlled generation to support new developments like Agentic RAG .Overall, many evaluation benchmarks are lagging behind rapid RAG+Reasoning advances, especially as LLMs grow more powerful.  Specifically, the current evaluation of RAG faces the following challenges.

\paragraph{Limited Challenge}

With improving LLM capabilities, many knowledge-based questions are no longer difficult, as they can be answered without external retrieval. Current multi-hop reasoning datasets, often built from artificial templates, offer limited challenge. There is an urgent need for more complex datasets reflecting real-world scenarios and practical use.

\paragraph{Lack of Specificity}

Existing evaluation tasks are still predominantly focused on factual assessment and knowledge retrieval, lacking evaluations that probe deeper analytical thinking. This constraint limits the ability to measure a model's capacity for profound reasoning and cognitive depth.

\paragraph{Task Uniformity}

The majority of benchmarks are overly dependent on QA tasks, focusing on reactive, question-and-answer-based interactions. There is a pressing need to introduce tasks aligned with real-world applications, such as active information retrieval tasks based on personal knowledge or proactive knowledge discovery.

\paragraph{Insufficient Dimensions}

Evaluations are primarily end-to-end, focusing solely on final outcomes. However, with the introduction of reasoning processes, RAG+Reasoning systems have become iterative, multi-step frameworks. Current evaluations are unable to assess intermediate reasoning steps or retrieval chains effectively. The absence of step-by-step supervision data limits both research and training of related methods. Furthermore, current evaluation methodologies lack comprehensive assessments of system performance trade-offs, such as computational cost and efficiency, which are critical for practical deployment.

This emergent landscape necessitates the creation of a new generation of evaluation frameworks that can address these shortcomings. Such frameworks must not only ensure the adaptability of retrieval and the controllability of generation but also integrate intermediate reasoning evaluation and efficiency metrics, paving the way for the development of more robust and efficient RAG systems suited to diverse real-world applications.

\subsection{New Tasks on RAG+Reasoning}
Recently, combining RAG with reasoning has significantly improved models’ ability to tackle more realistic and challenging tasks, raising the standards for evaluation methods. This subsection examines emerging tasks that assess their combined strengths, related tasks and datasets are shown in Table~\ref{tab:dataset}. Here, "emerging" refers not to entirely new tasks but to those with unprecedented complexity and demands. These include Deep Research tasks requiring multi-layered information integration and reasoning; PhD (Expert)-Level Complex Reasoning tasks targeting advanced scenario reasoning; and critical; domain-specific  decision support tasks like medical diagnosis and legal analysis. Such tasks demand not only external knowledge retrieval but also logical consistency, coherence, and depth in reasoning.

\begin{table*}[htbp]
\centering
\caption{Tasks and Datasets under the New Trend of RAG Combined with Reasoning}
\label{tab:dataset}
\resizebox{\textwidth}{!}{%
\begin{tabular}{@{}p{2.2cm}p{2.5cm}p{3cm}p{4cm}p{2.3cm}p{2.5cm}p{2.5cm}p{0.8cm}@{}}  
\toprule
\textbf{Task Type} & 
\textbf{Sub-Task} & 
\textbf{Dataset} & 
\multicolumn{1}{c}{\textbf{Description}} & 
\multicolumn{1}{c}{\textbf{Scale}} & 
\multicolumn{1}{c}{\textbf{Construction By}} & 
\multicolumn{1}{c}{\textbf{Evaluation}} & 
\textbf{Paper} \\ \midrule

\multirow{6}{*}{\begin{tabular}[c]{@{}c@{}}Deep Research \end{tabular}} & 
Deep Research & 
Agentic Reasoning Deep Research~\cite{agentic_reasoning} & 
PHD-level dataset covering finance, medicine, and law.& 
15-30 domains & 
PhD Experts & 
Expert pass rate & 
\cite{agentic_reasoning} \\ 

& Report Generation & 
WildSeek~\cite{WildSeek} & 
Info-seeking task–goal pairs for document generation. & 
100 samples & 
Rules/LLM/Manual & 
LLM& 
\cite{WriteHere} \\ 

& Report Generation & 
TELL ME A STORY~\cite{tell_me_a_story} & 
fiction writing evaluation dataset: detailed prompts and long-form narratives. & 
230 samples & 
Manual & 
LLM & 
\cite{WriteHere} \\ 

& Peer Review & 
Review-5k~\cite{CycleResearcher} & 
ICLR 2024 peer review dataset: paper metadata and structured reviewer feedback. & 
4,991 papers & 
OpenReview/arXiv & 
MSE/MAE/Acc & 
\cite{CycleResearcher} \\ 

& Report Generation & 
Research-14k~\cite{CycleResearcher} & 
2022–2024 Accepted ML papers: outlines, full texts, and cited abstracts. & 
14,911 papers & 
Semantic Scholar + arXiv & 
Simulated review scores & 
\cite{CycleResearcher} \\ 

& Report Generation & 
SolutionBench~\cite{li2025deepsolution} & 
Engineering benchmark: constrained solutions across 8 real-world domains. & 
1,050 datapoints & 
Manual/LLM extraction & 
Analytical/ Technical scores & 
\cite{li2025deepsolution} \\ 
\midrule

\multirow{12}{*}{\begin{tabular}[c]{@{}c@{}}Mathematics \&\\ Reasoning\end{tabular}} & 
Math Reasoning & 
GPQA~\cite{gpqa} & 
PHD-level MCQs in physics, chemistry, and biology. & 
744 sets & 
PhD Experts & 
Accuracy & 
\cite{agentic_reasoning} \\ 

& Math Reasoning & 
MATH500~\cite{MATH500} & 
500 math problems from the MATH test set. & 
500 problems & 
Public repos & 
Pass@K & 
\cite{search_o1} \\ 

& Programming & 
LiveCodeBench~\cite{LiveCodeBench} & 
Programming benchmark with easy, medium, and hard problems. & 
1,055 problems & 
Competition platforms & 
Pass@K & 
\cite{search_o1} \\ 

& Programming & 
USACO~\cite{USACO} & 
USA Computing Olympiad problems, testing algorithms and coding. & 
307 problems & 
USA Computing Olympiad & 
Pass@K& 
\cite{CR-planner} \\ 

& Math Reasoning & 
TheoremQA-Math~\cite{TheoremQA-Math} & 
BRIGHT subset: theorem-based math problems. & 
206 problems & 
STEM datasets & 
Accuracy & 
\cite{CR-planner} \\ 

& Programming & 
Gorilla~\cite{Gorilla} & 
API-aware code generation from HuggingFace, Torch Hub, TensorFlow Hub docs. & 
1,600 APIs & 
Manual  & 
AST matching & 
\cite{PORAG} \\ 

& Math Reasoning & 
OlympiadBench~\cite{olympiadbench}& 
Olympiad-level math competition problems. & 
1,000 problems & 
Competitions & 
Accuracy/F1 & 
\cite{RetrievalPRM} \\ 

& Complex Reasoning & 
ComplexWebQA~\cite{ComplexWebQA}& 
Multi-step reasoning over web queries with cross-document integration. & 
34,689 queries & 
Web snippets & 
Accuracy & 
\cite{MCTS-RAG} \\ 
\midrule

\multirow{7}{*}{\begin{tabular}[c]{@{}c@{}}Demanding \\ Retrieval\end{tabular}} & 
Domain Retrieval & 
StackEcon \& StackBio~\cite{TheoremQA-Math} & 
Biology and economics StackExchange questions for complex retrieval. & 
206 queries & 
StackExchange & 
nDCG@K & 
\cite{CR-planner} \\ 

& Active Retrieval & 
AR-Bench~\cite{UAR} & 
Active retrieval benchmark with four sub-tasks.& 
8k/sub-task & 
Synthetic & 
Accuracy & 
\cite{UAR} \\ 

& Real-time & 
TAQA~\cite{TAQA} & 
QA dataset with time-evolving answers. & 
10K-100K rows & 
Human-curated & 
LLM & 
\cite{UAR} \\ 

& Real-time  &
FreshQA~\cite{FreshQA} &
Dynamic fact QA benchmark with evolving answers &
600 samples &
Mixed sources  &
LLM &
\cite{UAR} \\
  
&  Domain Retrieval & 
PubMed~\cite{deepretrieval} & 
PICO-based medical search dataset linking reviews to PubMed. & 
21k+ samples & 
Systematic reviews & 
Recall@K & 
\cite{deepretrieval} \\ 

& Domain Retrieval &
Trial search~\cite{deepretrieval} &
PICO-based clinical trial search linked to ClinicalTrials.gov. &
7k+ samples &
Manually &
Recall@K &
\cite{deepretrieval} \\

& Domain Retrieval &
FinSearchBench-24~\cite{FinSearch} &
Financial retrieval benchmark covering stocks, rates, policy, trends. &
1,500 queries &
Manually &
Accuracy &
\cite{FinSearch} \\
\midrule

\multirow{5}{*}{\begin{tabular}[c]{@{}c@{}}Decision \&\\ QA\end{tabular}} & 
% Medical DX & 
% CMB-Clin & 
% Clinical diagnostic benchmark & 
% 208 Qs & 
% Textbooks & 
% GPT/human scores & 
% ~\cite{MRD-RAG} \\ 

% & Medical DX & 
% MM-Cases & 
% Synthetic patient cases & 
% 609 & 
% GPT + doctors & 
% GPT/human scores & 
% ~\cite{MRD-RAG} \\ 

Business &
DQA~\cite{planrag} &
Decision QA benchmark with business scenarios in enterprise settings. &
301 pairs &
video games &
Accuracy &
\cite{planrag} \\
 
& Medical &
CMB-Clin~\cite{CMB-Clin} &
CMB subset for clinical diagnosis reasoning in Chinese medical cases. &
74 cases &
Textbooks/diagnostic materials &
LLM/Expert &
\cite{MRD-RAG} \\

& Medical &
MM-Cases~\cite{MRD-RAG} &
Medicine cases generated by GPT-4o-mini, verified by doctors. &
609 cases &
LLM/doctor-reviewed &
LLM/Expert &
\cite{MRD-RAG} \\

& Medical &
TCM-Cases~\cite{MRD-RAG} &
TCM patient cases generated by GPT-4o-mini, verified by doctors. &
130 cases &
LLM/doctor-reviewed  &
LLM/Expert &
\cite{MRD-RAG} \\

\bottomrule
\end{tabular}%
}
\end{table*}

\subsubsection{Deep Research}

From the perspective of integrating RAG and reasoning, Deep Research tasks exemplify complex downstream applications. They require models to handle open-ended retrieval, produce long-form, structured text, and synthesize multi-source information through deep reasoning. This section analyzes their key features, evaluation datasets, and metrics.

At the core of Deep Research tasks lies the mission of addressing complex informational queries. These tasks are distinguished by several key attributes:

First, dynamic interactivity is essential. Models engage in iterative dialogue to uncover latent user needs or "unknown unknowns". For example, the Co-Storm~\cite{co-storm} framework enables collaboration with multiple language model agents to explore information gradually, easing user cognitive load and capturing unmet needs more accurately.

Second, integrating information from multiple sources is crucial. Models must consolidate diverse data to provide comprehensive coverage. For instance, uses dynamic mind maps to structure knowledge and produce cohesive reports, ensuring accuracy and completeness.

Third, expert-level accuracy is required. Many tasks demand domain expertise, expecting models to perform like human specialists. The Agentic Reasoning~\cite{agentic_reasoning} framework illustrates this with high-stakes scenarios like medical treatment design or legal analysis, where outputs are judged on correctness, depth, and coherence.

Fourth, multi-modal reasoning is often necessary. Deep Research tasks involve varied data types—text, code, knowledge graphs—and dynamic tool use such as web searches or code execution to enhance reasoning.

Finally, handling multiple real-world constraints is vital. Tasks may require generating practical solutions under specific conditions, like designing hospitals in challenging environments with factors like heavy rainfall and seismic activity, as seen in the DeepSolution framework. This ensures outputs are feasible and relevant.

To ensure the diversity and complexity of Deep Research tasks, their evaluation relies on datasets drawn from multiple domains. A few notable examples include:

WildSeek Dataset~\cite{WildSeek}: This dataset is constructed from real-world user information-seeking scenarios and comprises 100 data points covering 24 fields, including economics, computer science, and law. Each data point is characterized by a topic, user goal, and domain label. For example: "Domain: Economics; Topic: Development of a Shared Trading Currency; Goal: Investigate how a new shared currency could eliminate transaction costs". WildSeek effectively evaluates models' competence in dynamic interaction and multi-source information integration.

GAIA~\cite{gaia}. The GAIA Benchmark, developed jointly by Meta AI, Hugging Face, and others, is a comprehensive evaluation framework designed to assess general AI assistants’ ability to handle real-world problems. It features 466 carefully crafted tasks spanning language reasoning, visual perception, multi-agent collaboration, and adaptability, focusing on key skills like reasoning, multimodal processing, web browsing, and tool use. GAIA measures performance across dimensions such as task execution, adaptability, collaboration, generalization, and real-world reasoning with metrics like completion rate, response quality, efficiency, and robustness. Unlike traditional benchmarks, it emphasizes robustness and reliability in everyday scenarios, supports zero-shot evaluation, prevents data contamination, and is widely used in research and industry to guide AI development.

SolutionBench~\cite{li2025deepsolution}: This dataset spans eight engineering domains, including environmental, mining, and transportation engineering. Each instance presents a complex engineering problem with specific constraints. For example: "Design a safe and efficient hospital construction plan in a region with 3000mm annual rainfall, expansive soils, and frequent seismic activity."* SolutionBench evaluates models' ability to address multi-constraint problems and integrate specialized knowledge effectively.

The current evaluation system for DeepResearch faces the dual challenges of scarce specialized testing tasks and the difficulty of assessing complex, lengthy reports: On one hand, existing benchmark tests  only cover basic capabilities and lack systematic evaluation standards in specialized scenarios like business analysis and policy assessment; on the other hand, the multimodal integration, logical chain verification, and domain adaptability testing of long reports pose technical bottlenecks for traditional assessment methods, necessitating the development of new evaluation tools that integrate logic graphs, dynamic scenario simulation, and domain knowledge bases. 

In the future, the evaluation system will evolve into a multidimensional framework, including the construction of a three-level indicator matrix covering basic capabilities, reasoning levels, and application value. Overcoming these evaluation bottlenecks requires both technological innovation and joint standard-building efforts. This concerns not only the reliability validation of intelligent research tools but also the reshaping of research evaluation paradigms and industrial application boundaries.

% In conclusion, Deep Research tasks, positioned within the RAG and reasoning frameworks, epitomize complex problem-solving scenarios that combine dynamic interactivity, multi-source information integration, expert-level accuracy, multi-modal reasoning, and multi-constraint handling. Future research can explore optimizing workflows for these tasks, fostering advancements that enable more efficient and precise solutions to open-ended, high-stakes problems.

\subsubsection{PhD (Expert)-Level Complex Reasoning}

The integration of RAG with advanced reasoning has become essential for tackling expert-level, complex cognitive tasks, particularly at the PhD level. These tasks, including competitive programming, theorem-driven proof reasoning, and cross-disciplinary knowledge retrieval, require multi-layered logical inference and precise coordination between dynamic retrieval and domain-specific knowledge. PhD-level reasoning differs from standard evaluations across three dimensions: knowledge intensity, procedural rigor, and domain specificity. Knowledge intensity demands dynamic access to deep, specialized knowledge, such as analyzing dynamic programming time complexity or applying algebraic topology theorems—needs that surpass general corpora and call for domain-specific knowledge graphs and retrieval methods. Procedural rigor involves mathematical precision in multi-step proofs, requiring logical consistency in symbolic manipulation, theorem use, and counterexample refutation, as seen in international math competitions. Domain specificity reflects tailored reasoning methods, e.g., handling synchronization in concurrent programming or employing tensor calculus in quantum field theory.

Evaluation systems for such tasks are inherently multi-layered and multimodal. The USACO Benchmark~\cite{shi2024can} offers a graduated difficulty scale for programming reasoning, testing both correctness and algorithmic constraints like time complexity. TheoremQA-Math~\cite{theoremqa} links formalized math problems to theorem libraries, demanding verifiable mappings between theorem applications and calculations. Cross-disciplinary datasets like StackBio and StackEcon~\cite{li2024can} assess models’ ability to extract critical knowledge from dense, domain-rich documents, serving as strong tests for domain-oriented retrieval accuracy.

Modern evaluation surpasses traditional end-to-end tests by combining process and outcome validation. Frameworks like CR-Planner~\cite{CR-planner} use dual models—a Sub-Goal Critic to score reasoning chains and an Execution Critic to evaluate retrieval—allowing fine-grained step monitoring. For example, in dynamic programming, key steps like formulating state transitions and retrieving boundary conditions receive targeted feedback. Similarly, Search-O1~\cite{search_o1} quantifies knowledge completeness by tracking uncertainty indicators (e.g., tentative language), measuring confidence and accuracy. Outcome validation maintains strict correctness benchmarks in programming and combines metrics like F1 scores with expert review in open-domain scientific QA to ensure precise understanding of domain-specific terms.

\subsection{Challenges and Future Directions}
\subsubsection{Complex Domain Tasks}
Recent advances in RAG have provided novel solutions for more complex tasks in professional domains. These downstream tasks transcend the limitations of traditional question-answering models that rely solely on simple retrieval-generation patterns, involving challenges such as real-time information acquisition, integration of domain expertise, and dynamic decision-making support. The nature of these tasks can be characterized along three interrelated dimensions: (1) \textit{temporal dynamics}, emphasizing the rapid changes in data and reasoning environment; (2) \textit{domain specificity}, focusing on deep integration of industry knowledge and structured data; and (3) \textit{reasoning chain complexity}, reflecting requirements for multi-stage reasoning and fine-grained decomposition of queries. 

To rigorously evaluate such systems, innovative benchmarking approaches have been proposed. The FinSearchBench-24 dataset, for example, encompasses five months of market data variations, integrating multi-variable interactions across stock, policy, and industrial sectors, and includes over 1,500 multiple-choice questions, thereby surpassing the constraints of traditional static benchmarks. The evaluation adopts a hierarchical and quantitative methodology: the foundational level measures model accuracy and response latency; the intermediate layer assesses the temporal sensitivity of information relevance and the contribution of retrieval mechanisms to reasoning outcomes; and the advanced layer employs ablation studies to highlight performance variances under dynamic temporal decay. This multifaceted evaluation not only differentiates surface-level retrieval capabilities but also rigorously measures the synergy between reasoning quality and temporal context, furnishing theoretical and practical foundations for long-term stability and predictive accuracy in complex domain systems.

Experimental findings further reveal that establishing long-term evaluation protocols with temporal weighting functions is indispensable for adapting to realistic dynamic environments. Nonlinear declines in decision accuracy, observed when extending relevance windows from 72 to 168 hours, emphasize the importance of factoring temporal decay into assessment frameworks. Future work should extend these evaluation protocols to high-stakes domains such as medical diagnostics and legal consultation, where the standardization of interpretability metrics will critically support the evolution of RAG+ reasoning systems toward robust and trustworthy decision-assistance platforms.

\begin{figure*}[ht]
    \centering
    \includegraphics[width=1\linewidth]{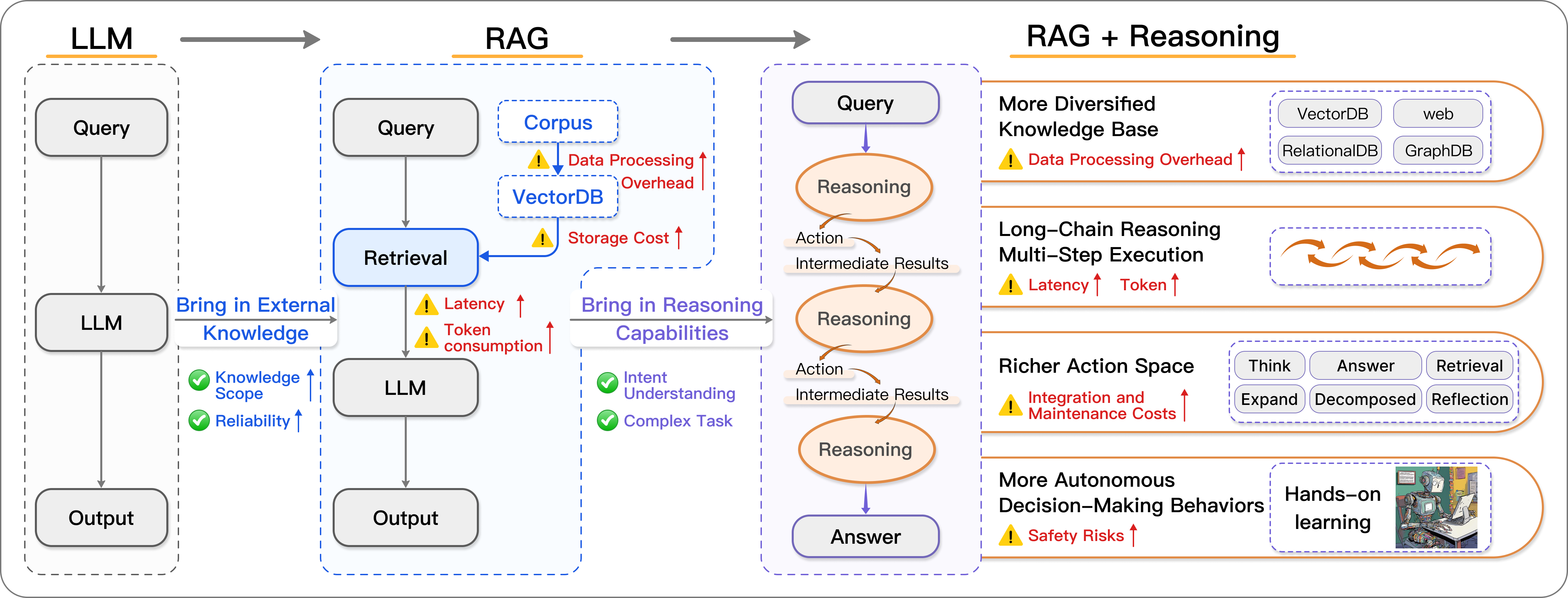}
    \caption{From LLM to RAG and then to RAG+Reasoning, performance improvement comes with additional cost.}
    \label{fig:cost}
\end{figure*}

\subsubsection{Decision Support and Active Retrieval}

The expansion of RAG+Reasoning frameworks into specialized tasks has fostered two complementary research paradigms: decision optimization and active retrieval. In the decision optimization category, systems must leverage heterogeneous structured data, rule bases, and objective functions to formulate optimal strategies. Representative systems like PlanRAG formalize \textit{Decision Question Answering} (Decision QA) tasks targeting enterprise-level scenarios including supply chain optimization, industrial resource allocation, and market price regulation. These tasks require planning multimodal reasoning paths where models iteratively retrieve data from relational and graph databases, integrate intricate business rules, and iteratively refine decision-making paths through replanning mechanisms. To evaluate such capabilities, the Decision QA (DQA) benchmark creates dual database versions (MySQL and Neo4j) derived from economic systems in strategy games, assessing cross-structured generalization. The evaluation consists of a three-tier framework: the core tier measures answer accuracy; the intermediate layer diagnoses error types to identify system bottlenecks; and the foundational tier focuses on retrieval efficiency and the impact of replanning frequency. This structured evaluation framework not only tracks performance but also offers actionable insights for system refinement.

Conversely, the active retrieval evaluation addresses the challenge of dynamically determining when and how to invoke retrieval under complex multimodal contexts.  Unlike rigid traditional RAG systems, UAR applies lightweight classifiers for fast, accurate triggers, improving performance in time-sensitive or creative tasks. Tested on AR-Bench, it combines binary trigger accuracy with GPT assessments, exact matches, and human reviews, boosting adaptability across diverse contexts.

Emerging trends in these evaluation paradigms indicate a shift from static, rule-based frameworks to dynamic system simulations, as exemplified by DQA's use of game engine-generated datasets to simulate realistic environments. Similarly, active retrieval tasks progress from simple retrieval trigger decisions toward collaborative multi-criteria decision-making. Evaluation methodologies are concurrently evolving from singular performance metrics to multidimensional matrices comprising core effectiveness, diagnostic error distributions, and economic cost measures. 

\section{Cost and Risk}
Integrating reasoning into RAG systems is neither effortless nor purely beneficial. Recent trends have exaggerated its advantages while downplaying the costs and risks. This trade-off between performance and cost is crucial. This section examines the expenses and misuse risks linked to adding reasoning to RAG systems. As shown in Figure~\ref{fig:cost}, the cost of moving from LLM to RAG, then to RAG + Reasoning, incurs an inevitable "invisible tax". Though often hidden by performance gains, this cost is vital in assessing these methods' overall practicality and efficiency.

The shift from LLM to RAG moves from simplicity to enhanced knowledge handling by incorporating external information. A basic LLM provides direct, efficient answers with low latency and token use but is limited to pre-trained knowledge, restricting complex or up-to-date queries. RAG overcomes this by adding a vector database for external retrieval, vastly expanding response scope and reliability. However, this requires substantial data processing, storage, and introduces higher latency and token costs due to data chunking, encoding, indexing, and retrieval overhead.

Advancing from RAG to RAG + Reasoning adds multi-step reasoning capabilities, enabling complex task handling, autonomous decisions, and more context-aware responses through intricate reasoning. This comes at the expense of increased delays, token consumption, processing demands, and greater complexity in system integration and maintenance. The reasoning layer’s autonomy also brings opaqueness, unpredictability, and heightened security and reliability risks. These challenges highlight the necessity of carefully balancing effectiveness against costs when adopting RAG + Reasoning in real-world applications.

\begin{figure*}[htbp]
    \centering
    \includegraphics[width=1\linewidth]{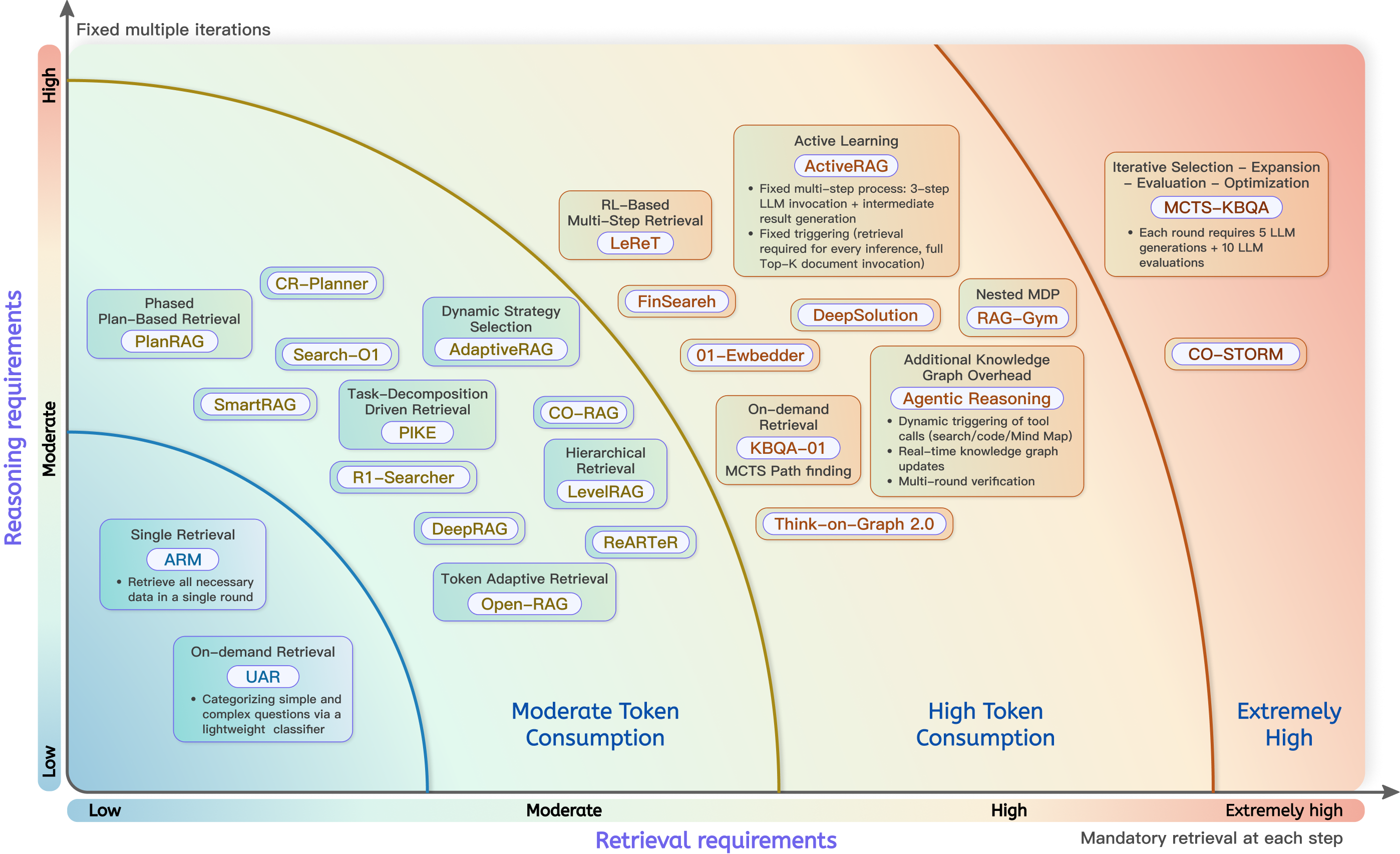}
    \caption{Cost quadrant diagram of retrieval and reasoning requirements}
    \label{fig:cost_paradigm}
\end{figure*}

\subsection{Cost Trade-off in RAG+Reasoning}

Figure~\ref{fig:cost_paradigm}  illustrates typical works combining RAG and Reasoning, showing retrieval and reasoning demands alongside token consumption. While integrating dynamic knowledge retrieval with multi-step reasoning greatly improves accuracy in more complex tasks, the resulting systemic costs are often underestimated in research and practice. These costs grow non-linearly, causing serious efficiency bottlenecks in real-world use. The tradeoff between effectiveness and efficiency stems from RAG+Reasoning’s architecture: multi-stage task decoupling, dynamic path planning, and intermediate state preservation. These features improve reasoning quality but trigger cascading increases in computational resources, token usage, and reduced retrieval efficiency. This section explores these implicit tradeoffs from the angles of resource use, token consumption, and retrieval efficiency.

\subsubsection{Non-Linear Growth of Computational Resources}
The RAG+Reasoning framework separates retrieval and reasoning into multiple stages, causing computational demands to grow non-linearly. Dynamic chain-of-reasoning methods execute multiple LLM generations and retrievals per inference, resulting in complexity far exceeding baseline models. Fixed-length reasoning chains trigger repeated retrieval and generation calls, increasing resource needs with task complexity. More advanced techniques like MCTS-guided methods add rounds of candidate path generation and evaluation, further multiplying runtime and memory usage on GPUs compared to linear methods. Even simpler multi-step planning tasks incur much higher overhead than single-stage retrieval models due to extra graph construction and analysis. While this resource intensity improves inference accuracy, it poses serious scalability challenges under limited resources as computational costs grow superlinearly with model size, retrieval chain length, and task complexity.

\subsubsection{Implicit Token Inflation}
Multi-step reasoning frameworks inherently cause significant token inflation through iterative intermediate processes like thought chains, retrieved documents, and verification feedback. Active learning setups consolidate multiple intermediate results—retrieved documents, counterfactuals, multi-round validations—leading to token usage well beyond typical limits. Chain-based retrieval also generates token bloat due to exhaustive candidate path exploration. Iterative reasoning path selection, expansion, and evaluation add heavy token overhead in tasks needing deep reasoning chains involving extensive sequence generation and evaluation. Token usage grows exponentially with task complexity and increases further when intermediate reasoning favors depth or breadth. This inflation raises API costs and memory demands, especially in long-text generation like Deep Research ~\cite{deepresearcher}.

\subsubsection{Marginal Decline in Retrieval Efficiency}

Dynamic retrieval improves knowledge precision but suffers diminishing efficiency as task complexity increases. Adaptive methods reduce retrievals for simple tasks but still require multiple iterations for complex ones, adding significant overhead compared to standard RAG. The tradeoff between retrieval quality and frequency further limits efficiency. High-accuracy retrieval methods incur heavy computational and time costs, negating their efficiency benefits. Even advanced retrieval-trigger optimizations can’t fully remove this overhead due to extra training and deployment costs~\cite{adaptiveRAG}. This natural efficiency ceiling highlights ongoing challenges in balancing retrieval accuracy and resource use, especially in large, complex tasks.

\subsubsection{Toward a Cost Model Framework}

Against this backdrop, the development of fine-grained cost models becomes a necessary precondition for balancing effectiveness and efficiency. Existing evaluation metrics, which often rely on single-task performance indicators (such as Exact Match or F1) or coarse-grained runtime statistics, lack the comprehensiveness to jointly model computational resources, token flow, and retrieval overhead. Consequently, they fail to quantify the true tradeoffs in reasoning mechanisms. For instance, while multi-hop reasoning may improve task accuracy, these improvements are frequently offset by exponential growth in token consumption and latency relative to baseline methods. A fine-grained cost model would enable researchers and practitioners to more accurately evaluate the real benefits of reasoning-centric frameworks while addressing the underexplored interplay between computational cost and task performance.

\subsection{Potential Risk of  Over-Thinking}

In the process of developing deep thinking models, "overthinking" poses a key risk to system efficiency and reliability~\cite{fan2025missing,sui2025stop,chen2024not,he2025can,wang2025don,cuadron2025danger}, and this issue is further amplified after combining with RAG. It appears as redundant reasoning steps, excessive validation of known conclusions, or unnecessarily broad retrieval scopes, wasting computational resources, increasing error propagation, and degrading performance. For example, in financial risk assessment, an LLM with RAG might retrieve multiple similar market reports and repeatedly verify the same economic indicators rather than focusing on core risks, leading to delayed decisions. This stems from an imbalance between reasoning and retrieval: after accessing external knowledge, the model can enter a "self-validation loop," repeatedly parsing overlapping or contradictory documents. The generation module, seeking reliability, may trigger further retrievals, creating a feedback loop that worsens inefficiency. This issue is critical in real-time systems like medical diagnosis, where over-retrieval of irrelevant literature can delay urgent decisions.

Case studies show the impact of overthinking~\cite{sui2025stop}. In legal document interpretation, early reasoning errors can amplify through the retrieval-generation loop, causing retrieval along incorrect paths and yielding illogical conclusions. This error propagation is evident in systems like the Search-o1~\cite{search_o1}, where flawed information extraction misguides subsequent reasoning. In industrial equipment manual interpretation, overextended reasoning with highly similar documents risks obscuring critical parameter differences, increasing procedural errors. These examples illustrate that overthinking not only hampers knowledge integration but also creates safety hazards in practical applications.

To mitigate these risks, researchers propose multiple optimization frameworks. ReaRAG~\cite{rearag} limits reasoning chain length and incorporates self-reflection to prune invalid branches. A simple and effective way is to use a two-stage filtering process, first narrowing documents by metadata, then validating fragment relevance, reducing redundant information—for instance, retrieving only relevant legal clauses rather than entire regulatory texts. The DeepSeek R1~\cite{deepseek_r1} applies reinforcement learning with distillation to penalize redundant steps, cutting repeated formula validation in math proofs by over 40\%. These approaches transform open-ended reasoning into controlled, goal-directed processes, using methods like attention weight analysis to measure information gain or confidence functions to evaluate reasoning paths.

Current research balances constraints with model creativity. Knowledge graph-guided reasoning is tested in clinical trials to prioritize key medical features over exhaustive literature retrieval~\cite{MRD-RAG}. Causal reasoning models aim to break error chains; for example, in financial forecasting, causal graphs restrict reasoning to logically relevant macroeconomic links. Adaptive stopping strategies adjust reasoning depth in customer service—simple queries use preset templates, complex issues activate multi-hop reasoning. These advances reshape retrieval-augmented reasoning, with the core challenge being to develop evaluation frameworks that avoid both "cognitive stagnation" from excessive constraints and "cognitive overload" from insufficient control.

Future progress will integrate cognitive science with computational modeling. By mimicking human "intuition-verification" decision-making, LLMs could switch seamlessly between rapid response and deep reasoning. In high-risk fields like industrial fault diagnosis, such hybrid models can quickly propose contingency plans after initial retrieval while verifying their validity through deeper analysis. This layered approach reduces overthinking risks and offers a safe, controllable path for applying LLMs in critical industries.

\section{Practical Guide}
\begin{figure*}[hbtp]
    \centering
    \includegraphics[width=1\linewidth]{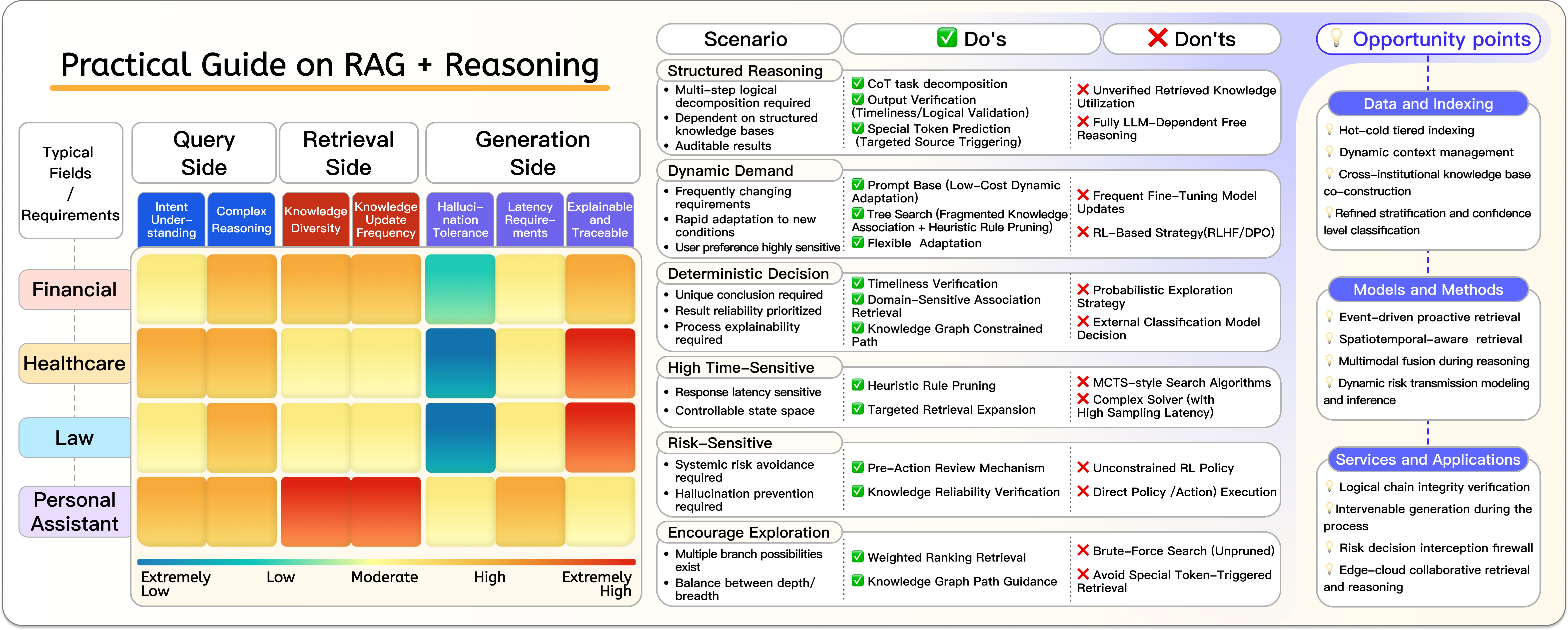}
    \caption{Practical guide to synergizing RAG and Reasoning}
    \label{fig:practical}
\end{figure*}

The combination of RAG and Reasoning is not a one-size-fits-all solution; it requires careful evaluation of each scenario's unique needs. As a rapidly evolving and relatively new field, practical applications are still limited, making best practices hard to define. This chapter abstracts and summarizes the key traits of typical RAG+Reasoning application domains and offers practical guidelines for system design based on these features. It provides recommendations on leveraging RAG’s strengths with Reasoning, highlighting priorities, pitfalls to avoid, and current opportunities (Figure~\ref{fig:practical}). The goal is to promote wider adoption and effective use of this technology in diverse, complex real-world settings.

\subsection{Domain characteristics}

As illustrated in the left part of Figure~\ref{fig:practical}, we develop a seven-dimensional feature system based on the three core stages of RAG—query, retrieval, and generation—to systematically analyze challenges and adaptation needs across various industries. The query stage emphasizes the complexity of intent understanding and the demand for advanced reasoning, recognizing that industries differ in query abstraction and specificity; some require quickly capturing implicit, deep intentions, while others need complex reasoning. Effective preservation of original semantic meaning during understanding and reasoning is key to improving RAG performance. Retrieval focuses on the system’s adaptability to diverse and dynamic knowledge sources, which vary from rich multi-domain data to rapidly updating information; frequent updates and fragmented knowledge present challenges that demand effective integration to ensure consistent support for generation. The generation stage requires high-quality outputs, with strict control over hallucinations—especially critical in sensitive fields like healthcare and law—along with varying latency requirements for real-time or delayed responses. Explainability and traceability at this stage are essential for system credibility and serve as key evaluation metrics. This comprehensive framework reveals technical bottlenecks and guides improvements, and is applied to analyze four representative domains: finance, healthcare, law, and personal assistants.

\subsubsection{Finance}

In the finance domain, user queries typically focus on structured needs like investment decisions and risk forecasting. While intent understanding is moderately complex, the system must perform advanced reasoning amid rapidly changing market conditions, relying heavily on external knowledge and frequent updates. For example, portfolio return forecasting integrates time series analysis, policy interpretation, and cross-market reasoning. Retrieval demands handling diverse data sources—real-time market data, annual reports, and regulatory filings—with update cycles often measured in minutes. During generation, strict latency and hallucination control are crucial, as outputs must include decision-making suggestions with full data traceability. Investment research reports, for instance, require annotated key indicators, their data sources, and computation logic to ensure transparency and regulatory compliance. High latency control and robust traceability are essential to maintain transparency and adherence to financial regulations.

\subsubsection{Healthcare}

Healthcare queries involve complex medical semantic parsing, often with ambiguous terms or incomplete symptoms. For example, "persistent chest pain with shortness of breath" requires multi-hop reasoning across cardiology, pulmonology, and emergency medicine. Retrieval must integrate electronic health records, medical imaging, and up-to-date clinical guidelines.  In generation, hallucination tolerance is minimal—errors in drug dosages or protocols risk malpractice. Therefore, accuracy, timeliness, and explainability are paramount, with every decision step traceable and verifiable.

\subsubsection{Legal Services}
Legal consultations often require interpreting statutes and citing cases, balancing precise legal terms with natural language nuances. Retrieval depends on structured, infrequently updated sources like case law databases and local regulations. Generation demands accuracy—for instance, drafting contract clauses must precisely cite specific statutes (e.g., Article 472 of the Civil Code) down to the paragraph level for traceability. Explainability is essential, with traceability usually above 95\%, and probabilistic language avoided to comply with strict judicial documentation standards.

\subsubsection{Personal Assistants}

This domain features diverse, dynamic user needs, including schedule management, real-time navigation, and open-domain conversations. Accurate intent disambiguation through contextual awareness is crucial. Retrieval integrates fragmented sources like user behavior logs, geolocation, and social media. Generation latency varies: weather updates require sub-second responses, while travel planning can tolerate 5+ seconds. Hallucination tolerance depends on context—creative outputs are acceptable for recipes but not for flight information, which demands full accuracy. This necessitates adaptive verification in the RAG system. Though intent complexity is lower than in healthcare or legal fields, the domain’s interaction diversity requires heavy reliance on external knowledge and dynamic balancing of latency and accuracy.

\subsection{Do’s and Don’ts}

Building on aforementioned domain characteristics, we further identify six common scenarios, and derive technical adaptation principles for each. This section outlines key optimization strategies (Do’s) and prohibitions (Don’ts) , to guide the co-design of RAG and reasoning.

\subsubsection{Structured Reasoning Scenarios}

For scenarios requiring multi-step logical decomposition and structured knowledge dependency, such as \emph{portfolio return prediction}, Chain-of-Thought (CoT) task decomposition and knowledge graph (KG)-driven graph reasoning approaches should be employed. Complex problems should be broken into verifiable sub-tasks, such as coupling market trend analysis with policy impact assessment, while leveraging knowledge graph constraints to ensure logical completeness and auditability. It is essential to incorporate a temporal validation layer to cross-check the consistency of timestamp-sensitive information (e.g., real-time market data or emergent regulatory policies) within a dynamic knowledge base. Approaches that exclude retrieval-based verification of salient features must be avoided, as they may lead to reasoning biases arising from the absence of structured knowledge anchors (e.g., critical indicators from financial statements). Furthermore, the reasoning space of LLMs should be constrained within domain-specific knowledge frameworks to prevent irrelevant or invalid deductions.

\subsubsection{Dynamic Demand-Responsive Scenarios}

For scenarios characterized by rapidly shifting demands and user preference variability, such as \emph{itinerary planning and multimodal interaction in personal assistant services}, a dynamic adaptation mechanism based on prompt engineering is recommended. By dynamically associating fragmented knowledge units (e.g., user behavior history and real-time traffic updates) with semantic templates and employing heuristic rules for search-space pruning (e.g., prioritizing locally updated information within the past 24 hours), the system can balance contextual adaptability with response speed. Model fine-tuning or reinforcement learning (RLHF/DPO)-based strategy updates should be avoided due to their lengthy iterative cycles and computational overhead, which cannot meet real-time responsiveness requirements, such as millisecond-grade reaction times for last-minute destination changes. Lightweight caching architectures should be implemented within the retrieval system, prioritizing frequently accessed knowledge fragments, such as operating hours of popular tourist attractions, to achieve an equilibrium between dynamism and stability.

\subsubsection{Deterministic Decision-Making Scenarios}

In scenarios requiring a single, reliable conclusion, such as \emph{clinical diagnosis generation in the healthcare domain}, a multi-level deterministic assurance system should be established. Time-validation layers can filter outdated knowledge (e.g., therapies no longer approved), while field-sensitive retrieval modules trigger predefined decision rules conforming to up-to-date clinical guidelines (e.g., those codified within the latest version of the International Classification of Diseases [ICD]). Knowledge graph path constraints should restrict the reasoning process to validated causal links within medical logic (e.g., linking symptom patterns to laboratory test results within corroborated diagnostic pathways), thereby minimizing the likelihood of deviations from standard protocols. Probabilistic exploration strategies that generate alternative hypotheses (e.g., speculative differential diagnoses for atypical pneumonia) should be strictly disallowed to avoid clinical misjudgments. Additionally, delegating decision-making authority to external classification models must be avoided to maintain end-to-end explainability and a clear causal link in the decision-making pipeline.

\subsubsection{Time-Sensitive Scenarios}

In tasks highly sensitive to response delays, such as \emph{real-time risk warnings and trading decisions in the financial sector}, heuristic rules should be employed to prioritize indexing of frequently queried knowledge units (e.g., volatility indices and liquidity indicators) at the top of the search hierarchy. Directed retrieval expansion strategies that preload potentially associated information (e.g., contractual clauses of derivative instruments tied to underlying assets) can further reduce latency in multi-turn interactions. Monte Carlo Tree Search (MCTS) and other sample-based algorithms are ill-suited for such scenarios due to the excessive computational complexity caused by branch expansion, rendering them infeasible within tight time constraints (e.g., milliseconds). Similarly, the invocation of complex mathematical solvers (e.g., numerical solutions for stochastic differential equations) can introduce uncontrollable delays and should be replaced with lightweight rule-based mechanisms (e.g., threshold-triggering mechanisms based on historical volatility ranges).

\subsubsection{Risk-Sensitive Scenarios}

For scenarios with minimal tolerance for errors, such as \emph{contract clause generation and citation of judicial interpretations in the legal sector}, a dual-layer defensive mechanism must be employed. A pre-action review layer should validate the compliance of generated content with statutory standards (e.g., ensuring consistency between liability clauses and Article 577 of the Civil Code), while a reliability validation layer performs cross-referencing validation across multiple sources (e.g., aligning Supreme Court precedents with regional court guidelines) to resolve potential conflicts. Retrieval systems must include version control modules to track and update legal references (e.g., automatically flagging repealed local statutes). Unconstrained reinforcement learning-based text generation methods must be avoided, as their exploratory nature risks violating the normative requirements of legal documents (e.g., generating presumptive liability terms unsupported by judicial interpretations). All decision-making actions must pass through deterministic rule engines to filter inadmissible outputs, and the system should never execute decision actions autonomously, such as generating legally binding arbitration notices without oversight.

\subsubsection{Complex Path Exploration Scenarios}

In exploration tasks involving multiple possible trajectories, such as \emph{differential diagnosis and therapeutic pathway optimization in medicine}, weighted ranking search algorithms should balance search depth and breadth. Knowledge graph topology can guide prioritization (e.g., standard treatment procedures for acute coronary syndrome), while Monte Carlo Tree Search can extend exploration into uncommon differential paths (e.g., rare genetic metabolic disorders). Dynamic pruning threshold functions should be designed (e.g., adjusting the scope of differential diagnosis based on patient history) to eliminate low-confidence hypotheses in real time, thereby controlling computational scale. Brute-force searching of all potential paths (e.g., concurrently testing hundreds of pathogens for nonspecific symptoms) should be avoided to prevent exponential computational scaling. Careful handling of specific token triggers during retrieval (e.g., avoiding spurious associations between "fever" and unrelated oncological hyperthermia research) is critical to maintaining logical coherence in diagnostic reasoning.

\subsection{Opportunity Points}

Based on the Do’s and Don’ts of current technologies analyzed in the previous section, there remain numerous directions with substantial academic value and application potential that have yet to be fully explored. This section systematically discusses several promising opportunity points across three dimensions: \emph{data and indexing}, \emph{models and methodologies}, and \emph{application services}.

\subsubsection{Data and Indexing}

\paragraph{Cold-Hot Tiered Indexing and Dynamic Context Management}  
The challenge of managing massive and highly heterogeneous data resources lies in devising an effective \emph{cold-hot tiered indexing} mechanism that prioritizes data according to their frequency of use and importance. Such a mechanism not only demands classification of data based on timeliness and access frequency but also requires integration with dynamic context management. This allows the system to intelligently retrieve the most relevant data according to the immediate context.

Moreover, a dynamically updated indexing mechanism can mitigate the loss of data timeliness, which often leads to deteriorated inference accuracy. By ensuring access to the most recent and task-appropriate data, this approach reduces redundancy and incorrect retrievals associated with static indexing. When combined with automated task scheduling and resource allocation strategies, fine-grained real-time inference support can be achieved, significantly enhancing the system's overall efficiency.

\paragraph{Cross-Institution Knowledge Base Construction}  
The construction of cross-institution or cross-domain knowledge bases offers new opportunities for advancing RAG+Reasoning research. At the core of large-scale cross-institutional knowledge bases lies the optimization of data integration and sharing mechanisms. This entails addressing challenges such as data security and privacy while adopting standardized data interfaces or leveraging federated learning paradigms to enable multidimensional data integration.

Through semantic alignment across multiple sources, entity resolution, and concept abstraction, cross-institutional knowledge can be transformed into authoritative and richly contextualized knowledge bases. These enhanced repositories provide robust contextual support for reasoning tasks and can deliver deeper insights in areas such as healthcare, finance, and urban management.

\paragraph{Fine-Grained Layering and Confidence Grading}  
In scenarios where retrieval and reasoning operate synchronously, the \emph{interpretability} and \emph{reliability} of generated outcomes are paramount. Fine-grained layering of data and indices, along with confidence grading of retrieval results, enables the system to selectively use the most trustworthy and relevant subsets of data during different stages of reasoning. This approach fosters transparency and traceability in final decisions or generative outputs.

For instance, in medical diagnosis scenarios, confidence grading can initiate additional verification or expert review in high-risk cases. In the legal domain, confidence layering systematically presents key evidence and identifies sources of uncertainty, reducing reasoning vulnerabilities and minimizing the risk of erroneous conclusions caused by information ambiguity.

\subsubsection{Models and Methodologies}

\paragraph{Event-Driven Active Retrieval}  
Traditional retrieval mechanisms are predominantly passive. However, \emph{event-driven active retrieval} presents a promising exploration avenue. By monitoring critical events, such as the injection of new data, user interactions, or changes in external sensors, event-triggered retrieval and reasoning processes can be initiated to capture and respond to potential risks and opportunities in real time. Integrating methodologies such as sequence-based event detection or multitask-learning-based intent recognition can facilitate automatic determination of \emph{when} and \emph{how} to trigger retrieval actions. Iteratively optimizing these processes contributes to a more efficient and continuous reasoning loop.

\paragraph{Spatiotemporal-Aware Retrieval and Association}  
Many applications, such as natural disaster monitoring, traffic flow prediction, and inventory management in retail, exhibit strong dependencies on temporal and spatial dimensions. By incorporating spatiotemporal-aware algorithms, retrieval processes can prioritize or emphasize crucial documents according to constraints tied to time and space. This not only enhances timeliness but also improves the purposefulness and accuracy of reasoning.

Furthermore, modeling the evolution of events within spatiotemporal dimensions—when combined with semantic indexing and vector-based retrieval mechanisms in RAG—can enable more precise characterization and utilization of complex spatiotemporal dynamics during reasoning.

\paragraph{Multimodal Fusion in Retrieval and Reasoning}  
Multimodal data (e.g., text, images, audio, video, and sensor data) collectively constitute a richer contextual environment, offering critical cues for reasoning tasks. However, existing studies are often limited to the retrieval of single or a few data modalities. Advancing research on multimodal fusion and reasoning mechanisms under the RAG+Reasoning framework has the potential to greatly enhance the system’s capacity for addressing complex queries.

The research focus lies in constructing cross-modal representation learning and alignment methods, enabling unified representations of the same entities or events across different modalities. During retrieval, confidence scores for each modality can be integrated into a comprehensive ranking process, culminating in multimodal-informed joint decision-making during reasoning. This approach not only improves contextual understanding in complex tasks but also broadens the application scope of RAG technologies in scenarios such as expert systems and autonomous driving, where sensory integration and interpretation are critical.

\paragraph{Dynamic Risk Propagation Modeling and Management}  
The tight coupling of retrieval and reasoning with multi-stage decision-making inevitably introduces risk propagation issues. Misjudgments of high-risk or low-confidence documents during upstream retrieval are often inherited by downstream reasoning processes, amplifying uncertainties and increasing error margins. To address this, dynamic risk modeling should be embedded within retrieval workflows, enabling risk quantification, tracking, and management at multiple stages. When necessary, risk mitigation mechanisms or process rollbacks can be triggered, creating a closed-loop correction framework.

Incorporating strategies for analyzing and managing risk propagation is not only a technical challenge but also a matter of system deployment and standardization. In high-stakes domains such as healthcare and financial risk management, establishing comprehensive safety standards and compliance protocols will be crucial. These protocols should treat dynamic risk propagation management as a critical component of evaluating and iterating knowledge retrieval and reasoning systems.

\subsubsection{Application Services}

\paragraph{Validation of Logical Chain Completeness}  
While RAG with Reasoning can provide partially interpretable reasoning outputs, verifying the completeness of logical chains remains a challenge. Future research could integrate formal verification or symbolic reasoning techniques to ensure consistency and completeness across key reasoning nodes and intermediate conclusions. This would prevent logical gaps or illogical leaps in reasoning, offering robust regulatory support for high-stakes industries such as law and finance.

\paragraph{Intervenable Generation During Reasoning}  
Contemporary Agentic  RAG often operate as "black boxes," rendering external interventions nearly impossible during generative reasoning tasks. However, providing mechanisms for human intervention—such as through visualization or interactive interfaces—could enable experts or users to perform manual corrections, initialize prior knowledge, or modify interim assumptions during the reasoning process. This would substantially enhance the system’s flexibility and safety.

Specifically, intervenable generation allows not only post hoc error corrections but also proactive identification and rectification of potential risks or biases at earlier stages. Interactive interpretable reasoning platforms or visualization tools grounded in knowledge graphs could empower users to scrutinize and influence reasoning workflows, thereby enhancing confidence and control in decision-making processes across diverse domains.

\paragraph{Risk Decision Interception Firewalls}  
In closed-loop automated tasks such as algorithmic trading or medical diagnostic decision-making, erroneous reasoning outputs can lead to catastrophic outcomes. To mitigate such risks, the system architecture should incorporate \emph{risk decision interception firewalls}, which perform multidimensional validations at critical reasoning nodes or prior to outputting decisions. When confidence levels or high-risk indicators breach thresholds, these firewalls can block decision outputs or escalate them for stricter human review.

This mechanism serves as a “final line of defense” for RAG+Reasoning systems, ensuring decision security in large-scale automated information networks. It also provides a robust foundation for compliance and regulatory auditing, enabling safer deployment in critical applications.

\paragraph{Edge-Cloud Collaborative Retrieval and Reasoning}  
With the rapid development of IoT and 5G technologies, many scenarios demand on-site data collection and preliminary processing on edge devices, followed by high-level retrieval and reasoning tasks on cloud platforms. Efficiently partitioning tasks, allocating resources, and maintaining consistency between indexes and models across the edge-cloud continuum represent critical research directions.

Leveraging techniques such as lightweight model compression, distributed index synchronization, and communication optimization can ensure fast reasoning while maximizing resource utilization. Edge-cloud collaborative solutions are particularly impactful for real-time industrial monitoring and smart city applications, reducing network latency and bandwidth bottlenecks while ensuring accurate and timely inference outputs.

In summary, RAG+Reasoning systems present many untapped opportunities across various dimensions.  Further research and practical validation could greatly improve their use in complex, high-risk scenarios while fueling new growth in GenAI.
\section{Future Trends}

In this chapter, we summarize four major trends in technological advancements based on current research, aiming to elucidate and guide the potential future directions of RAG.

\subsection{The Integration of RAG and Graph}

Recent developments have witnessed a growing synergy between RAG systems and graph-based approaches. The intrinsic benefits of graph structures, such as explicit logical relationships and knowledge indexing, have enabled new paradigms for addressing challenges in global reasoning, dynamic data management, and personalized services within RAG systems.

\textbf{Knowledge Organization.} 

Graph-structured knowledge organization frameworks offer a powerful alternative to traditional vector-based retrieval methods, excelling in modeling complex relationships and supporting global reasoning. For example, GraphRAG~\cite{graphrag} combines hierarchical graph indexing with community detection to extract entity relationship networks from text corpora, enabling large-scale thematic analysis through hierarchical summaries. Building on this, PIKE~\cite{pike} introduces a multi-level heterogeneous knowledge graph that organizes documents, semantic segments, and refined knowledge units into a three-layer hierarchy, improving extraction accuracy and multi-hop reasoning via atomized knowledge construction and task decomposition. For dynamic personalization, EMG-RAG~\cite{crafting} features a three-layer Editable Memory Graph architecture that structures memory data by ontology classification, subclass, and entity relationships, using reinforcement learning to enable real-time updates and multidimensional queries. Together, these advances leverage graph topologies to address the limitations of conventional RAG systems—such as one-dimensional representation and weak contextual links—enabling multilevel reasoning from local fact retrieval to global thematic summarization and forming a foundation for interpretable, adaptive RAG systems.

\textbf{Symbolic Reasoning.}
Graph-structured symbolic reasoning methods leverage the multi-hop reasoning power of Knowledge Graphs (KG) to better manage complex semantic and logical relationships. Frameworks like HippoRAG2 and the Think-on-Graph (ToG)~\cite{ToG2} series exemplify this. HippoRAG2~\cite{hippo2} builds open knowledge graphs and uses personalized PageRank with a dense-sparse coding approach inspired by brain memory, boosting performance in factual memory, semantic understanding, and multi-hop reasoning. Likewise, ToG-2 combines iterative retrieval of knowledge graphs and documents, using relationship discovery, entity pruning, and context-driven graph searches to integrate fine-grained information from unstructured text, enhancing implicit relationship detection.

\textbf{Task Planning.}
Graph-based task planning in RAG systems enhances complex problem-solving by overcoming the limitations of traditional linear workflows, which struggle with multi-step or multimodal reasoning. These approaches build dynamic knowledge graphs, like Mind Maps, to explicitly model logical dependencies and context. For instance, the Agentic Reasoning~\cite{agentic_reasoning} transforms reasoning chains into graph structures for entity extraction, relation identification, and community clustering, enabling dynamic path tracking and optimized retrieval, excelling in tasks like doctoral-level GPQA~\cite{gpqa}. Collaborative frameworks such as Co-STORM extend this to multi-agent scenarios, representing queries, tool calls, and knowledge integration as traversable graph nodes to support task decomposition and adaptive reasoning.

\textbf{Tool Usage and Management.}
Graph-enhanced approaches to tool management overcome limitations of traditional dependency modeling by effectively capturing complex relationships like parameter passing, functional collaboration, and resource management. Graph RAG-Tool Fusion~\cite{graph_tool} models tools as graph nodes within a dual-layer architecture of core system APIs and domain-specific tools, encoding direct and indirect dependencies as edges. It uses a two-stage retrieval process: vector-based tool retrieval followed by a graph-based depth-first search to assemble dependency-compliant toolsets.

\subsection{Multi-Model Collaboration}

Multi-model collaboration has emerged as a pivotal strategy for enhancing task complexity handling and domain adaptability in RAG systems~\cite{chen2025harnessing}. By integrating the strengths of different models, this approach achieves optimized performance. For example, the \textit{CR-Planner}~\cite{CR-planner} combines general-purpose generation models (e.g., GPT-4) with domain-specific critic models (e.g., Llama-3-8B). This hybrid system dynamically orchestrates subgoal planning and execution evaluation, utilizing MCTS to generate high-quality training data. Similarly, UAR~\cite{UAR}employs intent-aware and knowledge-requirement classifiers to dynamically trigger retrieval, decoupling lightweight classification tasks from resource-intensive decoding operations of LLMs. Furthermore, \textit{Adaptive-RAG} ~\cite{adaptiveRAG} deploys small-complexity classifiers to route queries into different levels of processing strategies, balancing response speed for simple queries with deep reasoning for complex ones. These strategies form a closed "generation-evaluation"loop, leveraging complementary strengths across models to achieve improved accuracy and computational efficiency.

\subsection{Multi-Modal Collaboration}

The breakthrough in Chain-of-Thought (CoT) capabilities of language models has catalyzed the transition of multimodal reasoning from perceptual-level integration to cognitive-level reasoning, promoting Multimodal Collaborative Reasoning as a key trend~\cite{bi2025reasoning} By deeply integrating the logical reasoning capabilities of language models with the spatial-semantic representation of multimodal data, it significantly enhances information synthesis in complex scenarios~\cite{abootorabi2025ask}. For instance, in the medical domain, multimodal RAG systems such as \textit{MedCoT~\cite{medcot}} utilize hierarchical expert systems to integrate CT imaging and pathology reports, enabling knowledge graph validation of diagnostic hypotheses and reducing misdiagnosis risks. Future research will likely focus on robust cross-modal knowledge alignment, progressive knowledge distillation, and adaptive reasoning frameworks.

\subsection{Customized Reinforcement Learning}

The application of reinforcement learning (RL) in RAG systems has become instrumental in improving module coordination and enhancing overall efficiency. Recent studies focus on designing reward mechanisms tailored to the specific needs of RAG systems. Frameworks such as \textit{RAG-Gym~\cite{rag-gym}} and \textit{DeepRAG~\cite{deeprag}} model reasoning processes using Markov Decision Processes and introduce fine-grained process supervision mechanisms. Additionally, \textit{ReARTeR~\cite{rearag}} and \textit{SmartRAG~\cite{smartrag}} incorporate trust-aware reward strategies and end-to-end policy optimization to achieve superior accuracy and robustness. Opportunities remain for further exploring automated reward modeling with LLMs to facilitate fine-grained supervision.

\section{Conclusion}

This paper has systematically reviewed the synergistic integration of Retrieval-Augmented Generation (RAG) and reasoning, providing a formal definition of reasoning within the RAG framework as a structured, multi-step, goal-driven process that dynamically combines parametric and retrieved knowledge to address complex problems.

We presented a comprehensive taxonomy covering the purposes, collaboration paradigms, and implementation methods underlying RAG+Reasoning systems. The synergy enables more precise retrieval informed by logical analysis and enhances reasoning with contextually relevant, up-to-date knowledge beyond parametric limitations.

While the enhanced reasoning capabilities allow tackling complex knowledge-intensive tasks such as deep research, expert-level problem solving, and domain-specific decision support, practical challenges remain. These include computational and token costs that grow non-linearly, risks of overthinking leading to inefficiency and error propagation, and the lack of evaluation frameworks that effectively assess intermediate reasoning quality alongside final results.

To bridge the gap from theory to real-world application, we proposed practical design guidelines tailored to diverse domains like finance, healthcare, law, and personal assistants, emphasizing adaptability to heterogeneous, dynamic knowledge sources and strict requirements for output reliability and traceability.

Finally, we identified promising directions for future research, including graph-structured knowledge integration, multimodal and multi-model collaborative reasoning architectures, and advanced reinforcement learning techniques for optimizing retrieval-reasoning workflows.

Overall, this work establishes both a theoretical foundation and practical roadmap to drive the development of next-generation RAG+Reasoning systems capable of robust, transparent, and efficient cognition, paving the way for impactful applications across academia and industry.

\bibliographystyle{ACM-Reference-Format}
\bibliography{RAR}


\begin{thebibliography}{110}

%%% the Web address.

\ifx \showCODEN    \undefined \def \showCODEN     #1{\unskip}     \fi
\ifx \showISBNx    \undefined \def \showISBNx     #1{\unskip}     \fi
\ifx \showISBNxiii \undefined \def \showISBNxiii  #1{\unskip}     \fi
\ifx \showISSN     \undefined \def \showISSN      #1{\unskip}     \fi
\ifx \showLCCN     \undefined \def \showLCCN      #1{\unskip}     \fi
\ifx \shownote     \undefined \def \shownote      #1{#1}          \fi
\ifx \showarticletitle \undefined \def \showarticletitle #1{#1}   \fi
\ifx \showURL      \undefined \def \showURL       {\relax}        \fi
% invisible to TeX
\providecommand\bibfield[2]{#2}
\providecommand\bibinfo[2]{#2}
\providecommand\natexlab[1]{#1}
\providecommand\showeprint[2][]{arXiv:#2}

\bibitem[Abdallah et~al\mbox{.}(2025)]%
        {rankify}
\bibfield{author}{\bibinfo{person}{Abdelrahman Abdallah}, \bibinfo{person}{Bhawna Piryani}, \bibinfo{person}{Jamshid Mozafari}, \bibinfo{person}{Mohammed Ali}, {and} \bibinfo{person}{Adam Jatowt}.} \bibinfo{year}{2025}\natexlab{}.
\newblock \showarticletitle{Rankify: A comprehensive python toolkit for retrieval, re-ranking, and retrieval-augmented generation}.
\newblock \bibinfo{journal}{\emph{arXiv preprint arXiv:2502.02464}} (\bibinfo{year}{2025}).
\newblock


\bibitem[Abootorabi et~al\mbox{.}(2025)]%
        {abootorabi2025ask}
\bibfield{author}{\bibinfo{person}{Mohammad~Mahdi Abootorabi}, \bibinfo{person}{Amirhosein Zobeiri}, \bibinfo{person}{Mahdi Dehghani}, \bibinfo{person}{Mohammadali Mohammadkhani}, \bibinfo{person}{Bardia Mohammadi}, \bibinfo{person}{Omid Ghahroodi}, \bibinfo{person}{Mahdieh~Soleymani Baghshah}, {and} \bibinfo{person}{Ehsaneddin Asgari}.} \bibinfo{year}{2025}\natexlab{}.
\newblock \showarticletitle{Ask in Any Modality: A Comprehensive Survey on Multimodal Retrieval-Augmented Generation}.
\newblock \bibinfo{journal}{\emph{arXiv preprint arXiv:2502.08826}} (\bibinfo{year}{2025}).
\newblock


\bibitem[Asai et~al\mbox{.}(2023)]%
        {self_rag}
\bibfield{author}{\bibinfo{person}{Akari Asai}, \bibinfo{person}{Zeqiu Wu}, \bibinfo{person}{Yizhong Wang}, \bibinfo{person}{Avirup Sil}, {and} \bibinfo{person}{Hannaneh Hajishirzi}.} \bibinfo{year}{2023}\natexlab{}.
\newblock \showarticletitle{Self-rag: Learning to retrieve, generate, and critique through self-reflection}. In \bibinfo{booktitle}{\emph{The Twelfth International Conference on Learning Representations}}.
\newblock


\bibitem[Bi et~al\mbox{.}(2025b)]%
        {bi2025reasoning}
\bibfield{author}{\bibinfo{person}{Jing Bi}, \bibinfo{person}{Susan Liang}, \bibinfo{person}{Xiaofei Zhou}, \bibinfo{person}{Pinxin Liu}, \bibinfo{person}{Junjia Guo}, \bibinfo{person}{Yunlong Tang}, \bibinfo{person}{Luchuan Song}, \bibinfo{person}{Chao Huang}, \bibinfo{person}{Guangyu Sun}, \bibinfo{person}{Jinxi He}, {et~al\mbox{.}}} \bibinfo{year}{2025}\natexlab{b}.
\newblock \showarticletitle{Why Reasoning Matters? A Survey of Advancements in Multimodal Reasoning (v1)}.
\newblock \bibinfo{journal}{\emph{arXiv preprint arXiv:2504.03151}} (\bibinfo{year}{2025}).
\newblock


\bibitem[Bi et~al\mbox{.}(2025a)]%
        {StePO-Rec}
\bibfield{author}{\bibinfo{person}{Yuxi Bi}, \bibinfo{person}{Yunfan Gao}, {and} \bibinfo{person}{Haofen Wang}.} \bibinfo{year}{2025}\natexlab{a}.
\newblock \showarticletitle{StePO-Rec: Towards Personalized Outfit Styling Assistant via Knowledge-Guided Multi-Step Reasoning}.
\newblock \bibinfo{journal}{\emph{arXiv preprint arXiv:2504.09915}} (\bibinfo{year}{2025}).
\newblock


\bibitem[Chen et~al\mbox{.}(2025b)]%
        {ReSearch}
\bibfield{author}{\bibinfo{person}{Mingyang Chen}, \bibinfo{person}{Tianpeng Li}, \bibinfo{person}{Haoze Sun}, \bibinfo{person}{Yijie Zhou}, \bibinfo{person}{Chenzheng Zhu}, \bibinfo{person}{Fan Yang}, \bibinfo{person}{Zenan Zhou}, \bibinfo{person}{Weipeng Chen}, \bibinfo{person}{Haofen Wang}, \bibinfo{person}{Jeff~Z Pan}, {et~al\mbox{.}}} \bibinfo{year}{2025}\natexlab{b}.
\newblock \showarticletitle{Learning to Reason with Search for LLMs via Reinforcement Learning}.
\newblock \bibinfo{journal}{\emph{arXiv preprint arXiv:2503.19470}} (\bibinfo{year}{2025}).
\newblock


\bibitem[Chen et~al\mbox{.}(2025f)]%
        {ARM}
\bibfield{author}{\bibinfo{person}{Peter~Baile Chen}, \bibinfo{person}{Yi Zhang}, \bibinfo{person}{Michael Cafarella}, {and} \bibinfo{person}{Dan Roth}.} \bibinfo{year}{2025}\natexlab{f}.
\newblock \showarticletitle{Can we Retrieve Everything All at Once? ARM: An Alignment-Oriented LLM-based Retrieval Method}.
\newblock \bibinfo{journal}{\emph{arXiv preprint arXiv:2501.18539}} (\bibinfo{year}{2025}).
\newblock


\bibitem[Chen et~al\mbox{.}(2025c)]%
        {long_cot}
\bibfield{author}{\bibinfo{person}{Qiguang Chen}, \bibinfo{person}{Libo Qin}, \bibinfo{person}{Jinhao Liu}, \bibinfo{person}{Dengyun Peng}, \bibinfo{person}{Jiannan Guan}, \bibinfo{person}{Peng Wang}, \bibinfo{person}{Mengkang Hu}, \bibinfo{person}{Yuhang Zhou}, \bibinfo{person}{Te Gao}, {and} \bibinfo{person}{Wangxiang Che}.} \bibinfo{year}{2025}\natexlab{c}.
\newblock \showarticletitle{Towards reasoning era: A survey of long chain-of-thought for reasoning large language models}.
\newblock \bibinfo{journal}{\emph{arXiv preprint arXiv:2503.09567}} (\bibinfo{year}{2025}).
\newblock


\bibitem[Chen et~al\mbox{.}(2023)]%
        {theoremqa}
\bibfield{author}{\bibinfo{person}{Wenhu Chen}, \bibinfo{person}{Ming Yin}, \bibinfo{person}{Max Ku}, \bibinfo{person}{Pan Lu}, \bibinfo{person}{Yixin Wan}, \bibinfo{person}{Xueguang Ma}, \bibinfo{person}{Jianyu Xu}, \bibinfo{person}{Xinyi Wang}, {and} \bibinfo{person}{Tony Xia}.} \bibinfo{year}{2023}\natexlab{}.
\newblock \showarticletitle{Theoremqa: A theorem-driven question answering dataset}.
\newblock \bibinfo{journal}{\emph{arXiv preprint arXiv:2305.12524}} (\bibinfo{year}{2023}).
\newblock


\bibitem[Chen et~al\mbox{.}(2024)]%
        {chen2024not}
\bibfield{author}{\bibinfo{person}{Xingyu Chen}, \bibinfo{person}{Jiahao Xu}, \bibinfo{person}{Tian Liang}, \bibinfo{person}{Zhiwei He}, \bibinfo{person}{Jianhui Pang}, \bibinfo{person}{Dian Yu}, \bibinfo{person}{Linfeng Song}, \bibinfo{person}{Qiuzhi Liu}, \bibinfo{person}{Mengfei Zhou}, \bibinfo{person}{Zhuosheng Zhang}, {et~al\mbox{.}}} \bibinfo{year}{2024}\natexlab{}.
\newblock \showarticletitle{Do not think that much for 2+ 3=? on the overthinking of o1-like llms}.
\newblock \bibinfo{journal}{\emph{arXiv preprint arXiv:2412.21187}} (\bibinfo{year}{2024}).
\newblock


\bibitem[Chen et~al\mbox{.}(2025d)]%
        {MRD-RAG}
\bibfield{author}{\bibinfo{person}{Yixiang Chen}, \bibinfo{person}{Penglei Sun}, \bibinfo{person}{Xiang Li}, {and} \bibinfo{person}{Xiaowen Chu}.} \bibinfo{year}{2025}\natexlab{d}.
\newblock \showarticletitle{MRD-RAG: Enhancing Medical Diagnosis with Multi-Round Retrieval-Augmented Generation}.
\newblock \bibinfo{journal}{\emph{arXiv preprint arXiv:2504.07724}} (\bibinfo{year}{2025}).
\newblock


\bibitem[Chen et~al\mbox{.}(2025e)]%
        {mmoa_rag}
\bibfield{author}{\bibinfo{person}{Yiqun Chen}, \bibinfo{person}{Lingyong Yan}, \bibinfo{person}{Weiwei Sun}, \bibinfo{person}{Xinyu Ma}, \bibinfo{person}{Yi Zhang}, \bibinfo{person}{Shuaiqiang Wang}, \bibinfo{person}{Dawei Yin}, \bibinfo{person}{Yiming Yang}, {and} \bibinfo{person}{Jiaxin Mao}.} \bibinfo{year}{2025}\natexlab{e}.
\newblock \showarticletitle{Improving Retrieval-Augmented Generation through Multi-Agent Reinforcement Learning}.
\newblock \bibinfo{journal}{\emph{arXiv preprint arXiv:2501.15228}} (\bibinfo{year}{2025}).
\newblock


\bibitem[Chen et~al\mbox{.}(2025a)]%
        {chen2025harnessing}
\bibfield{author}{\bibinfo{person}{Zhijun Chen}, \bibinfo{person}{Jingzheng Li}, \bibinfo{person}{Pengpeng Chen}, \bibinfo{person}{Zhuoran Li}, \bibinfo{person}{Kai Sun}, \bibinfo{person}{Yuankai Luo}, \bibinfo{person}{Qianren Mao}, \bibinfo{person}{Dingqi Yang}, \bibinfo{person}{Hailong Sun}, {and} \bibinfo{person}{Philip~S Yu}.} \bibinfo{year}{2025}\natexlab{a}.
\newblock \showarticletitle{Harnessing Multiple Large Language Models: A Survey on LLM Ensemble}.
\newblock \bibinfo{journal}{\emph{arXiv preprint arXiv:2502.18036}} (\bibinfo{year}{2025}).
\newblock


\bibitem[Cheng et~al\mbox{.}(2024)]%
        {UAR}
\bibfield{author}{\bibinfo{person}{Qinyuan Cheng}, \bibinfo{person}{Xiaonan Li}, \bibinfo{person}{Shimin Li}, \bibinfo{person}{Qin Zhu}, \bibinfo{person}{Zhangyue Yin}, \bibinfo{person}{Yunfan Shao}, \bibinfo{person}{Linyang Li}, \bibinfo{person}{Tianxiang Sun}, \bibinfo{person}{Hang Yan}, {and} \bibinfo{person}{Xipeng Qiu}.} \bibinfo{year}{2024}\natexlab{}.
\newblock \showarticletitle{Unified active retrieval for retrieval augmented generation}.
\newblock \bibinfo{journal}{\emph{arXiv preprint arXiv:2406.12534}} (\bibinfo{year}{2024}).
\newblock


\bibitem[Cuadron et~al\mbox{.}(2025)]%
        {cuadron2025danger}
\bibfield{author}{\bibinfo{person}{Alejandro Cuadron}, \bibinfo{person}{Dacheng Li}, \bibinfo{person}{Wenjie Ma}, \bibinfo{person}{Xingyao Wang}, \bibinfo{person}{Yichuan Wang}, \bibinfo{person}{Siyuan Zhuang}, \bibinfo{person}{Shu Liu}, \bibinfo{person}{Luis~Gaspar Schroeder}, \bibinfo{person}{Tian Xia}, \bibinfo{person}{Huanzhi Mao}, {et~al\mbox{.}}} \bibinfo{year}{2025}\natexlab{}.
\newblock \showarticletitle{The Danger of Overthinking: Examining the Reasoning-Action Dilemma in Agentic Tasks}.
\newblock \bibinfo{journal}{\emph{arXiv preprint arXiv:2502.08235}} (\bibinfo{year}{2025}).
\newblock


\bibitem[Dao and Le(2025)]%
        {ReZero}
\bibfield{author}{\bibinfo{person}{Alan Dao} {and} \bibinfo{person}{Thinh Le}.} \bibinfo{year}{2025}\natexlab{}.
\newblock \bibinfo{title}{ReZero: Enhancing LLM search ability by trying one-more-time}.
\newblock
\showeprint[arxiv]{2504.11001}~[cs.CL]
\urldef\tempurl%
\url{https://arxiv.org/abs/2504.11001}
\showURL{%
\tempurl}


\bibitem[Dettmers et~al\mbox{.}(2023)]%
        {qlora}
\bibfield{author}{\bibinfo{person}{Tim Dettmers}, \bibinfo{person}{Artidoro Pagnoni}, \bibinfo{person}{Ari Holtzman}, {and} \bibinfo{person}{Luke Zettlemoyer}.} \bibinfo{year}{2023}\natexlab{}.
\newblock \showarticletitle{Qlora: Efficient finetuning of quantized llms}.
\newblock \bibinfo{journal}{\emph{Advances in neural information processing systems}}  \bibinfo{volume}{36} (\bibinfo{year}{2023}), \bibinfo{pages}{10088--10115}.
\newblock


\bibitem[Edge et~al\mbox{.}(2024)]%
        {graphrag}
\bibfield{author}{\bibinfo{person}{Darren Edge}, \bibinfo{person}{Ha Trinh}, \bibinfo{person}{Newman Cheng}, \bibinfo{person}{Joshua Bradley}, \bibinfo{person}{Alex Chao}, \bibinfo{person}{Apurva Mody}, \bibinfo{person}{Steven Truitt}, \bibinfo{person}{Dasha Metropolitansky}, \bibinfo{person}{Robert~Osazuwa Ness}, {and} \bibinfo{person}{Jonathan Larson}.} \bibinfo{year}{2024}\natexlab{}.
\newblock \showarticletitle{From local to global: A graph rag approach to query-focused summarization}.
\newblock \bibinfo{journal}{\emph{arXiv preprint arXiv:2404.16130}} (\bibinfo{year}{2024}).
\newblock


\bibitem[Fan et~al\mbox{.}(2025)]%
        {fan2025missing}
\bibfield{author}{\bibinfo{person}{Chenrui Fan}, \bibinfo{person}{Ming Li}, \bibinfo{person}{Lichao Sun}, {and} \bibinfo{person}{Tianyi Zhou}.} \bibinfo{year}{2025}\natexlab{}.
\newblock \showarticletitle{Missing Premise exacerbates Overthinking: Are Reasoning Models losing Critical Thinking Skill?}
\newblock \bibinfo{journal}{\emph{arXiv preprint arXiv:2504.06514}} (\bibinfo{year}{2025}).
\newblock


\bibitem[Gao et~al\mbox{.}(2024a)]%
        {smartrag}
\bibfield{author}{\bibinfo{person}{Jingsheng Gao}, \bibinfo{person}{Linxu Li}, \bibinfo{person}{Weiyuan Li}, \bibinfo{person}{Yuzhuo Fu}, {and} \bibinfo{person}{Bin Dai}.} \bibinfo{year}{2024}\natexlab{a}.
\newblock \showarticletitle{SmartRAG: Jointly Learn RAG-Related Tasks From the Environment Feedback}.
\newblock \bibinfo{journal}{\emph{arXiv preprint arXiv:2410.18141}} (\bibinfo{year}{2024}).
\newblock


\bibitem[Gao et~al\mbox{.}(2023)]%
        {retrieval_survey}
\bibfield{author}{\bibinfo{person}{Yunfan Gao}, \bibinfo{person}{Yun Xiong}, \bibinfo{person}{Xinyu Gao}, \bibinfo{person}{Kangxiang Jia}, \bibinfo{person}{Jinliu Pan}, \bibinfo{person}{Yuxi Bi}, \bibinfo{person}{Yi Dai}, \bibinfo{person}{Jiawei Sun}, {and} \bibinfo{person}{Haofen Wang}.} \bibinfo{year}{2023}\natexlab{}.
\newblock \showarticletitle{Retrieval-augmented generation for large language models: A survey}.
\newblock \bibinfo{journal}{\emph{arXiv preprint arXiv:2312.10997}} (\bibinfo{year}{2023}).
\newblock


\bibitem[Gao et~al\mbox{.}(2024b)]%
        {modularRAG}
\bibfield{author}{\bibinfo{person}{Yunfan Gao}, \bibinfo{person}{Yun Xiong}, \bibinfo{person}{Meng Wang}, {and} \bibinfo{person}{Haofen Wang}.} \bibinfo{year}{2024}\natexlab{b}.
\newblock \showarticletitle{Modular rag: Transforming rag systems into lego-like reconfigurable frameworks}.
\newblock \bibinfo{journal}{\emph{arXiv preprint arXiv:2407.21059}} (\bibinfo{year}{2024}).
\newblock


\bibitem[Gao et~al\mbox{.}(2025)]%
        {frag}
\bibfield{author}{\bibinfo{person}{Zengyi Gao}, \bibinfo{person}{Yukun Cao}, \bibinfo{person}{Hairu Wang}, \bibinfo{person}{Ao Ke}, \bibinfo{person}{Yuan Feng}, \bibinfo{person}{Xike Xie}, {and} \bibinfo{person}{S~Kevin Zhou}.} \bibinfo{year}{2025}\natexlab{}.
\newblock \showarticletitle{FRAG: A Flexible Modular Framework for Retrieval-Augmented Generation based on Knowledge Graphs}.
\newblock \bibinfo{journal}{\emph{arXiv preprint arXiv:2501.09957}} (\bibinfo{year}{2025}).
\newblock


\bibitem[Guan et~al\mbox{.}(2025)]%
        {deeprag}
\bibfield{author}{\bibinfo{person}{Xinyan Guan}, \bibinfo{person}{Jiali Zeng}, \bibinfo{person}{Fandong Meng}, \bibinfo{person}{Chunlei Xin}, \bibinfo{person}{Yaojie Lu}, \bibinfo{person}{Hongyu Lin}, \bibinfo{person}{Xianpei Han}, \bibinfo{person}{Le Sun}, {and} \bibinfo{person}{Jie Zhou}.} \bibinfo{year}{2025}\natexlab{}.
\newblock \showarticletitle{DeepRAG: Thinking to Retrieval Step by Step for Large Language Models}.
\newblock \bibinfo{journal}{\emph{arXiv preprint arXiv:2502.01142}} (\bibinfo{year}{2025}).
\newblock


\bibitem[Guo et~al\mbox{.}(2025a)]%
        {deepseek-r1}
\bibfield{author}{\bibinfo{person}{Daya Guo}, \bibinfo{person}{Dejian Yang}, \bibinfo{person}{Haowei Zhang}, \bibinfo{person}{Junxiao Song}, \bibinfo{person}{Ruoyu Zhang}, \bibinfo{person}{Runxin Xu}, \bibinfo{person}{Qihao Zhu}, \bibinfo{person}{Shirong Ma}, \bibinfo{person}{Peiyi Wang}, \bibinfo{person}{Xiao Bi}, {et~al\mbox{.}}} \bibinfo{year}{2025}\natexlab{a}.
\newblock \showarticletitle{Deepseek-r1: Incentivizing reasoning capability in llms via reinforcement learning}.
\newblock \bibinfo{journal}{\emph{arXiv preprint arXiv:2501.12948}} (\bibinfo{year}{2025}).
\newblock


\bibitem[Guo et~al\mbox{.}(2025b)]%
        {deepseek_r1}
\bibfield{author}{\bibinfo{person}{Daya Guo}, \bibinfo{person}{Dejian Yang}, \bibinfo{person}{Haowei Zhang}, \bibinfo{person}{Junxiao Song}, \bibinfo{person}{Ruoyu Zhang}, \bibinfo{person}{Runxin Xu}, \bibinfo{person}{Qihao Zhu}, \bibinfo{person}{Shirong Ma}, \bibinfo{person}{Peiyi Wang}, \bibinfo{person}{Xiao Bi}, {et~al\mbox{.}}} \bibinfo{year}{2025}\natexlab{b}.
\newblock \showarticletitle{Deepseek-r1: Incentivizing reasoning capability in llms via reinforcement learning}.
\newblock \bibinfo{journal}{\emph{arXiv preprint arXiv:2501.12948}} (\bibinfo{year}{2025}).
\newblock


\bibitem[Guo et~al\mbox{.}(2024)]%
        {lightrag}
\bibfield{author}{\bibinfo{person}{Zirui Guo}, \bibinfo{person}{Lianghao Xia}, \bibinfo{person}{Yanhua Yu}, \bibinfo{person}{Tu Ao}, {and} \bibinfo{person}{Chao Huang}.} \bibinfo{year}{2024}\natexlab{}.
\newblock \showarticletitle{Lightrag: Simple and fast retrieval-augmented generation}.
\newblock  (\bibinfo{year}{2024}).
\newblock


\bibitem[Guti{\'e}rrez et~al\mbox{.}(2025)]%
        {hippo2}
\bibfield{author}{\bibinfo{person}{Bernal~Jim{\'e}nez Guti{\'e}rrez}, \bibinfo{person}{Yiheng Shu}, \bibinfo{person}{Weijian Qi}, \bibinfo{person}{Sizhe Zhou}, {and} \bibinfo{person}{Yu Su}.} \bibinfo{year}{2025}\natexlab{}.
\newblock \showarticletitle{From RAG to Memory: Non-Parametric Continual Learning for Large Language Models}.
\newblock \bibinfo{journal}{\emph{arXiv preprint arXiv:2502.14802}} (\bibinfo{year}{2025}).
\newblock


\bibitem[He et~al\mbox{.}(2024)]%
        {olympiadbench}
\bibfield{author}{\bibinfo{person}{Chaoqun He}, \bibinfo{person}{Renjie Luo}, \bibinfo{person}{Yuzhuo Bai}, \bibinfo{person}{Shengding Hu}, \bibinfo{person}{Zhen~Leng Thai}, \bibinfo{person}{Junhao Shen}, \bibinfo{person}{Jinyi Hu}, \bibinfo{person}{Xu Han}, \bibinfo{person}{Yujie Huang}, \bibinfo{person}{Yuxiang Zhang}, {et~al\mbox{.}}} \bibinfo{year}{2024}\natexlab{}.
\newblock \showarticletitle{Olympiadbench: A challenging benchmark for promoting agi with olympiad-level bilingual multimodal scientific problems}.
\newblock \bibinfo{journal}{\emph{arXiv preprint arXiv:2402.14008}} (\bibinfo{year}{2024}).
\newblock


\bibitem[He et~al\mbox{.}(2025)]%
        {he2025can}
\bibfield{author}{\bibinfo{person}{Yancheng He}, \bibinfo{person}{Shilong Li}, \bibinfo{person}{Jiaheng Liu}, \bibinfo{person}{Weixun Wang}, \bibinfo{person}{Xingyuan Bu}, \bibinfo{person}{Ge Zhang}, \bibinfo{person}{Zhongyuan Peng}, \bibinfo{person}{Zhaoxiang Zhang}, \bibinfo{person}{Zhicheng Zheng}, \bibinfo{person}{Wenbo Su}, {et~al\mbox{.}}} \bibinfo{year}{2025}\natexlab{}.
\newblock \showarticletitle{Can Large Language Models Detect Errors in Long Chain-of-Thought Reasoning?}
\newblock \bibinfo{journal}{\emph{arXiv preprint arXiv:2502.19361}} (\bibinfo{year}{2025}).
\newblock


\bibitem[Ho et~al\mbox{.}(2020)]%
        {2WikiMultiHopQA}
\bibfield{author}{\bibinfo{person}{Xanh Ho}, \bibinfo{person}{Anh-Khoa~Duong Nguyen}, \bibinfo{person}{Saku Sugawara}, {and} \bibinfo{person}{Akiko Aizawa}.} \bibinfo{year}{2020}\natexlab{}.
\newblock \showarticletitle{Constructing a multi-hop qa dataset for comprehensive evaluation of reasoning steps}.
\newblock \bibinfo{journal}{\emph{arXiv preprint arXiv:2011.01060}} (\bibinfo{year}{2020}).
\newblock


\bibitem[Hong et~al\mbox{.}(2025)]%
        {FG-RAG}
\bibfield{author}{\bibinfo{person}{Yubin Hong}, \bibinfo{person}{Chaofan Li}, \bibinfo{person}{Jingyi Zhang}, {and} \bibinfo{person}{Yingxia Shao}.} \bibinfo{year}{2025}\natexlab{}.
\newblock \showarticletitle{FG-RAG: Enhancing Query-Focused Summarization with Context-Aware Fine-Grained Graph RAG}.
\newblock \bibinfo{journal}{\emph{arXiv preprint arXiv:2504.07103}} (\bibinfo{year}{2025}).
\newblock


\bibitem[Hongjin et~al\mbox{.}(2024)]%
        {TheoremQA-Math}
\bibfield{author}{\bibinfo{person}{SU Hongjin}, \bibinfo{person}{Howard Yen}, \bibinfo{person}{Mengzhou Xia}, \bibinfo{person}{Weijia Shi}, \bibinfo{person}{Niklas Muennighoff}, \bibinfo{person}{Han-yu Wang}, \bibinfo{person}{Liu Haisu}, \bibinfo{person}{Quan Shi}, \bibinfo{person}{Zachary~S Siegel}, \bibinfo{person}{Michael Tang}, {et~al\mbox{.}}} \bibinfo{year}{2024}\natexlab{}.
\newblock \showarticletitle{BRIGHT: A Realistic and Challenging Benchmark for Reasoning-Intensive Retrieval}. In \bibinfo{booktitle}{\emph{The Thirteenth International Conference on Learning Representations}}.
\newblock


\bibitem[Hsu et~al\mbox{.}(2024)]%
        {LeReT}
\bibfield{author}{\bibinfo{person}{Sheryl Hsu}, \bibinfo{person}{Omar Khattab}, \bibinfo{person}{Chelsea Finn}, {and} \bibinfo{person}{Archit Sharma}.} \bibinfo{year}{2024}\natexlab{}.
\newblock \showarticletitle{Grounding by trying: Llms with reinforcement learning-enhanced retrieval}.
\newblock \bibinfo{journal}{\emph{arXiv preprint arXiv:2410.23214}} (\bibinfo{year}{2024}).
\newblock


\bibitem[Hu(2025)]%
        {reinforce++}
\bibfield{author}{\bibinfo{person}{Jian Hu}.} \bibinfo{year}{2025}\natexlab{}.
\newblock \showarticletitle{REINFORCE++: A Simple and Efficient Approach for Aligning Large Language Models}.
\newblock \bibinfo{journal}{\emph{arXiv preprint arXiv:2501.03262}} (\bibinfo{year}{2025}).
\newblock


\bibitem[Hu et~al\mbox{.}(2025)]%
        {MCTS-RAG}
\bibfield{author}{\bibinfo{person}{Yunhai Hu}, \bibinfo{person}{Yilun Zhao}, \bibinfo{person}{Chen Zhao}, {and} \bibinfo{person}{Arman Cohan}.} \bibinfo{year}{2025}\natexlab{}.
\newblock \showarticletitle{MCTS-RAG: Enhancing Retrieval-Augmented Generation with Monte Carlo Tree Search}.
\newblock \bibinfo{journal}{\emph{arXiv preprint arXiv:2503.20757}} (\bibinfo{year}{2025}).
\newblock


\bibitem[Huot et~al\mbox{.}(2024)]%
        {tell_me_a_story}
\bibfield{author}{\bibinfo{person}{Fantine Huot}, \bibinfo{person}{Reinald~Kim Amplayo}, \bibinfo{person}{Jennimaria Palomaki}, \bibinfo{person}{Alice~Shoshana Jakobovits}, \bibinfo{person}{Elizabeth Clark}, {and} \bibinfo{person}{Mirella Lapata}.} \bibinfo{year}{2024}\natexlab{}.
\newblock \showarticletitle{Agents' Room: Narrative Generation through Multi-step Collaboration}.
\newblock \bibinfo{journal}{\emph{arXiv preprint arXiv:2410.02603}} (\bibinfo{year}{2024}).
\newblock


\bibitem[Islam et~al\mbox{.}(2024)]%
        {openrag}
\bibfield{author}{\bibinfo{person}{Shayekh~Bin Islam}, \bibinfo{person}{Md~Asib Rahman}, \bibinfo{person}{KSM Hossain}, \bibinfo{person}{Enamul Hoque}, \bibinfo{person}{Shafiq Joty}, {and} \bibinfo{person}{Md~Rizwan Parvez}.} \bibinfo{year}{2024}\natexlab{}.
\newblock \showarticletitle{Open-rag: Enhanced retrieval-augmented reasoning with open-source large language models}.
\newblock \bibinfo{journal}{\emph{arXiv preprint arXiv:2410.01782}} (\bibinfo{year}{2024}).
\newblock


\bibitem[Jaech et~al\mbox{.}(2024)]%
        {O1}
\bibfield{author}{\bibinfo{person}{Aaron Jaech}, \bibinfo{person}{Adam Kalai}, \bibinfo{person}{Adam Lerer}, \bibinfo{person}{Adam Richardson}, \bibinfo{person}{Ahmed El-Kishky}, \bibinfo{person}{Aiden Low}, \bibinfo{person}{Alec Helyar}, \bibinfo{person}{Aleksander Madry}, \bibinfo{person}{Alex Beutel}, \bibinfo{person}{Alex Carney}, {et~al\mbox{.}}} \bibinfo{year}{2024}\natexlab{}.
\newblock \showarticletitle{Openai o1 system card}.
\newblock \bibinfo{journal}{\emph{arXiv preprint arXiv:2412.16720}} (\bibinfo{year}{2024}).
\newblock


\bibitem[Jain et~al\mbox{.}(2024)]%
        {LiveCodeBench}
\bibfield{author}{\bibinfo{person}{Naman Jain}, \bibinfo{person}{King Han}, \bibinfo{person}{Alex Gu}, \bibinfo{person}{Wen-Ding Li}, \bibinfo{person}{Fanjia Yan}, \bibinfo{person}{Tianjun Zhang}, \bibinfo{person}{Sida Wang}, \bibinfo{person}{Armando Solar-Lezama}, \bibinfo{person}{Koushik Sen}, {and} \bibinfo{person}{Ion Stoica}.} \bibinfo{year}{2024}\natexlab{}.
\newblock \showarticletitle{Livecodebench: Holistic and contamination free evaluation of large language models for code}.
\newblock \bibinfo{journal}{\emph{arXiv preprint arXiv:2403.07974}} (\bibinfo{year}{2024}).
\newblock


\bibitem[Jeong et~al\mbox{.}(2024)]%
        {adaptiveRAG}
\bibfield{author}{\bibinfo{person}{Soyeong Jeong}, \bibinfo{person}{Jinheon Baek}, \bibinfo{person}{Sukmin Cho}, \bibinfo{person}{Sung~Ju Hwang}, {and} \bibinfo{person}{Jong~C Park}.} \bibinfo{year}{2024}\natexlab{}.
\newblock \showarticletitle{Adaptive-rag: Learning to adapt retrieval-augmented large language models through question complexity}.
\newblock \bibinfo{journal}{\emph{arXiv preprint arXiv:2403.14403}} (\bibinfo{year}{2024}).
\newblock


\bibitem[Jiang(2025)]%
        {deepretrieval}
\bibfield{author}{\bibinfo{person}{Pengcheng Jiang}.} \bibinfo{year}{2025}\natexlab{}.
\newblock \showarticletitle{DeepRetrieval: Powerful Query Generation for Information Retrieval with Reinforcement Learning}.
\newblock \bibinfo{journal}{\emph{arXiv preprint arXiv:2503.00223}} (\bibinfo{year}{2025}).
\newblock


\bibitem[Jiang et~al\mbox{.}(2024a)]%
        {co-storm}
\bibfield{author}{\bibinfo{person}{Yucheng Jiang}, \bibinfo{person}{Yijia Shao}, \bibinfo{person}{Dekun Ma}, \bibinfo{person}{Sina~J Semnani}, {and} \bibinfo{person}{Monica~S Lam}.} \bibinfo{year}{2024}\natexlab{a}.
\newblock \showarticletitle{Into the unknown unknowns: Engaged human learning through participation in language model agent conversations}.
\newblock \bibinfo{journal}{\emph{arXiv preprint arXiv:2408.15232}} (\bibinfo{year}{2024}).
\newblock


\bibitem[Jiang et~al\mbox{.}(2024b)]%
        {WildSeek}
\bibfield{author}{\bibinfo{person}{Yucheng Jiang}, \bibinfo{person}{Yijia Shao}, \bibinfo{person}{Dekun Ma}, \bibinfo{person}{Sina~J Semnani}, {and} \bibinfo{person}{Monica~S Lam}.} \bibinfo{year}{2024}\natexlab{b}.
\newblock \showarticletitle{Into the unknown unknowns: Engaged human learning through participation in language model agent conversations}.
\newblock \bibinfo{journal}{\emph{arXiv preprint arXiv:2408.15232}} (\bibinfo{year}{2024}).
\newblock


\bibitem[Jiang et~al\mbox{.}(2023)]%
        {Flare}
\bibfield{author}{\bibinfo{person}{Zhengbao Jiang}, \bibinfo{person}{Frank~F Xu}, \bibinfo{person}{Luyu Gao}, \bibinfo{person}{Zhiqing Sun}, \bibinfo{person}{Qian Liu}, \bibinfo{person}{Jane Dwivedi-Yu}, \bibinfo{person}{Yiming Yang}, \bibinfo{person}{Jamie Callan}, {and} \bibinfo{person}{Graham Neubig}.} \bibinfo{year}{2023}\natexlab{}.
\newblock \showarticletitle{Active retrieval augmented generation}. In \bibinfo{booktitle}{\emph{Proceedings of the 2023 Conference on Empirical Methods in Natural Language Processing}}. \bibinfo{pages}{7969--7992}.
\newblock


\bibitem[Joshi et~al\mbox{.}(2024)]%
        {reaper}
\bibfield{author}{\bibinfo{person}{Ashutosh Joshi}, \bibinfo{person}{Sheikh~Muhammad Sarwar}, \bibinfo{person}{Samarth Varshney}, \bibinfo{person}{Sreyashi Nag}, \bibinfo{person}{Shrivats Agrawal}, {and} \bibinfo{person}{Juhi Naik}.} \bibinfo{year}{2024}\natexlab{}.
\newblock \showarticletitle{REAPER: Reasoning based retrieval planning for complex RAG systems}. In \bibinfo{booktitle}{\emph{Proceedings of the 33rd ACM International Conference on Information and Knowledge Management}}. \bibinfo{pages}{4621--4628}.
\newblock


\bibitem[Kwiatkowski et~al\mbox{.}(2019)]%
        {nq}
\bibfield{author}{\bibinfo{person}{Tom Kwiatkowski}, \bibinfo{person}{Jennimaria Palomaki}, \bibinfo{person}{Olivia Redfield}, \bibinfo{person}{Michael Collins}, \bibinfo{person}{Ankur Parikh}, \bibinfo{person}{Chris Alberti}, \bibinfo{person}{Danielle Epstein}, \bibinfo{person}{Illia Polosukhin}, \bibinfo{person}{Jacob Devlin}, \bibinfo{person}{Kenton Lee}, {et~al\mbox{.}}} \bibinfo{year}{2019}\natexlab{}.
\newblock \showarticletitle{Natural questions: a benchmark for question answering research}.
\newblock \bibinfo{journal}{\emph{Transactions of the Association for Computational Linguistics}}  \bibinfo{volume}{7} (\bibinfo{year}{2019}), \bibinfo{pages}{453--466}.
\newblock


\bibitem[Lee et~al\mbox{.}(2024)]%
        {planrag}
\bibfield{author}{\bibinfo{person}{Myeonghwa Lee}, \bibinfo{person}{Seonho An}, {and} \bibinfo{person}{Min-Soo Kim}.} \bibinfo{year}{2024}\natexlab{}.
\newblock \showarticletitle{PlanRAG: A plan-then-retrieval augmented generation for generative large language models as decision makers}. In \bibinfo{booktitle}{\emph{Proceedings of the 2024 Conference of the North American Chapter of the Association for Computational Linguistics: Human Language Technologies (Volume 1: Long Papers)}}. \bibinfo{pages}{6537--6555}.
\newblock


\bibitem[Lee et~al\mbox{.}(2025)]%
        {rearag}
\bibfield{author}{\bibinfo{person}{Zhicheng Lee}, \bibinfo{person}{Shulin Cao}, \bibinfo{person}{Jinxin Liu}, \bibinfo{person}{Jiajie Zhang}, \bibinfo{person}{Weichuan Liu}, \bibinfo{person}{Xiaoyin Che}, \bibinfo{person}{Lei Hou}, {and} \bibinfo{person}{Juanzi Li}.} \bibinfo{year}{2025}\natexlab{}.
\newblock \showarticletitle{ReaRAG: Knowledge-guided Reasoning Enhances Factuality of Large Reasoning Models with Iterative Retrieval Augmented Generation}.
\newblock \bibinfo{journal}{\emph{arXiv preprint arXiv:2503.21729}} (\bibinfo{year}{2025}).
\newblock


\bibitem[Li et~al\mbox{.}(2024c)]%
        {FinSearch}
\bibfield{author}{\bibinfo{person}{Jinzheng Li}, \bibinfo{person}{Jingshu Zhang}, \bibinfo{person}{Hongguang Li}, {and} \bibinfo{person}{Yiqing Shen}.} \bibinfo{year}{2024}\natexlab{c}.
\newblock \showarticletitle{An Agent Framework for Real-Time Financial Information Searching with Large Language Models}.
\newblock \bibinfo{journal}{\emph{arXiv preprint arXiv:2502.15684}} (\bibinfo{year}{2024}).
\newblock


\bibitem[Li et~al\mbox{.}(2025a)]%
        {search_o1}
\bibfield{author}{\bibinfo{person}{Xiaoxi Li}, \bibinfo{person}{Guanting Dong}, \bibinfo{person}{Jiajie Jin}, \bibinfo{person}{Yuyao Zhang}, \bibinfo{person}{Yujia Zhou}, \bibinfo{person}{Yutao Zhu}, \bibinfo{person}{Peitian Zhang}, {and} \bibinfo{person}{Zhicheng Dou}.} \bibinfo{year}{2025}\natexlab{a}.
\newblock \showarticletitle{Search-o1: Agentic search-enhanced large reasoning models}.
\newblock \bibinfo{journal}{\emph{arXiv preprint arXiv:2501.05366}} (\bibinfo{year}{2025}).
\newblock


\bibitem[Li et~al\mbox{.}(2024a)]%
        {CR-planner}
\bibfield{author}{\bibinfo{person}{Xingxuan Li}, \bibinfo{person}{Weiwen Xu}, \bibinfo{person}{Ruochen Zhao}, \bibinfo{person}{Fangkai Jiao}, \bibinfo{person}{Shafiq Joty}, {and} \bibinfo{person}{Lidong Bing}.} \bibinfo{year}{2024}\natexlab{a}.
\newblock \showarticletitle{Can We Further Elicit Reasoning in LLMs? Critic-Guided Planning with Retrieval-Augmentation for Solving Challenging Tasks}.
\newblock \bibinfo{journal}{\emph{arXiv preprint arXiv:2410.01428}} (\bibinfo{year}{2024}).
\newblock


\bibitem[Li et~al\mbox{.}(2024b)]%
        {li2024can}
\bibfield{author}{\bibinfo{person}{Xingxuan Li}, \bibinfo{person}{Weiwen Xu}, \bibinfo{person}{Ruochen Zhao}, \bibinfo{person}{Fangkai Jiao}, \bibinfo{person}{Shafiq Joty}, {and} \bibinfo{person}{Lidong Bing}.} \bibinfo{year}{2024}\natexlab{b}.
\newblock \showarticletitle{Can We Further Elicit Reasoning in LLMs? Critic-Guided Planning with Retrieval-Augmentation for Solving Challenging Tasks}.
\newblock \bibinfo{journal}{\emph{arXiv preprint arXiv:2410.01428}} (\bibinfo{year}{2024}).
\newblock


\bibitem[Li et~al\mbox{.}(2025b)]%
        {li2025deepsolution}
\bibfield{author}{\bibinfo{person}{Zhuoqun Li}, \bibinfo{person}{Haiyang Yu}, \bibinfo{person}{Xuanang Chen}, \bibinfo{person}{Hongyu Lin}, \bibinfo{person}{Yaojie Lu}, \bibinfo{person}{Fei Huang}, \bibinfo{person}{Xianpei Han}, \bibinfo{person}{Yongbin Li}, {and} \bibinfo{person}{Le Sun}.} \bibinfo{year}{2025}\natexlab{b}.
\newblock \showarticletitle{Deepsolution: Boosting complex engineering solution design via tree-based exploration and bi-point thinking}.
\newblock \bibinfo{journal}{\emph{arXiv preprint arXiv:2502.20730}} (\bibinfo{year}{2025}).
\newblock


\bibitem[Lightman et~al\mbox{.}(2023)]%
        {MATH500}
\bibfield{author}{\bibinfo{person}{Hunter Lightman}, \bibinfo{person}{Vineet Kosaraju}, \bibinfo{person}{Yuri Burda}, \bibinfo{person}{Harrison Edwards}, \bibinfo{person}{Bowen Baker}, \bibinfo{person}{Teddy Lee}, \bibinfo{person}{Jan Leike}, \bibinfo{person}{John Schulman}, \bibinfo{person}{Ilya Sutskever}, {and} \bibinfo{person}{Karl Cobbe}.} \bibinfo{year}{2023}\natexlab{}.
\newblock \showarticletitle{Let's verify step by step}. In \bibinfo{booktitle}{\emph{The Twelfth International Conference on Learning Representations}}.
\newblock


\bibitem[Liu et~al\mbox{.}(2024)]%
        {medcot}
\bibfield{author}{\bibinfo{person}{Jiaxiang Liu}, \bibinfo{person}{Yuan Wang}, \bibinfo{person}{Jiawei Du}, \bibinfo{person}{Joey~Tianyi Zhou}, {and} \bibinfo{person}{Zuozhu Liu}.} \bibinfo{year}{2024}\natexlab{}.
\newblock \showarticletitle{Medcot: Medical chain of thought via hierarchical expert}.
\newblock \bibinfo{journal}{\emph{arXiv preprint arXiv:2412.13736}} (\bibinfo{year}{2024}).
\newblock


\bibitem[Lumer et~al\mbox{.}(2025)]%
        {graph_tool}
\bibfield{author}{\bibinfo{person}{Elias Lumer}, \bibinfo{person}{Pradeep~Honaganahalli Basavaraju}, \bibinfo{person}{Myles Mason}, \bibinfo{person}{James~A Burke}, {and} \bibinfo{person}{Vamse~Kumar Subbiah}.} \bibinfo{year}{2025}\natexlab{}.
\newblock \showarticletitle{Graph RAG-Tool Fusion}.
\newblock \bibinfo{journal}{\emph{arXiv preprint arXiv:2502.07223}} (\bibinfo{year}{2025}).
\newblock


\bibitem[Luo et~al\mbox{.}(2025)]%
        {kbqa_o1}
\bibfield{author}{\bibinfo{person}{Haoran Luo}, \bibinfo{person}{Yikai Guo}, \bibinfo{person}{Qika Lin}, \bibinfo{person}{Xiaobao Wu}, \bibinfo{person}{Xinyu Mu}, \bibinfo{person}{Wenhao Liu}, \bibinfo{person}{Meina Song}, \bibinfo{person}{Yifan Zhu}, \bibinfo{person}{Luu~Anh Tuan}, {et~al\mbox{.}}} \bibinfo{year}{2025}\natexlab{}.
\newblock \showarticletitle{KBQA-o1: Agentic Knowledge Base Question Answering with Monte Carlo Tree Search}.
\newblock \bibinfo{journal}{\emph{arXiv preprint arXiv:2501.18922}} (\bibinfo{year}{2025}).
\newblock


\bibitem[Lyu et~al\mbox{.}(2025)]%
        {crud}
\bibfield{author}{\bibinfo{person}{Yuanjie Lyu}, \bibinfo{person}{Zhiyu Li}, \bibinfo{person}{Simin Niu}, \bibinfo{person}{Feiyu Xiong}, \bibinfo{person}{Bo Tang}, \bibinfo{person}{Wenjin Wang}, \bibinfo{person}{Hao Wu}, \bibinfo{person}{Huanyong Liu}, \bibinfo{person}{Tong Xu}, {and} \bibinfo{person}{Enhong Chen}.} \bibinfo{year}{2025}\natexlab{}.
\newblock \showarticletitle{Crud-rag: A comprehensive chinese benchmark for retrieval-augmented generation of large language models}.
\newblock \bibinfo{journal}{\emph{ACM Transactions on Information Systems}} \bibinfo{volume}{43}, \bibinfo{number}{2} (\bibinfo{year}{2025}), \bibinfo{pages}{1--32}.
\newblock


\bibitem[Ma et~al\mbox{.}(2024)]%
        {ToG2}
\bibfield{author}{\bibinfo{person}{Shengjie Ma}, \bibinfo{person}{Chengjin Xu}, \bibinfo{person}{Xuhui Jiang}, \bibinfo{person}{Muzhi Li}, \bibinfo{person}{Huaren Qu}, \bibinfo{person}{Cehao Yang}, \bibinfo{person}{Jiaxin Mao}, {and} \bibinfo{person}{Jian Guo}.} \bibinfo{year}{2024}\natexlab{}.
\newblock \showarticletitle{Think-on-Graph 2.0: Deep and Faithful Large Language Model Reasoning with Knowledge-guided Retrieval Augmented Generation}.
\newblock \bibinfo{journal}{\emph{arXiv preprint arXiv:2407.10805}} (\bibinfo{year}{2024}).
\newblock


\bibitem[Ma et~al\mbox{.}(2023)]%
        {rrr}
\bibfield{author}{\bibinfo{person}{Xinbei Ma}, \bibinfo{person}{Yeyun Gong}, \bibinfo{person}{Pengcheng He}, \bibinfo{person}{Hai Zhao}, {and} \bibinfo{person}{Nan Duan}.} \bibinfo{year}{2023}\natexlab{}.
\newblock \showarticletitle{Query rewriting in retrieval-augmented large language models}. In \bibinfo{booktitle}{\emph{Proceedings of the 2023 Conference on Empirical Methods in Natural Language Processing}}. \bibinfo{pages}{5303--5315}.
\newblock


\bibitem[Mialon et~al\mbox{.}(2023)]%
        {gaia}
\bibfield{author}{\bibinfo{person}{Gr{\'e}goire Mialon}, \bibinfo{person}{Cl{\'e}mentine Fourrier}, \bibinfo{person}{Thomas Wolf}, \bibinfo{person}{Yann LeCun}, {and} \bibinfo{person}{Thomas Scialom}.} \bibinfo{year}{2023}\natexlab{}.
\newblock \showarticletitle{Gaia: a benchmark for general ai assistants}. In \bibinfo{booktitle}{\emph{The Twelfth International Conference on Learning Representations}}.
\newblock


\bibitem[Muennighoff et~al\mbox{.}(2025)]%
        {s1}
\bibfield{author}{\bibinfo{person}{Niklas Muennighoff}, \bibinfo{person}{Zitong Yang}, \bibinfo{person}{Weijia Shi}, \bibinfo{person}{Xiang~Lisa Li}, \bibinfo{person}{Li Fei-Fei}, \bibinfo{person}{Hannaneh Hajishirzi}, \bibinfo{person}{Luke Zettlemoyer}, \bibinfo{person}{Percy Liang}, \bibinfo{person}{Emmanuel Cand{\`e}s}, {and} \bibinfo{person}{Tatsunori Hashimoto}.} \bibinfo{year}{2025}\natexlab{}.
\newblock \showarticletitle{s1: Simple test-time scaling}.
\newblock \bibinfo{journal}{\emph{arXiv preprint arXiv:2501.19393}} (\bibinfo{year}{2025}).
\newblock


\bibitem[Patil et~al\mbox{.}(2024)]%
        {Gorilla}
\bibfield{author}{\bibinfo{person}{Shishir~G Patil}, \bibinfo{person}{Tianjun Zhang}, \bibinfo{person}{Xin Wang}, {and} \bibinfo{person}{Joseph~E Gonzalez}.} \bibinfo{year}{2024}\natexlab{}.
\newblock \showarticletitle{Gorilla: Large language model connected with massive apis}.
\newblock \bibinfo{journal}{\emph{Advances in Neural Information Processing Systems}}  \bibinfo{volume}{37} (\bibinfo{year}{2024}), \bibinfo{pages}{126544--126565}.
\newblock


\bibitem[Petroni et~al\mbox{.}(2020)]%
        {kilt}
\bibfield{author}{\bibinfo{person}{Fabio Petroni}, \bibinfo{person}{Aleksandra Piktus}, \bibinfo{person}{Angela Fan}, \bibinfo{person}{Patrick Lewis}, \bibinfo{person}{Majid Yazdani}, \bibinfo{person}{Nicola De~Cao}, \bibinfo{person}{James Thorne}, \bibinfo{person}{Yacine Jernite}, \bibinfo{person}{Vladimir Karpukhin}, \bibinfo{person}{Jean Maillard}, {et~al\mbox{.}}} \bibinfo{year}{2020}\natexlab{}.
\newblock \showarticletitle{KILT: a benchmark for knowledge intensive language tasks}.
\newblock \bibinfo{journal}{\emph{arXiv preprint arXiv:2009.02252}} (\bibinfo{year}{2020}).
\newblock


\bibitem[Pezeshkpour and Hruschka(2025)]%
        {Insight-RAG}
\bibfield{author}{\bibinfo{person}{Pouya Pezeshkpour} {and} \bibinfo{person}{Estevam Hruschka}.} \bibinfo{year}{2025}\natexlab{}.
\newblock \showarticletitle{Insight-RAG: Enhancing LLMs with Insight-Driven Augmentation}.
\newblock \bibinfo{journal}{\emph{arXiv preprint arXiv:2504.00187}} (\bibinfo{year}{2025}).
\newblock


\bibitem[Rein et~al\mbox{.}(2024)]%
        {gpqa}
\bibfield{author}{\bibinfo{person}{David Rein}, \bibinfo{person}{Betty~Li Hou}, \bibinfo{person}{Asa~Cooper Stickland}, \bibinfo{person}{Jackson Petty}, \bibinfo{person}{Richard~Yuanzhe Pang}, \bibinfo{person}{Julien Dirani}, \bibinfo{person}{Julian Michael}, {and} \bibinfo{person}{Samuel~R Bowman}.} \bibinfo{year}{2024}\natexlab{}.
\newblock \showarticletitle{Gpqa: A graduate-level google-proof q\&a benchmark}. In \bibinfo{booktitle}{\emph{First Conference on Language Modeling}}.
\newblock


\bibitem[Shao et~al\mbox{.}(2023)]%
        {iter-retgen}
\bibfield{author}{\bibinfo{person}{Zhihong Shao}, \bibinfo{person}{Yeyun Gong}, \bibinfo{person}{Yelong Shen}, \bibinfo{person}{Minlie Huang}, \bibinfo{person}{Nan Duan}, {and} \bibinfo{person}{Weizhu Chen}.} \bibinfo{year}{2023}\natexlab{}.
\newblock \showarticletitle{Enhancing retrieval-augmented large language models with iterative retrieval-generation synergy}.
\newblock \bibinfo{journal}{\emph{arXiv preprint arXiv:2305.15294}} (\bibinfo{year}{2023}).
\newblock


\bibitem[Shao et~al\mbox{.}(2024)]%
        {grpo}
\bibfield{author}{\bibinfo{person}{Zhihong Shao}, \bibinfo{person}{Peiyi Wang}, \bibinfo{person}{Qihao Zhu}, \bibinfo{person}{Runxin Xu}, \bibinfo{person}{Junxiao Song}, \bibinfo{person}{Xiao Bi}, \bibinfo{person}{Haowei Zhang}, \bibinfo{person}{Mingchuan Zhang}, \bibinfo{person}{YK Li}, \bibinfo{person}{Y Wu}, {et~al\mbox{.}}} \bibinfo{year}{2024}\natexlab{}.
\newblock \showarticletitle{Deepseekmath: Pushing the limits of mathematical reasoning in open language models}.
\newblock \bibinfo{journal}{\emph{arXiv preprint arXiv:2402.03300}} (\bibinfo{year}{2024}).
\newblock


\bibitem[Shi et~al\mbox{.}(2024a)]%
        {USACO}
\bibfield{author}{\bibinfo{person}{Quan Shi}, \bibinfo{person}{Michael Tang}, \bibinfo{person}{Karthik Narasimhan}, {and} \bibinfo{person}{Shunyu Yao}.} \bibinfo{year}{2024}\natexlab{a}.
\newblock \showarticletitle{Can Language Models Solve Olympiad Programming?}
\newblock \bibinfo{journal}{\emph{arXiv preprint arXiv:2404.10952}} (\bibinfo{year}{2024}).
\newblock


\bibitem[Shi et~al\mbox{.}(2024b)]%
        {shi2024can}
\bibfield{author}{\bibinfo{person}{Quan Shi}, \bibinfo{person}{Michael Tang}, \bibinfo{person}{Karthik Narasimhan}, {and} \bibinfo{person}{Shunyu Yao}.} \bibinfo{year}{2024}\natexlab{b}.
\newblock \showarticletitle{Can Language Models Solve Olympiad Programming?}
\newblock \bibinfo{journal}{\emph{arXiv preprint arXiv:2404.10952}} (\bibinfo{year}{2024}).
\newblock


\bibitem[Song et~al\mbox{.}(2025)]%
        {R1-Searcher}
\bibfield{author}{\bibinfo{person}{Huatong Song}, \bibinfo{person}{Jinhao Jiang}, \bibinfo{person}{Yingqian Min}, \bibinfo{person}{Jie Chen}, \bibinfo{person}{Zhipeng Chen}, \bibinfo{person}{Wayne~Xin Zhao}, \bibinfo{person}{Lei Fang}, {and} \bibinfo{person}{Ji-Rong Wen}.} \bibinfo{year}{2025}\natexlab{}.
\newblock \showarticletitle{R1-Searcher: Incentivizing the Search Capability in LLMs via Reinforcement Learning}.
\newblock \bibinfo{journal}{\emph{arXiv preprint arXiv:2503.05592}} (\bibinfo{year}{2025}).
\newblock


\bibitem[Srinivas and Runkana(2025)]%
        {PORAG}
\bibfield{author}{\bibinfo{person}{Sakhinana~Sagar Srinivas} {and} \bibinfo{person}{Venkataramana Runkana}.} \bibinfo{year}{2025}\natexlab{}.
\newblock \showarticletitle{Scaling Test-Time Inference with Policy-Optimized, Dynamic Retrieval-Augmented Generation via KV Caching and Decoding}.
\newblock \bibinfo{journal}{\emph{arXiv preprint arXiv:2504.01281}} (\bibinfo{year}{2025}).
\newblock


\bibitem[Sui et~al\mbox{.}(2025)]%
        {sui2025stop}
\bibfield{author}{\bibinfo{person}{Yang Sui}, \bibinfo{person}{Yu-Neng Chuang}, \bibinfo{person}{Guanchu Wang}, \bibinfo{person}{Jiamu Zhang}, \bibinfo{person}{Tianyi Zhang}, \bibinfo{person}{Jiayi Yuan}, \bibinfo{person}{Hongyi Liu}, \bibinfo{person}{Andrew Wen}, \bibinfo{person}{Hanjie Chen}, \bibinfo{person}{Xia Hu}, {et~al\mbox{.}}} \bibinfo{year}{2025}\natexlab{}.
\newblock \showarticletitle{Stop overthinking: A survey on efficient reasoning for large language models}.
\newblock \bibinfo{journal}{\emph{arXiv preprint arXiv:2503.16419}} (\bibinfo{year}{2025}).
\newblock


\bibitem[Sun et~al\mbox{.}(2025)]%
        {rearter}
\bibfield{author}{\bibinfo{person}{Zhongxiang Sun}, \bibinfo{person}{Qipeng Wang}, \bibinfo{person}{Weijie Yu}, \bibinfo{person}{Xiaoxue Zang}, \bibinfo{person}{Kai Zheng}, \bibinfo{person}{Jun Xu}, \bibinfo{person}{Xiao Zhang}, \bibinfo{person}{Song Yang}, {and} \bibinfo{person}{Han Li}.} \bibinfo{year}{2025}\natexlab{}.
\newblock \showarticletitle{ReARTeR: Retrieval-Augmented Reasoning with Trustworthy Process Rewarding}.
\newblock \bibinfo{journal}{\emph{arXiv preprint arXiv:2501.07861}} (\bibinfo{year}{2025}).
\newblock


\bibitem[Talmor and Berant(2018)]%
        {ComplexWebQA}
\bibfield{author}{\bibinfo{person}{Alon Talmor} {and} \bibinfo{person}{Jonathan Berant}.} \bibinfo{year}{2018}\natexlab{}.
\newblock \showarticletitle{The web as a knowledge-base for answering complex questions}.
\newblock \bibinfo{journal}{\emph{arXiv preprint arXiv:1803.06643}} (\bibinfo{year}{2018}).
\newblock


\bibitem[Tran et~al\mbox{.}(2024)]%
        {rare}
\bibfield{author}{\bibinfo{person}{Hieu Tran}, \bibinfo{person}{Zonghai Yao}, \bibinfo{person}{Junda Wang}, \bibinfo{person}{Yifan Zhang}, \bibinfo{person}{Zhichao Yang}, {and} \bibinfo{person}{Hong Yu}.} \bibinfo{year}{2024}\natexlab{}.
\newblock \showarticletitle{RARE: Retrieval-Augmented Reasoning Enhancement for Large Language Models}.
\newblock \bibinfo{journal}{\emph{arXiv preprint arXiv:2412.02830}} (\bibinfo{year}{2024}).
\newblock


\bibitem[Trivedi et~al\mbox{.}(2022a)]%
        {IRCoT}
\bibfield{author}{\bibinfo{person}{Harsh Trivedi}, \bibinfo{person}{Niranjan Balasubramanian}, \bibinfo{person}{Tushar Khot}, {and} \bibinfo{person}{Ashish Sabharwal}.} \bibinfo{year}{2022}\natexlab{a}.
\newblock \showarticletitle{Interleaving retrieval with chain-of-thought reasoning for knowledge-intensive multi-step questions}.
\newblock \bibinfo{journal}{\emph{arXiv preprint arXiv:2212.10509}} (\bibinfo{year}{2022}).
\newblock


\bibitem[Trivedi et~al\mbox{.}(2022b)]%
        {musique}
\bibfield{author}{\bibinfo{person}{Harsh Trivedi}, \bibinfo{person}{Niranjan Balasubramanian}, \bibinfo{person}{Tushar Khot}, {and} \bibinfo{person}{Ashish Sabharwal}.} \bibinfo{year}{2022}\natexlab{b}.
\newblock \showarticletitle{MuSiQue: Multihop Questions via Single-hop Question Composition}.
\newblock \bibinfo{journal}{\emph{Transactions of the Association for Computational Linguistics}}  \bibinfo{volume}{10} (\bibinfo{year}{2022}), \bibinfo{pages}{539--554}.
\newblock


\bibitem[Vu et~al\mbox{.}(2023)]%
        {FreshQA}
\bibfield{author}{\bibinfo{person}{Tu Vu}, \bibinfo{person}{Mohit Iyyer}, \bibinfo{person}{Xuezhi Wang}, \bibinfo{person}{Noah Constant}, \bibinfo{person}{Jerry Wei}, \bibinfo{person}{Jason Wei}, \bibinfo{person}{Chris Tar}, \bibinfo{person}{Yun-Hsuan Sung}, \bibinfo{person}{Denny Zhou}, \bibinfo{person}{Quoc Le}, {et~al\mbox{.}}} \bibinfo{year}{2023}\natexlab{}.
\newblock \showarticletitle{Freshllms: Refreshing large language models with search engine augmentation}.
\newblock \bibinfo{journal}{\emph{arXiv preprint arXiv:2310.03214}} (\bibinfo{year}{2023}).
\newblock


\bibitem[Wang et~al\mbox{.}(2025c)]%
        {wang2025don}
\bibfield{author}{\bibinfo{person}{Ante Wang}, \bibinfo{person}{Linfeng Song}, \bibinfo{person}{Ye Tian}, \bibinfo{person}{Dian Yu}, \bibinfo{person}{Haitao Mi}, \bibinfo{person}{Xiangyu Duan}, \bibinfo{person}{Zhaopeng Tu}, \bibinfo{person}{Jinsong Su}, {and} \bibinfo{person}{Dong Yu}.} \bibinfo{year}{2025}\natexlab{c}.
\newblock \showarticletitle{Don't Get Lost in the Trees: Streamlining LLM Reasoning by Overcoming Tree Search Exploration Pitfalls}.
\newblock \bibinfo{journal}{\emph{arXiv preprint arXiv:2502.11183}} (\bibinfo{year}{2025}).
\newblock


\bibitem[Wang et~al\mbox{.}(2025b)]%
        {pike}
\bibfield{author}{\bibinfo{person}{Jinyu Wang}, \bibinfo{person}{Jingjing Fu}, \bibinfo{person}{Rui Wang}, \bibinfo{person}{Lei Song}, {and} \bibinfo{person}{Jiang Bian}.} \bibinfo{year}{2025}\natexlab{b}.
\newblock \showarticletitle{PIKE-RAG: sPecIalized KnowledgE and Rationale Augmented Generation}.
\newblock \bibinfo{journal}{\emph{arXiv preprint arXiv:2501.11551}} (\bibinfo{year}{2025}).
\newblock


\bibitem[Wang et~al\mbox{.}(2025a)]%
        {CoRAG}
\bibfield{author}{\bibinfo{person}{Liang Wang}, \bibinfo{person}{Haonan Chen}, \bibinfo{person}{Nan Yang}, \bibinfo{person}{Xiaolong Huang}, \bibinfo{person}{Zhicheng Dou}, {and} \bibinfo{person}{Furu Wei}.} \bibinfo{year}{2025}\natexlab{a}.
\newblock \showarticletitle{Chain-of-Retrieval Augmented Generation}.
\newblock \bibinfo{journal}{\emph{arXiv preprint arXiv:2501.14342}} (\bibinfo{year}{2025}).
\newblock


\bibitem[Wang et~al\mbox{.}(2024e)]%
        {DeepNote}
\bibfield{author}{\bibinfo{person}{Ruobing Wang}, \bibinfo{person}{Daren Zha}, \bibinfo{person}{Shi Yu}, \bibinfo{person}{Qingfei Zhao}, \bibinfo{person}{Yuxuan Chen}, \bibinfo{person}{Yixuan Wang}, \bibinfo{person}{Shuo Wang}, \bibinfo{person}{Yukun Yan}, \bibinfo{person}{Zhenghao Liu}, \bibinfo{person}{Xu Han}, {et~al\mbox{.}}} \bibinfo{year}{2024}\natexlab{e}.
\newblock \showarticletitle{Retriever-and-Memory: Towards Adaptive Note-Enhanced Retrieval-Augmented Generation}.
\newblock \bibinfo{journal}{\emph{arXiv preprint arXiv:2410.08821}} (\bibinfo{year}{2024}).
\newblock


\bibitem[Wang et~al\mbox{.}(2024b)]%
        {wang2024decoding}
\bibfield{author}{\bibinfo{person}{Siqi Wang}, \bibinfo{person}{Chao Liang}, \bibinfo{person}{Yunfan Gao}, \bibinfo{person}{Yang Liu}, \bibinfo{person}{Jing Li}, {and} \bibinfo{person}{Haofen Wang}.} \bibinfo{year}{2024}\natexlab{b}.
\newblock \showarticletitle{Decoding Urban Industrial Complexity: Enhancing Knowledge-Driven Insights via IndustryScopeGPT}. In \bibinfo{booktitle}{\emph{Proceedings of the 32nd ACM International Conference on Multimedia}}. \bibinfo{pages}{4757--4765}.
\newblock


\bibitem[Wang et~al\mbox{.}(2024c)]%
        {domainrag}
\bibfield{author}{\bibinfo{person}{Shuting Wang}, \bibinfo{person}{Jiongnan Liu}, \bibinfo{person}{Shiren Song}, \bibinfo{person}{Jiehan Cheng}, \bibinfo{person}{Yuqi Fu}, \bibinfo{person}{Peidong Guo}, \bibinfo{person}{Kun Fang}, \bibinfo{person}{Yutao Zhu}, {and} \bibinfo{person}{Zhicheng Dou}.} \bibinfo{year}{2024}\natexlab{c}.
\newblock \showarticletitle{Domainrag: A chinese benchmark for evaluating domain-specific retrieval-augmented generation}.
\newblock \bibinfo{journal}{\emph{arXiv preprint arXiv:2406.05654}} (\bibinfo{year}{2024}).
\newblock


\bibitem[Wang et~al\mbox{.}(2023)]%
        {CMB-Clin}
\bibfield{author}{\bibinfo{person}{Xidong Wang}, \bibinfo{person}{Guiming~Hardy Chen}, \bibinfo{person}{Dingjie Song}, \bibinfo{person}{Zhiyi Zhang}, \bibinfo{person}{Zhihong Chen}, \bibinfo{person}{Qingying Xiao}, \bibinfo{person}{Feng Jiang}, \bibinfo{person}{Jianquan Li}, \bibinfo{person}{Xiang Wan}, \bibinfo{person}{Benyou Wang}, {et~al\mbox{.}}} \bibinfo{year}{2023}\natexlab{}.
\newblock \showarticletitle{Cmb: A comprehensive medical benchmark in chinese}.
\newblock \bibinfo{journal}{\emph{arXiv preprint arXiv:2308.08833}} (\bibinfo{year}{2023}).
\newblock


\bibitem[Wang et~al\mbox{.}(2024d)]%
        {searching}
\bibfield{author}{\bibinfo{person}{Xiaohua Wang}, \bibinfo{person}{Zhenghua Wang}, \bibinfo{person}{Xuan Gao}, \bibinfo{person}{Feiran Zhang}, \bibinfo{person}{Yixin Wu}, \bibinfo{person}{Zhibo Xu}, \bibinfo{person}{Tianyuan Shi}, \bibinfo{person}{Zhengyuan Wang}, \bibinfo{person}{Shizheng Li}, \bibinfo{person}{Qi Qian}, {et~al\mbox{.}}} \bibinfo{year}{2024}\natexlab{d}.
\newblock \showarticletitle{Searching for best practices in retrieval-augmented generation}.
\newblock \bibinfo{journal}{\emph{arXiv preprint arXiv:2407.01219}} (\bibinfo{year}{2024}).
\newblock


\bibitem[Wang et~al\mbox{.}(2024a)]%
        {crafting}
\bibfield{author}{\bibinfo{person}{Zheng Wang}, \bibinfo{person}{Zhongyang Li}, \bibinfo{person}{Zeren Jiang}, \bibinfo{person}{Dandan Tu}, {and} \bibinfo{person}{Wei Shi}.} \bibinfo{year}{2024}\natexlab{a}.
\newblock \showarticletitle{Crafting Personalized Agents through Retrieval-Augmented Generation on Editable Memory Graphs}.
\newblock \bibinfo{journal}{\emph{arXiv preprint arXiv:2409.19401}} (\bibinfo{year}{2024}).
\newblock


\bibitem[Wang et~al\mbox{.}(2025d)]%
        {rare2}
\bibfield{author}{\bibinfo{person}{Zhengren Wang}, \bibinfo{person}{Jiayang Yu}, \bibinfo{person}{Dongsheng Ma}, \bibinfo{person}{Zhe Chen}, \bibinfo{person}{Yu Wang}, \bibinfo{person}{Zhiyu Li}, \bibinfo{person}{Feiyu Xiong}, \bibinfo{person}{Yanfeng Wang}, \bibinfo{person}{Linpeng Tang}, \bibinfo{person}{Wentao Zhang}, {et~al\mbox{.}}} \bibinfo{year}{2025}\natexlab{d}.
\newblock \showarticletitle{RARE: Retrieval-Augmented Reasoning Modeling}.
\newblock \bibinfo{journal}{\emph{arXiv preprint arXiv:2503.23513}} (\bibinfo{year}{2025}).
\newblock


\bibitem[Weng et~al\mbox{.}(2024)]%
        {CycleResearcher}
\bibfield{author}{\bibinfo{person}{Yixuan Weng}, \bibinfo{person}{Minjun Zhu}, \bibinfo{person}{Guangsheng Bao}, \bibinfo{person}{Hongbo Zhang}, \bibinfo{person}{Jindong Wang}, \bibinfo{person}{Yue Zhang}, {and} \bibinfo{person}{Linyi Yang}.} \bibinfo{year}{2024}\natexlab{}.
\newblock \showarticletitle{Cycleresearcher: Improving automated research via automated review}.
\newblock \bibinfo{journal}{\emph{arXiv preprint arXiv:2411.00816}} (\bibinfo{year}{2024}).
\newblock


\bibitem[Wu et~al\mbox{.}(2025b)]%
        {agentic_reasoning}
\bibfield{author}{\bibinfo{person}{Junde Wu}, \bibinfo{person}{Jiayuan Zhu}, {and} \bibinfo{person}{Yuyuan Liu}.} \bibinfo{year}{2025}\natexlab{b}.
\newblock \showarticletitle{Agentic Reasoning: Reasoning LLMs with Tools for the Deep Research}.
\newblock \bibinfo{journal}{\emph{arXiv preprint arXiv:2502.04644}} (\bibinfo{year}{2025}).
\newblock


\bibitem[Wu et~al\mbox{.}(2025a)]%
        {KG-RAR}
\bibfield{author}{\bibinfo{person}{Wenjie Wu}, \bibinfo{person}{Yongcheng Jing}, \bibinfo{person}{Yingjie Wang}, \bibinfo{person}{Wenbin Hu}, {and} \bibinfo{person}{Dacheng Tao}.} \bibinfo{year}{2025}\natexlab{a}.
\newblock \showarticletitle{Graph-augmented reasoning: Evolving step-by-step knowledge graph retrieval for llm reasoning}.
\newblock \bibinfo{journal}{\emph{arXiv preprint arXiv:2503.01642}} (\bibinfo{year}{2025}).
\newblock


\bibitem[Xi et~al\mbox{.}(2025)]%
        {omnithink}
\bibfield{author}{\bibinfo{person}{Zekun Xi}, \bibinfo{person}{Wenbiao Yin}, \bibinfo{person}{Jizhan Fang}, \bibinfo{person}{Jialong Wu}, \bibinfo{person}{Runnan Fang}, \bibinfo{person}{Ningyu Zhang}, \bibinfo{person}{Jiang Yong}, \bibinfo{person}{Pengjun Xie}, \bibinfo{person}{Fei Huang}, {and} \bibinfo{person}{Huajun Chen}.} \bibinfo{year}{2025}\natexlab{}.
\newblock \showarticletitle{OmniThink: Expanding Knowledge Boundaries in Machine Writing through Thinking}.
\newblock \bibinfo{journal}{\emph{arXiv preprint arXiv:2501.09751}} (\bibinfo{year}{2025}).
\newblock


\bibitem[Xiao et~al\mbox{.}(2025)]%
        {RetroRAG}
\bibfield{author}{\bibinfo{person}{Liang Xiao}, \bibinfo{person}{Wen Dai}, \bibinfo{person}{Shuai Chen}, \bibinfo{person}{Bin Qin}, \bibinfo{person}{Chongyang Shi}, \bibinfo{person}{Haopeng Jing}, {and} \bibinfo{person}{Tianyu Guo}.} \bibinfo{year}{2025}\natexlab{}.
\newblock \showarticletitle{Retrieval-Augmented Generation by Evidence Retroactivity in LLMs}.
\newblock \bibinfo{journal}{\emph{arXiv preprint arXiv:2501.05475}} (\bibinfo{year}{2025}).
\newblock


\bibitem[Xiong et~al\mbox{.}(2025b)]%
        {rag-gym}
\bibfield{author}{\bibinfo{person}{Guangzhi Xiong}, \bibinfo{person}{Qiao Jin}, \bibinfo{person}{Xiao Wang}, \bibinfo{person}{Yin Fang}, \bibinfo{person}{Haolin Liu}, \bibinfo{person}{Yifan Yang}, \bibinfo{person}{Fangyuan Chen}, \bibinfo{person}{Zhixing Song}, \bibinfo{person}{Dengyu Wang}, \bibinfo{person}{Minjia Zhang}, {et~al\mbox{.}}} \bibinfo{year}{2025}\natexlab{b}.
\newblock \showarticletitle{Rag-gym: Optimizing reasoning and search agents with process supervision}.
\newblock \bibinfo{journal}{\emph{arXiv preprint arXiv:2502.13957}} (\bibinfo{year}{2025}).
\newblock


\bibitem[Xiong et~al\mbox{.}(2025c)]%
        {mcts-KBQA}
\bibfield{author}{\bibinfo{person}{Guanming Xiong}, \bibinfo{person}{Haochen Li}, {and} \bibinfo{person}{Wen Zhao}.} \bibinfo{year}{2025}\natexlab{c}.
\newblock \showarticletitle{MCTS-KBQA: Monte Carlo Tree Search for Knowledge Base Question Answering}.
\newblock \bibinfo{journal}{\emph{arXiv preprint arXiv:2502.13428}} (\bibinfo{year}{2025}).
\newblock


\bibitem[Xiong et~al\mbox{.}(2025a)]%
        {WriteHere}
\bibfield{author}{\bibinfo{person}{Ruibin Xiong}, \bibinfo{person}{Yimeng Chen}, \bibinfo{person}{Dmitrii Khizbullin}, {and} \bibinfo{person}{J{\"u}rgen Schmidhuber}.} \bibinfo{year}{2025}\natexlab{a}.
\newblock \showarticletitle{Beyond Outlining: Heterogeneous Recursive Planning for Adaptive Long-form Writing with Language Models}.
\newblock \bibinfo{journal}{\emph{arXiv preprint arXiv:2503.08275}} (\bibinfo{year}{2025}).
\newblock


\bibitem[Xu et~al\mbox{.}(2025)]%
        {lrm_survey}
\bibfield{author}{\bibinfo{person}{Fengli Xu}, \bibinfo{person}{Qianyue Hao}, \bibinfo{person}{Zefang Zong}, \bibinfo{person}{Jingwei Wang}, \bibinfo{person}{Yunke Zhang}, \bibinfo{person}{Jingyi Wang}, \bibinfo{person}{Xiaochong Lan}, \bibinfo{person}{Jiahui Gong}, \bibinfo{person}{Tianjian Ouyang}, \bibinfo{person}{Fanjin Meng}, {et~al\mbox{.}}} \bibinfo{year}{2025}\natexlab{}.
\newblock \showarticletitle{Towards Large Reasoning Models: A Survey of Reinforced Reasoning with Large Language Models}.
\newblock \bibinfo{journal}{\emph{arXiv preprint arXiv:2501.09686}} (\bibinfo{year}{2025}).
\newblock


\bibitem[Xu et~al\mbox{.}(2024)]%
        {activerag}
\bibfield{author}{\bibinfo{person}{Zhipeng Xu}, \bibinfo{person}{Zhenghao Liu}, \bibinfo{person}{Yukun Yan}, \bibinfo{person}{Shuo Wang}, \bibinfo{person}{Shi Yu}, \bibinfo{person}{Zheni Zeng}, \bibinfo{person}{Chaojun Xiao}, \bibinfo{person}{Zhiyuan Liu}, \bibinfo{person}{Ge Yu}, {and} \bibinfo{person}{Chenyan Xiong}.} \bibinfo{year}{2024}\natexlab{}.
\newblock \showarticletitle{ActiveRAG: Autonomously Knowledge Assimilation and Accommodation through Retrieval-Augmented Agents}.
\newblock \bibinfo{journal}{\emph{arXiv preprint arXiv:2402.13547}} (\bibinfo{year}{2024}).
\newblock


\bibitem[Yan et~al\mbox{.}(2025)]%
        {o1_embedder}
\bibfield{author}{\bibinfo{person}{Ruiran Yan}, \bibinfo{person}{Zheng Liu}, {and} \bibinfo{person}{Defu Lian}.} \bibinfo{year}{2025}\natexlab{}.
\newblock \showarticletitle{O1 embedder: Let retrievers think before action}.
\newblock \bibinfo{journal}{\emph{arXiv preprint arXiv:2502.07555}} (\bibinfo{year}{2025}).
\newblock


\bibitem[Zhang et~al\mbox{.}(2024)]%
        {HiRAG}
\bibfield{author}{\bibinfo{person}{Xiaoming Zhang}, \bibinfo{person}{Ming Wang}, \bibinfo{person}{Xiaocui Yang}, \bibinfo{person}{Daling Wang}, \bibinfo{person}{Shi Feng}, {and} \bibinfo{person}{Yifei Zhang}.} \bibinfo{year}{2024}\natexlab{}.
\newblock \showarticletitle{Hierarchical Retrieval-Augmented Generation Model with Rethink for Multi-hop Question Answering}.
\newblock \bibinfo{journal}{\emph{arXiv preprint arXiv:2408.11875}} (\bibinfo{year}{2024}).
\newblock


\bibitem[Zhang et~al\mbox{.}(2025)]%
        {levelrag}
\bibfield{author}{\bibinfo{person}{Zhuocheng Zhang}, \bibinfo{person}{Yang Feng}, {and} \bibinfo{person}{Min Zhang}.} \bibinfo{year}{2025}\natexlab{}.
\newblock \showarticletitle{LevelRAG: Enhancing Retrieval-Augmented Generation with Multi-hop Logic Planning over Rewriting Augmented Searchers}.
\newblock \bibinfo{journal}{\emph{arXiv preprint arXiv:2502.18139}} (\bibinfo{year}{2025}).
\newblock


\bibitem[Zhao et~al\mbox{.}(2024)]%
        {TAQA}
\bibfield{author}{\bibinfo{person}{Bowen Zhao}, \bibinfo{person}{Zander Brumbaugh}, \bibinfo{person}{Yizhong Wang}, \bibinfo{person}{Hannaneh Hajishirzi}, {and} \bibinfo{person}{Noah~A Smith}.} \bibinfo{year}{2024}\natexlab{}.
\newblock \showarticletitle{Set the clock: Temporal alignment of pretrained language models}.
\newblock \bibinfo{journal}{\emph{arXiv preprint arXiv:2402.16797}} (\bibinfo{year}{2024}).
\newblock


\bibitem[Zhao et~al\mbox{.}(2025)]%
        {medrag}
\bibfield{author}{\bibinfo{person}{Xuejiao Zhao}, \bibinfo{person}{Siyan Liu}, \bibinfo{person}{Su-Yin Yang}, {and} \bibinfo{person}{Chunyan Miao}.} \bibinfo{year}{2025}\natexlab{}.
\newblock \showarticletitle{MedRAG: Enhancing Retrieval-augmented Generation with Knowledge Graph-Elicited Reasoning for Healthcare Copilot}.
\newblock \bibinfo{journal}{\emph{arXiv preprint arXiv:2502.04413}} (\bibinfo{year}{2025}).
\newblock


\bibitem[Zheng et~al\mbox{.}(2025)]%
        {deepresearcher}
\bibfield{author}{\bibinfo{person}{Yuxiang Zheng}, \bibinfo{person}{Dayuan Fu}, \bibinfo{person}{Xiangkun Hu}, \bibinfo{person}{Xiaojie Cai}, \bibinfo{person}{Lyumanshan Ye}, \bibinfo{person}{Pengrui Lu}, {and} \bibinfo{person}{Pengfei Liu}.} \bibinfo{year}{2025}\natexlab{}.
\newblock \showarticletitle{DeepResearcher: Scaling Deep Research via Reinforcement Learning in Real-world Environments}.
\newblock \bibinfo{journal}{\emph{arXiv preprint arXiv:2504.03160}} (\bibinfo{year}{2025}).
\newblock


\bibitem[Zhong et~al\mbox{.}(2025)]%
        {zhong2025meta}
\bibfield{author}{\bibinfo{person}{Yijie Zhong}, \bibinfo{person}{Feifan Wu}, \bibinfo{person}{Mengying Guo}, \bibinfo{person}{Xiaolian Zhang}, \bibinfo{person}{Meng Wang}, {and} \bibinfo{person}{Haofen Wang}.} \bibinfo{year}{2025}\natexlab{}.
\newblock \showarticletitle{Meta-PKE: Memory-Enhanced Task-Adaptive Personal Knowledge Extraction in Daily Life}.
\newblock \bibinfo{journal}{\emph{Information Processing \& Management}} \bibinfo{volume}{62}, \bibinfo{number}{4} (\bibinfo{year}{2025}), \bibinfo{pages}{104097}.
\newblock


\bibitem[Zhou et~al\mbox{.}(2024)]%
        {metaRAG}
\bibfield{author}{\bibinfo{person}{Yujia Zhou}, \bibinfo{person}{Zheng Liu}, \bibinfo{person}{Jiajie Jin}, \bibinfo{person}{Jian-Yun Nie}, {and} \bibinfo{person}{Zhicheng Dou}.} \bibinfo{year}{2024}\natexlab{}.
\newblock \showarticletitle{Metacognitive retrieval-augmented large language models}. In \bibinfo{booktitle}{\emph{Proceedings of the ACM Web Conference 2024}}. \bibinfo{pages}{1453--1463}.
\newblock


\bibitem[Zhu et~al\mbox{.}(2025b)]%
        {RetrievalPRM}
\bibfield{author}{\bibinfo{person}{Jiachen Zhu}, \bibinfo{person}{Congmin Zheng}, \bibinfo{person}{Jianghao Lin}, \bibinfo{person}{Kounianhua Du}, \bibinfo{person}{Ying Wen}, \bibinfo{person}{Yong Yu}, \bibinfo{person}{Jun Wang}, {and} \bibinfo{person}{Weinan Zhang}.} \bibinfo{year}{2025}\natexlab{b}.
\newblock \showarticletitle{Retrieval-Augmented Process Reward Model for Generalizable Mathematical Reasoning}.
\newblock \bibinfo{journal}{\emph{arXiv preprint arXiv:2502.14361}} (\bibinfo{year}{2025}).
\newblock


\bibitem[Zhu et~al\mbox{.}(2025a)]%
        {ChainRAG}
\bibfield{author}{\bibinfo{person}{Rongzhi Zhu}, \bibinfo{person}{Xiangyu Liu}, \bibinfo{person}{Zequn Sun}, \bibinfo{person}{Yiwei Wang}, {and} \bibinfo{person}{Wei Hu}.} \bibinfo{year}{2025}\natexlab{a}.
\newblock \showarticletitle{Mitigating Lost-in-Retrieval Problems in Retrieval Augmented Multi-Hop Question Answering}.
\newblock \bibinfo{journal}{\emph{arXiv preprint arXiv:2502.14245}} (\bibinfo{year}{2025}).
\newblock


\end{thebibliography}
\appendix
\section*{Appendix}

\subsection*{Agentic RAG Symbol Reference System}
\label{sec:symbols}

The following table presents a complete symbol reference system with formally defined mathematical notations for all core concepts.
\begin{table*}[htbp]
\centering
\caption{Basic states and system components}
\label{tab:basic_states}
\resizebox{0.8\linewidth}{!}{%
\begin{tabular}{@{}lll@{}}
\toprule
\textbf{Symbol} & \textbf{Type} & \textbf{Definition \& Description} \\ 
\midrule
\( S_t = (H_t, C_t) \) & Composite state & Complete system state at timestep \( t \), containing historical information and context vectors \\
\( H_t \) & Vector/Set & Historical information aggregation \\
\( C_t \) & Vector & Contextual embedding vectors \\
\( q_t \) & Vector & Vector representation of current query at step \( t \) \\
\( \mathcal{K}_t \) & Set & Dynamic knowledge base (Initialized as \( \mathcal{K}_0 = \emptyset \)) \\
\bottomrule
\end{tabular}%
}

\end{table*}

\begin{table*}[htbp]
\centering
\caption{Action space and policy definitions}
\label{tab:actions}
\resizebox{0.8\linewidth}{!}{%
\begin{tabular}{@{}lll@{}}
\toprule
\textbf{Symbol} & \textbf{Type} & \textbf{Definition \& Description} \\ 
\midrule
\( \mathcal{A} \) & Set & Action space, e.g., \( \mathcal{A} = \{\text{Retrieve}, \text{Generate}, \text{Verify}, \text{Terminate}\} \) \\
\( a_t \) & Scalar & Selected action at timestep \( t \) (\( a_t \in \mathcal{A} \)) \\
\( \pi(S_t; \Theta) \) & Function & Policy function with parameters \( \Theta \), mapping states to action probability distributions (\( \pi: \mathcal{S} \to \Delta(\mathcal{A}) \)) \\
\bottomrule
\end{tabular}%
}

\end{table*}

\begin{table*}[htbp]
\caption{State transition mechanisms}
\resizebox{0.8\linewidth}{!}{%
\begin{tabular}{@{}lll@{}}
\toprule
\textbf{Symbol} & \textbf{Type} & \textbf{Definition \& Description} \\ 
\midrule
\( \delta \) & Function & \textbf{State transition function}, update rule \( S_{t+1} = \delta(S_t, \cdot) \) \\
\( \mathcal{T}_a \) & Function & Low-level state transition operation for action \( a \) (e.g., \( \mathcal{T}_{\text{Retrieve}} \) denotes retrieval) \\
\( \mathcal{R} \) & Function & \textbf{Retrieval function}, \( \mathcal{R}(S_t) \) returns retrieval results \\
\( \circ \) & Operator & Function composition operator (e.g., \( f \circ g(x) = f(g(x)) \)) \\
\bottomrule
\end{tabular}%
}
\label{tab:transitions}
\end{table*}

\begin{table*}[htbp]
\centering
\caption{Feedback and optimization components}
\label{tab:feedback}
\resizebox{0.7\linewidth}{!}{%
\begin{tabular}{@{}lll@{}}
\toprule
\textbf{Symbol} & \textbf{Type} & \textbf{Definition \& Description} \\ 
\midrule
\( R(S_t, a_t, S_{t+1}) \) & Function & \textbf{Reward function}, outputs reward value \( r_t \) \\
\( \mathbb{I}(\cdot) \) & Function & Indicator function (returns 1 if condition holds, else 0) \\
\( \nabla_\theta J(\theta) \) & Operator & Policy gradient for optimizing policy parameters \( \Theta \) \\
\( \gamma \) & Scalar & Discount factor for cumulative reward calculation \\
\bottomrule
\end{tabular}%
}

\end{table*}

\begin{table*}[htbp]
\centering
\caption{Submodule-specific symbols}
\label{tab:submodules}
\resizebox{0.8\linewidth}{!}{%
\begin{tabular}{@{}lll@{}}
\toprule
\textbf{Symbol} & \textbf{Type} & \textbf{Definition \& Description} \\ 
\midrule
\( \Psi \) & Function & \textbf{Reasoning function}, generates intermediate reasoning results \\
\( \Gamma \) & Function & \textbf{Decision function}, produces final outputs (e.g., answers) \\
\( \psi(\cdot) \) & Function & \textbf{Branch selector} for reflective reasoning path selection \\
\( \phi(\cdot) \) & Function & Confidence mapping function (evaluations to scalar confidence) \\
\( \tau \) & Scalar & Decision threshold for triggering specific operations (e.g., verification/termination) \\
\bottomrule
\end{tabular}%
}

\end{table*}

\subsection*{Symbol Design Hierarchy}
\begin{itemize}
\item \textbf{Base states/actions}: Standard font (\( S_t, a_t \))
\item \textbf{Sets/spaces}: Calligraphic font (\( \mathcal{A}, \mathcal{K}_t \))
\item \textbf{Core mechanism functions}: Uppercase Greek (\( \Psi, \Gamma \))
\item \textbf{Operational functions}: Calligraphic font (\( \mathcal{R}, \mathcal{T}_a \))
\item \textbf{Auxiliary functions}: Lowercase Greek (\( \delta, \phi \)) or blackboard bold (\( \mathbb{I} \))
\end{itemize}

\subsection*{Annotation Guidelines}
\begin{itemize}
\item \textbf{Symbol disambiguation}:
  \begin{itemize}
  \item \( \mathcal{R} \) strictly denotes retrieval function (vs. reward \( R \))
  \item \( \delta \) exclusively represents state transitions (vs. branch selector \( \psi \))
  \end{itemize}
  
\item \textbf{Dynamic extensions}:
  \begin{itemize}
  \item Action space \( \mathcal{A} \) and knowledge base \( \mathcal{K}_t \) support incremental updates:
  \( \mathcal{K}_{t+1} = \mathcal{K}_t \oplus \text{Retrieve}(q_t) \)
  \end{itemize}
\end{itemize}

\end{document}